\documentclass[11pt,a4paper]{elsarticle}
\usepackage{amsmath,amssymb,amsfonts,a4wide,graphicx,bm,times,mathtools} 

\usepackage{tikz}
\usetikzlibrary{calc,decorations.markings}
\usepackage{braket}
 
\usepackage{booktabs}

\def \h {\hbar}
\def \th {\theta}
\def \L {\Lambda}
\def \ve {\varepsilon}

\def \vp {\varphi}
\def \g {\gamma}
\def \CP {\mathcal{P}}

\usepackage{hyperref}
\hypersetup{
    colorlinks,
    citecolor=green,
    filecolor=blue,
    linkcolor=blue,
    urlcolor=purple
}
\hypersetup{linktocpage}
\usepackage{lscape}

\def \ba {\begin{aligned}}
\def \ea {\end{aligned}}

\makeatletter \let\old@startsection=\@startsection
\renewcommand{\@startsection}[6]
{\old@startsection{#1}{#2}{#3}{#4}{#5}{#6\mathversion{bold}}}
\makeatother

\makeatletter

\@addtoreset{equation}{section}
\makeatother

\newcommand{\be}{\begin{equation}}
\newcommand{\ee}{\end{equation}}

 \def\Re{{\rm Re ~}}
 \def\Im{{\rm Im ~}}
 \def\h{\hbar}
  \def\th{\theta}
 \def \L {\Lambda}
 \usepackage[spanish,english]{babel}
\usepackage[utf8]{inputenc}
\usepackage{comment}
\usepackage{xcolor}
\numberwithin{figure}{section}
\numberwithin{table}{section}

\usepackage{changepage}

\begin{document}

\begin{titlepage}
\begin{adjustwidth}{1.25cm}{1.25cm}
\begin{flushright}
\end{flushright}
\vspace{0.5cm}
\begin{center}
{\Large \bf Integrability, susy $SU(2)$ matter gauge theories and black holes}
\lineskip .75em
\vskip 2.5cm
{Davide Fioravanti $^{a}$, Daniele Gregori $^{a}$ and Hongfei Shu $^{b,c}$}
\vskip 2.5em
 {\normalsize\it 
$^{a}$ Sezione INFN di Bologna,\\ Dipartimento di Fisica e Astronomia,
Universit\`a di Bologna} \\
{\normalsize\it Via Irnerio 46, 40126 Bologna, Italy}\\
\texttt{ fioravanti .at. bo.infn.it ,\quad daniele.gregori6 .at. unibo.it}\\
[.4cm]
{\normalsize\it $^{b}$ Beijing Institute of Mathematical Sciences and Applications (BIMSA), Beijing, 101408, China\\
$^{c}$ Yau Mathematical Sciences Center (YMSC), Tsinghua University, Beijing, 100084, China}\\
{\texttt{shuphy124 .at. gmail.com}}

\vskip 4.0em
\textbf{Abstract}\\
\vskip 1.0em
    
\end{center}
    We show that previous correspondence between some (integrable) statistical field theory quantities and periods of $SU(2)$ $\mathcal{N}=2$ deformed gauge theory still holds if we add $N_f=1,2$ flavours of matter. Moreover, the correspondence entails a new non-perturbative solution to the theory. Eventually, we use this solution to give exact results on quasinormal modes of black branes and holes.
    \vfill
   \end{adjustwidth}
\end{titlepage}
 

\tableofcontents

\section{Preliminaries and overview} \label{intro}

From over three decades, many effective low energy $\mathcal{N}=2$ supersymmetric gauge theories have been \textit{solved exactly} by Seiberg-Witten (SW) theory~\cite{SeibergWitten:1994pure,SeibergWitten:1994QCD}. A crucial feature is a weak-strong coupling duality, which allows us to compute the full effective action at any coupling. In practice, this theory prescribes to compute the effective prepotential $\mathcal{F}^{(0)}$ by means of peculiar periods $a^{(0)},a^{(0)}_D$, defined as periods of a differential $\lambda$, living on a elliptic (or hyperelliptic) curve $y_{SW}$. In later developments, a suitable regularisation of the ADHM procedure \cite{ADHM} was devised, to compute instanton contributions to the prepotential $\mathcal{F}$. It requires a deformation of spacetime through two complex parameters $\epsilon_1,\epsilon_2$ (the so-called $\Omega$-background). Then $\mathcal{F}$ can be computed order by order in the instanton (exponential) coupling $\Lambda$, through combinatorial calculus on Young diagrams of the gauge group representations~\cite{Flume:2002az,Nekrasov2003,NekrasovOkounkov:2003,NekrasovOkounkov2006}. In the case of only one deformation with $\epsilon_2=0$, called Nekrasov-Shatashvili (NS) regime~\cite{NekrasovShatashvili:2009}, the SW curve becomes \textit{deformed} or \textit{quantised} as an ordinary differential equation (ODE). In the case of $SU(2)$ gauge group, this is a (time-independent) Schr\"odinger-like equation in which $\epsilon_1=\hbar$ plays the r\^{o}le of Planck constant\footnote{The original \textit{classical} SW elliptic curve is recovered simply as the leading order of the WKB asymptotic expansion as $\hbar \to 0$.}. For instance, for the $SU(2)$ $N_f=0$ flavours (pure) gauge theory, the ODE is the (modified) Mathieu equation~\cite{MironovMorozov:2009,BasarDunne:2015,Kashani-PoorTroost:2015}:
\be \label{ODE0}
- \hbar^2  \frac{d^2}{dy^2} \psi(y) + [\Lambda_0^2 \cosh y + u ] \psi(y) = 0\,,
\ee
where $u$ is the Coulomb branch modulus and $\Lambda_0$ the scale. The r\^ole of the deformed or quantum SW periods may be played by the cycle integrals of the quantum momentum $\mathcal{P}(y)= - i \frac{d}{dy} \ln \psi(y)$~\cite{MironovMorozov:2009}
\begin{equation} \label{qperint}
\begin{pmatrix}a(\hbar,u, \Lambda_0)\\ a_D(\hbar,u, \Lambda_0)
\end{pmatrix}= \oint_{A,B} \mathcal{P}(y,\hbar, u, \Lambda_0)\, d y =2\pi i\sum_{n}\text{Res}\mathcal{P}(y)\biggr |_{y_n^{A,B}}\,\, . 
 \end{equation} 
Even if the quantum SW periods can be defined and computed also in other ways\footnote{As well known, the WKB method gives rise to the $\hbar \to 0$ asymptotic expansion of the quantum periods~\cite{MironovMorozov:2009,ItoKannoOkubo:2017}, which then requires an exact resummation, difficult to perform in practice ({\it cf.}~\cite{GrassiGuMarino} and references therein). Moreover, one can use the above converging series around $\Lambda_0=0$, which however may be very difficult to use at intermediate and large $\Lambda_0$. The interested reader is invited to~\cite{GrassiHaoNeitzke:2021} and references therein for a thorough comparison of different computation methods for the quantum gauge periods. In this paper, we give a new method to interpret and compute the integral definition~\eqref{qperint} ({\it cf.} also the details in appendix~\ref{appProof}).}, in our work starting from definition \eqref{qperint} is essential for showing that they satisfy suitable functional and integral equations. In some cases these can be derived from known integrable Lagrangians, but more generally we can think of them as \emph{defining} integrable structures or integrable models (IMs). This approach is similar in spirit to the $S$-matrix \cite{Mussardo:1992}.

In a nutshell, we analyse the solution to the ODEs, which both physical theories share, and formulate an extension of the so-called \textit{ODE/IM} correpondence~\cite{DoreyTateo1998,BazhanovLukyanovZamolodchikov2001}. In this way, we have found a novel correspondence between $4D$ deformed $\mathcal{N}=2$ supersymmetric Yang-Mills (SYM) and $2D$ IMs, as outlined for the $SU(2)$ $N_f=0$ flavours theory in the work~\cite{FioravantiGregori:2019}. Importantly, we have shown that the quantum gauge periods $a$ and $a_D$ are directly connected to the Baxter's $Q$ and $T$ functions of the self-dual Liouville IM\footnote{Or CFT with central charge $c=25$.}, by the simple relations:
\be  \label{QTcycles0}
Q(\th,p) = \exp  \left\{\frac{2 \pi i }{\hbar}\,a_D(\hbar ,u,\L_0)  \right\} \qquad T(\th,p) = 2 \cos \left\{\frac{2 \pi }{\hbar}a(\hbar,u,\Lambda_0)\right\}\,,
\ee 
with the map between the gauge and integrability parameters:
\be 
\frac{\hbar}{\Lambda_0}=\frac{1}{\sqrt{2}} e^{-\theta} \qquad \frac{u}{\Lambda_0^2} =\frac{1}{2} p^2 e^{-2\theta} \,,
\label{gau-int-map}
\ee 
where $p$ is the Liouville vacuum momentum and $\theta$ the rapidity.
Extending~\cite{DoreyTateo1999}, the $Q,T$ functions above (and in the following) appear in the ODE/IM correspondence as connection coefficients of certain solutions of the ODEs, called sometimes {\it radial} and  {\it lateral}, respectively. Then, they have been proven to satisfy~\eqref{QTcycles0}, as explained in appendix~\ref{appProof}. Besides, they expand asymptotically at $\th \to +\infty$ in terms of the eigenvalues of the local integrals of motion (LIMs) $\mathbb{I}_{2n-1}(p)$, $n=1,2,\dots$ of the self-dual Liouville field theory; then, these key objects of integrability can be resummed\footnote{Under the double limit $\th \to +\infty$, $p \to + \infty$, corresponding to the WKB expansion $\hbar \to 0$, $u/\Lambda_0^2 \neq 0 $.} into the gauge periods~\cite{FioravantiGregori:2019}. Nonetheless, $Q$ and $T$ functions, as well as the $Y$ functions derived from them, satisfy certain exact functional relations, which differ in the integrability or SYM side as a consequence of the map (\ref{gau-int-map})~\cite{FioravantiGregori:2019}. In particular, those for the $Y$s may also be inverted into the Thermodynamic Bethe Ansatz (TBA) non-linear integral equations, which can be concretely solved both on the integrability~\cite{Zamolodchikov:2000,ZamolodchikovMemorial} and on the gauge side~\cite{FioravantiGregori:2019}.\footnote{For some understanding on the \textit{physical origin} of this apparently casual correspondence between ODEs and IMs, besides the SW geometrical motivation deepened here, we refer to our other previous works~\cite{FioravantiRossiShu:2020,FioravantiRossi:2021}. In these actually, we outline the way for deriving the ODE from the quantum system. Actually, without mention to the full integrable structure~\cite{FioravantiRossiShu:2020}, some work on exact TBA equations for periods of other gauge theories has been developed from an arguably more conjectural framework in ~\cite{GaiottoMooreNeitzke:2008,GaiottoMooreNeitzke:2009,GaiottoOpers}. In particular, the latter contains some numerical work on particular cases of ours and is rather inspiring.} It has been soon clear that this gauge-integrability construction holds much more in general, as proposed in~\cite{FioravantiGregori:2019}. Already in the subsequent work~\cite{FioravantiPoghossian:2019}, the same correspondence has been found between the $SU(3)$ colour group gauge theory (with $N_f=0$) and the $A_2$ Toda CFT (central charge $c=98$). In fact this extension has found its completion in the present paper, where we show the gauge-integrability correspondence to hold upon adding $N_f=1$ and $N_f=2$ matter multiplets to the $SU(2)$ colour group.

Furthermore, other very interesting articles opened new research ways, as the very same NS-deformed $\mathcal{N}=2$ $SU(2)$ gauge theories found new applications to black holes (BHs) physics. Specifically for the perturbation theory which models the ringdown (final) phase of BHs merging, it was first found that (Bohr-Sommerfeld like) quantisations conditions on the quantum gauge periods $a_D,a$ provide a new analytic exact characterisation of quasinormal modes (QNMs)\footnote{QNMs are the characteristic frequencies of the gravitational wave signal in ringdown (after merging) phase. } and could be used also to compute them~\cite{AminovGrassiHatsuda:2020,BianchiConsoliGrilloMorales:2021,BonelliIossaLichtigTanzini:2021,BianchiConsoliGrilloMorales:2021b}. Then, exploiting the AGT duality~\cite{AldayGaiottoTachikawa:2010,Gaiotto:2009} between four dimensional $\mathcal{N}=2$ gauge theories and two dimensional Conformal Field Theories (CFTs), also the latter kind of theories found applications to BHs~\cite{BonelliIossaLichtigTanzini:2021}\footnote{These CFTs are different from ours. In fact, we relate to $N_f=0$ gauge theory the $c=25$ self-dual Liouville, rather then the $c \to +\infty$ Liouville as AGT does for the NS limit~\cite{AldayGaiottoTachikawa:2010}. Further investigations on the relation between these two Liouville models would be interesting.}, allowing access to other BHs observables such as the greybody factor and Love numbers\footnote{The greybody factor, or absorption coefficient, is associated to Hawking radiation, while Love numbers describe tidal deformations of BHs.} ~\cite{BonelliIossaLichtigTanzini:2021,BonelliIossaLichtigTanzini:2022,ConsoliFucitoMoralesPoghossian:2022}. From these many other applications and new results followed, for instance: an isospectral simpler equation to the perturbation ODE~\cite{Hatsuda:2020Teukolsky}; improved theoretical proofs of BHs stability~\cite{CasalsCosta:2021}; a simpler interpretation of Chandrasekhar transformation as exchange of gauge mass parameters~\cite{NakajimaLin:2021}; precise determination of the conditions of invariance under (Couch-Torrence) transformations which exchange inner horizon and null infinity~\cite{BianchiDiRusso:2022}; an exact formula for the thermal scalar two-point function in four-dimensional holographic conformal field theories~\cite{DodelsonGrassiIossaLichtigZhiboedov:2022}. We emphasise that the BHs suitable to a study through these new formal methods are also very ``real", that is they can connect to astrophysics and are interesting also for gravitation phenomenology and the search for deviations from General Relativity~\cite{AminovGrassiHatsuda:2020,Mayerson:2020}.\footnote{For instance, if real BHs possessed horizon-scale structure, forbidden by General Relativity (GR) but allowed by modified theories of gravity or String Theory, it would manifest itself as echoes in the gravitational wave signal in the later ringdown phase and would be accessible to future high precision detectors~\cite{LIGOScientific:2016,CardosoPani:2017,BianchiConsoliGrilloMorales:2021}.}

We have been also able to connect integrability to this new research field in our previous work~\cite{FioravantiGregori:2021}. Specifically, still using the ODE/IM correspondence, we related the mathematically precise definition of QNMs~\cite{Nollert:1999} to quantisation conditions on various Baxter's functions. In particular, it turned out that QNMs $\omega_n$ are nothing but the zeros of the Baxter's $Q$ function (\textit{Bethe roots} condition):
\be 
Q(\omega_n) = 0 \qquad \omega_n \propto e^{\theta_n} \qquad n \in \mathbb{N}\,,
\ee 
and can be computed very efficiently with a new method from integrability: the Thermodynamic Bethe Ansatz (TBA). This is a type of nonlinear integral equation (in the rapidity $\theta$) amenable to exact solution (that is, for any $\theta$), which overcomes the limitations of the perturbative gauge theory approach, holding in the $\theta\lesssim 0$ regime. In the present work we aim at generalizing the application of ODE/IM to these topics, by considering also the $N_f=1$ gauge theory, corresponding to a generalization of extremal Reissner-Nordstr\"om (RN) BHs (in the null entropy limit). We give also full details of the derivations and carry out extensive numerical tests. All this contributes to prove and clarify the fundamental heuristic result of~\cite{AminovGrassiHatsuda:2020} on the new gauge-gravity connection, through the further connection we find of the very same $\mathcal N=2$ gauge theories to quantum integrable models\footnote{For other explanations through AGT duality or (conjecturally) M-Theory see~\cite{BonelliIossaLichtigTanzini:2021,BianchiConsoliGrilloMorales:2021b}}. 

This paper is structured as follows. In section~\ref{ODE/IM} we derive the integrability structures for the $SU(2)$ $N_f=1$ and $N_f=2$ theory. In sections~\ref{secY} and~\ref{secT} we connect the gauge periods $a,a_D$ to the integrability $Y$ and $T$ functions. In section~\ref{applications} we show some new results for both gauge theory and integrability, deriving from the above link. In section~\ref{gravity} we make explicit the gravity counterpart of the gauge and integrability theories involved and find other applications. Finally in section~\ref{conclusions} we give some conclusions, point out present limitations of our method and many future possible developments. Several technical appendixes are also added.

\section{ODE/IM correspondence for gauge theory} \label{ODE/IM}

\subsection{Gauge-Integrability dictionary}

The quantum Seiberg-Witten curves for $SU(2)$ $N_f=1,2$ $\mathcal{N}=2$ gauge theory, deformed in the Nekrasov-Shatashvili limit $\epsilon_2 \to 0$, $\epsilon_1 = \hbar \neq 0$, can be constructed from the corresponding classical curves, as explained in appendix~\ref{sec:qSWNf}. Eventually they become the following ODEs: for $N_f=1$ 
\be \label{ODEgau1}
-\hbar^2 \frac{d^2}{dy^2} \psi(y) + \left[ \frac{\Lambda_1^2}{4} (e^{2y}+e^{-y}) + \Lambda_1 m e^y + u\right] \psi(y) = 0\,,
\ee 
and for $N_f=2$ \footnote{With the first realization $N_+=1$, \textit{cf.} appendix~\ref{sec:qSWNf}.}
\be  \label{ODEgau2}
-\hbar^2 \frac{d^2}{dy^2} \psi(y) + \left[ \frac{\Lambda_2^2}{8} \cosh(2y) + \frac{1}{2}\Lambda_2 m_1 e^y + \frac{1}{2}\Lambda_2 m_2 e^{-y} + u\right] \psi(y) = 0\,,
\ee
where $u$ is the moduli parameter, $\Lambda_1,\Lambda_2$ are the instanton coupling parameters, $m,m_1,m_2$ are masses of the flavour hypermultiplets~\cite{ItoKannoOkubo:2017}. We notice that both equations are of the Doubly Confluent Heun kind~\cite{Ronveaux:1995}, that is with two irregular singularities at $y \to \pm \infty$ (\textit{cf.} appendix~\ref{appDCHE}). 

The first physical observation we can make is that \eqref{ODEgau1} and \eqref{ODEgau2} can be mapped into the ODEs for the Integrable Perturbed Hairpin Model (IPHM)~\cite{FateevLukyanov:2005}
\be \label{ODEint1}
-\frac{d^2}{dy^2} \psi(y) + [e^{2\th} (e^{2y}+e^{-y}) +2 e^\theta q  e^y + p^2] \psi(y)  = 0 \,,
\ee 
and its generalization\footnote{In details, \eqref{ODEint1} corresponds to IPHM for positive parameter $n=1$~\cite{FateevLukyanov:2005}, while \eqref{ODEint2} generalizes the IPHM with parameter $n=2$ through the additional parameter $q_2$. Generalizing \eqref{ODEint2} to arbitrary $n$ would require the replacement $ 2 \cosh (2 y) \to e^{2y}+ e^{-2n y}$. This is clearly a possible ODE/IM construction, but it does not have a $\mathcal N=2$ gauge theory counterpart.}
\be \label{ODEint2}
-\frac{d^2}{dy^2} \psi(y) + [2e^{2\th} \cosh (2y) + 2e^\theta q_1 e^y + 2e^\theta q_2 e^{-y}+ p^2] \psi(y)  = 0 \,,
\ee 
where $\th$ is the TBA rapidity, for $N_f=1$ $p,q$ parametrize the Fock vacuum of the IPHM and for $N_f=2$ $p,q_1,q_2$ generalize it. In details, the gauge-integrability parameter dictionary is the following
\be \label{DictGau1}
\frac{\hbar}{\Lambda_1} = \frac{1}{2} e^{-\th} \qquad \frac{u}{\Lambda_1^2} = \frac{1}{4} p^2 e^{-2\th} \qquad \frac{m}{\Lambda_1} = \frac{1}{2} q e^{-\th}\,,
\ee 
\be \label{DictGau2}
\frac{\hbar}{\Lambda_2} = \frac{1}{4} e^{-\th} \qquad \frac{u}{\Lambda_2^2} = \frac{1}{16} p^2 e^{-2\th} \qquad \frac{m_{1,2}}{\Lambda_2} = \frac{1}{4} q_{1,2} e^{-\th}\,,
\ee 
or also
\be 
\frac{u}{\hbar^2} = p^2  \qquad \frac{m}{\hbar} = q\,,
\ee 
\be 
\frac{u}{\hbar^2} = p^2  \qquad \frac{m_1}{\hbar} = q_1 \qquad \frac{m_2}{\hbar} = q_2\,.
\ee
We notice that, on one hand, in~\cite{FateevLukyanov:2005}, $p$ and $q$ were considered fixed; on the other hand, in the gauge theory it is natural to keep $\Lambda_1$, $u$ and $m$ fixed. The mixed dependence on $\th$ then gives a nontrivial map, producing for instance different integrable structures in different parameters.

As a special case, for $q=0$, equation~\eqref{ODEint1} can be related to the ODE (Generalized Mathieu equation) associated to the Integrable Liouville model with coupling $b=\sqrt{2}$~\cite{GrassiHaoNeitzke:2021,FioravantiGregori:2019,ZamolodchikovMemorial}.

\subsection{Integrability functional relations}

The integrability equations \eqref{ODEint1} and \eqref{ODEint2} are invariant under the following discrete symmetries. For $N_f=1$
\be  \label{OmegaSym1}
\ba 
\Omega_+ &: \,\,y \to y +2\pi i/3 \qquad \theta \to \theta+i \pi/3\qquad q \to - q\,, \\
\Omega_-  &: \,\,y \to y -2\pi i/3 \qquad \theta \to \theta+2\pi i /3\qquad q \to  q \,,
\ea 
\ee 
and for $N_f=2$
\be \label{OmegaSym2}
\ba 
\Omega_{+} &: \,\, y \to y+ i \pi/2\qquad \th \to \th+i \pi/2 \qquad q_1 \to - q_1\qquad q_2 \to + q_2\,,  \\
\Omega_{-} &: \,\, y \to y- i \pi /2\qquad \th \to \th+i \pi/2\qquad q_1 \to q_1 \qquad q_2 \to - q_2\,.\
\ea 
\ee 
This symmetry is spontaneously broken by the regular solutions for $y \to \pm \infty$, defined by the asymptotics, for $N_f=1$:
\be 
\ba\label{asyreg1}
\psi_{+,0}(y)&\simeq 2^{-\frac{1}{2}-q} e^{-(\frac{1}{2}+q)\th-\left(\frac{1}{2}+q\right)y-e^{\th+y}} \qquad &y \to + \infty  \\
\psi_{-,0}(y) &\simeq 2^{-\frac{1}{2}} e^{-\frac{1}{2}\theta +\frac{1}{4} y -2e^{\th -y/2} }\qquad &y \to - \infty \,,
\ea
\ee 
and for $N_f=2$:
\be  \label{asyreg2} 
\ba 
\psi_{+,0}(y) &\simeq 2^{-\frac{1}{2}-q_1 }e^{-(\frac{1}{2}+q_1)\th-(\frac{1}{2}+q_1)y}e^{-e^{\th+y}}\, \quad y\to + \infty \,\\
\psi_{-,0}(y) &\simeq 2^{-\frac{1}{2}-q_2}e^{-(\frac{1}{2}+q_2)\th+(\frac{1}{2}+q_2)y}e^{- e^{\th- y}}\quad y\to - \infty\,.
\ea 
\ee 
The solutions $(\psi_{+,0},\psi_{-,0})$ form a basis, of course. However, we can generate other independent solutions by using the symmetries as follows
\be
\psi_{+,k}=\Omega_{+}^{k}\psi_{+} \,, \qquad  \psi_{-,k}=\Omega_{-}^{k}\psi_{-}\, \qquad k \in \mathbb{Z}\,.
\ee
For $k \neq 0$ such solutions are in general diverging, for $y \to \pm \infty$. A basis of solutions is then given also, for instance, by $(\psi_{+,0},\psi_{+,1})$. Importantly, the solutions $\psi_\pm$ are invariant under the symmetry $\Omega_\mp$, respectively: 
\be 
\Omega_+\psi_{-,k}=\psi_{-,k} \qquad \Omega_-\psi_{+,k}=\psi_{+,k}\,.
\ee
We choose the normalization so that we have the following Wronskians for next neighbour $k$-$k+1$ solutions. For $N_f=1$
\be
W[\psi_{+,k+1},\psi_{+,k}]=i e^{(-1)^k i \pi q} \qquad  W[\psi_{-,k+1},\psi_{-,k}]=-i\,\,,
\ee
and for $N_f=2$
\be
W[\psi_{+,k+1},\psi_{+,k}]=i e^{(-1)^k i \pi q_1} \qquad  W[\psi_{-,k+1},\psi_{-,k}]=-i e^{(-1)^k i \pi q_2}\,\,.
\ee

As usual in ODE/IM correspondence, we can define the integrability Baxter's $Q$ function as the Wronskian of the regular solutions at different singular points $y \to \pm \infty$:
\be 
Q = W[\psi_{+,0} ,\psi_{-,0}] \,.
\ee 
Mathematically, this quantity is called also the central connection coefficient, since it appears in the connection relations for solutions at different singular points $y \to \pm \infty$. To write such relations, it is convenient to introduce the notation, for $N_f=1$:
\be \label{Qsignalt1}
Q_\pm(\th) =W[\psi_{+,0} ,\psi_{-,0}](\th,p,\pm q)\,,
\ee 
and for $N_f=2$:
\be  \label{Qsignalt2}
Q_{\pm,\pm}(\th) =W[\psi_{+,0} ,\psi_{-,0}](\th,p,\pm q_1,\pm q_2)\qquad Q_{\pm,\mp}(\th) =W[\psi_{+,0} ,\psi_{-,0}](\th,p,\pm q_1,\mp q_2)\,.
\ee 
We have to expand the solutions $(\psi_{-,0},\psi_{-,1})$ in terms of $(\psi_{+,0},\psi_{+,1})$, with coefficients obtained very simply by taking the Wronskians of both sides of the relations and using the symmetries $\Omega_{\pm}$ to change the parameters of $Q$. Thus we obtain, for $N_f=1$
\begin{align}\label{conn1-+}
i e^{i \pi q}\psi_{-,0}&=Q_-(\theta + i \frac{\pi}{3}) \psi_{+,0}-Q_+(\theta )\psi_{+,1}\\
i e^{i \pi q}\psi_{-,1}&=Q_-(\theta + i \pi ) \psi_{+,0}-Q_+(\theta + i\frac{2 \pi }{3})\psi_{+,1}\,,
\end{align}
and for $N_f=2$
\be 
\ba  \label{conn2-+}
i e^{i \pi q_1}\psi_{-,0}&=Q_{-,+}(\th+i\frac{\pi}{2})\psi_{+,0}- Q_{+,+}(\th)\psi_{+,1} \\
i e^{i \pi q_1}\psi_{-,1}&=Q_{-,-}(\th+i\pi)\psi_{+,0}-Q_{+,-}(\th+i\frac{\pi}{2})\psi_{+,1}\,.
\ea 
\ee
By taking the Wronskian of the first line with the second line (and also shifting $\th$ and flipping the sign of $q$), we obtain the first integrability structure, that is the $QQ$ system. For $N_f=1$
\be  \label{QQ1}
Q_+(\th+i\frac{\pi}{2})Q_-(\th-i\frac{\pi}{2})=e^{-i \pi q}+Q_+(\th-i\frac{\pi}{6})Q_-(\th+i\frac{\pi}{6})\,.
\ee 
and for $N_f=2$
\be  \label{QQ2}
\ba
Q_{+,-}(\th+\frac{i \pi}{2})Q_{-,+}(\th-\frac{i \pi}{2})
= e^{-i\pi (q_1-q_2)} +Q_{-,-}(\th)Q_{+,+}(\th)\,.
\ea
\ee 
For this particular ODEs with two irregular singularities, it is possible to define also an integrability $Y$ function and obtain a $Y$ system relation starting directly from the $Q$ function and $QQ$ system relation\footnote{Rather than from the $T$ functions and $T$ system as in~\cite{DoreyTateo1999}.}. So we define the following $Y$ functions: for $N_f=1$\footnote{We notice that for the $N_f=1$ theory, albeit corresponding to one hypermultiplet less, in the $Y$ function the $Q$ functions appear with different $\th$ arguments and this will lead to several technical complications.}
\be \label{intYdef1}
Y_\pm(\th)=e^{\pm i \pi q}Q_\pm(\th-i\frac{\pi}{6})Q_\mp(\th+i\frac{\pi}{6})\,,
\ee 
and for $N_f=2$
\be  \label{intYdef2}
Y_{+,\pm}(\th) =e^{ i \pi( q_1 \mp q_2)}Q_{+,\pm}(\th)Q_{-,\mp }(\th)\, \qquad Y_{-,\pm}(\th) =e^{ i \pi( -q_1 \mp q_2)}Q_{-,\pm}(\th)Q_{+,\mp }(\th)\,.
\ee
Equivalent definitions are obtained by the $QQ$ systems as, for $N_f=1$:
\be  \label{QQY1}
e^{\pm i \pi q}Q_\pm(\th+i\frac{\pi}{2})Q_\mp(\th-i\frac{\pi}{2}) = 1 + Y_\pm(\th)\,,
\ee 
and for $N_f=2$:
\be \label{QQY2}
e^{i\pi (q_1-q_2)} Q_{+,-}(\th+\frac{i \pi}{2})Q_{-,+}(\th-\frac{i \pi}{2})
= 1+Y_{+,+}(\th)\,.
\ee
The $Y$ systems can be now obtained by taking a product of the $QQ$ system with itself with suitable parameters so to obtain a close relation in terms of $Y$ functions. For $N_f=1$
\be 
\label{int-Ysys1}
Y_\pm(\th+i\frac{\pi}{2})Y_\mp(\th-i\frac{\pi}{2})=\left[1+Y_\mp(\th+i\frac{\pi}{6})\right]\left[1+Y_\pm(\th-i\frac{\pi}{6})\right] \,,
\ee  
and for $N_f=2$
\be 
\ba \label{Ysyst2}
Y_{+,-}(\th+\frac{i \pi}{2} )Y_{-,+}(\th-\frac{i \pi}{2} ) = [1 + Y_{+,+}(\th)][1 + Y_{-,-}(\th)]\,.
\ea
\ee 
Now, the presence of the irregular singularities of ODEs \eqref{ODEint1}-\eqref{ODEint2} at $y \to + \infty$ (Stokes phenomenon) plays a r\^ole for defining the $T$ functions, for $N_f=1$  
\be  \label{Tdef1}
T_{+}(\th) = - i W[\psi_{-,-1},\psi_{-,1}]\,,\qquad \,\, \tilde{T}_{+}(\th) =  i W[\psi_{+,-1},\psi_{+,1}]\,.
\ee 
and for $N_f=2$
\be  \label{Tdef2}
T_{+,+}(\th) = - i W[\psi_{-,-1},\psi_{-,1}]\,,\qquad \,\, \tilde{T}_{+,+}(\th) =  i W[\psi_{+,-1},\psi_{+,1}]\,.
\ee 
(with of course $T_-$ $T_{\mp,\pm}$ defined with the flipped masses as in \eqref{Qsignalt1},~\eqref{Qsignalt2}.)
By expanding $\psi_{\pm,1}$ in terms of $\psi_{\pm,0}$, $\psi_{\pm,-1}$, for $N_f=1$
\be \label{connLatNf1}
 \psi_{+,1} = - e^{2i \pi q}\psi_{+,-1} + e^{i \pi q} \tilde{T}_{+,+}(\th)  \psi_{+,0} \qquad \psi_{-,1} = - \psi_{-,-1} +T_{+,+}(\th) \psi_{-,0}\,,
\ee 
or for $N_f=2$
\be  	\label{connLatNf2}
\psi_{+,1} = -e^{2 i \pi q_1} \psi_{+,-1} + e^{i \pi q_1} \tilde{T}_{+,+}(\th) \psi_{+,0} \qquad \psi_{-,1} = -e^{2i \pi q_2} \psi_{-,-1} + T_{+,+}(\th) e^{i \pi q_2}\psi_{-,0}\,,
\ee 
we obtain the $TQ$ relations, for $N_f=1$
\be  \label{TQ1}
\ba
T_{\pm}(\th) Q_\pm(\th) &= Q_\pm(\th-i\frac{2\pi }{3})+Q_\pm(\th+i \frac{2\pi }{3})\,\\
\tilde{T}_{\pm}(\th) Q_{\pm}(\th) &= e^{\pm i \pi q_1} Q_\mp(\th-\frac{i \pi}{3})+e^{\mp i \pi q_1} Q_\mp(\th+\frac{i \pi}{3})\,,
\ea
\ee 
or for $N_f=2$
\be 
\ba \label{TQ2}
T_{+,+}(\th) Q_{+,+}(\th) &= e^{i \pi q_2} Q_{+,-}(\th-\frac{i \pi}{2})+e^{-i \pi q_2} Q_{+,-}(\th+\frac{i \pi}{2})\\
\tilde{T}_{+,+}(\th) Q_{+,+}(\th) &= e^{i \pi q_1} Q_{-,+}(\th-\frac{i \pi}{2})+e^{-i \pi q_1} Q_{-,+}(\th+\frac{i \pi}{2})\,.
\ea
\ee  
By applying the $\Omega_+$ and $\Omega_-$ symmetries to the $T$ and $\tilde{T}$ functions it is immediate to obtain also the periodicity relations, for $N_f=1$
\be  \label{Tper1}
T_\pm(\th+  i\frac{\pi}{3})=T_\mp (\th)\qquad \tilde{T}_\pm(\th+ i \frac{2\pi }{3})=\tilde{T}_\pm (\th)\,,
\ee
and for $N_f=2$
\be  \label{Tper2}
T_{-,+}(\th+  i\frac{\pi}{2})=T_{+,+} (\th)\qquad \tilde{T}_{+,-}(\th+ i \frac{\pi }{2})=\tilde{T}_{+,+}(\th)\,.
\ee

\subsection{$Q$ function's exact expressions and asymptotic expansion}\label{sec:QIntExact}

From the previous ODE/IM analysis, namely equations~\eqref{conn1-+}-\eqref{conn2-+}, we find a limit formula Baxter's $Q$ function as, for $N_f=1$
\be \label{Qdefpsi}
Q_+(\th)=-i e^{i\pi q}\lim_{y\to + \infty}\frac{\psi_{-,0}(y,\th)}{\psi_{+,1}(y,\th)}\,,
\ee
or for $N_f=2$
\be
Q_{+,+}(\th)=-i e^{i\pi q_1}\lim_{y\to + \infty}\frac{\psi_{-,0}(y,\th)}{\psi_{+,1}(y,\th)}\,.
\ee
From these limit formulae we can obtain other ones as integrals, which allow to concretely compute $Q$. However, to do that, it is convenient first to transform the second order linear ODEs~\eqref{ODEint1}-\eqref{ODEint2} for $\psi$ into their equivalent first order nonlinear Riccati equations, for the logarithmic derivative of $\psi$. Besides, since we will later need to asymptotically expand the solution for $y \to \pm \infty$ and $\th \to \infty$, it is convenient to change variable so to single out the leading order behaviour in $y,\th$ and simplify higher orders calculations. So, we change independent variables as 
\be 
d w = \sqrt{\phi} \,d y \qquad \phi = \begin{cases} 
-e^{2y}-e^{-y} \qquad N_f=1\\
-2 \cosh(2y) \qquad N_f=2
\end{cases}\,.
\ee 
To keep the ODE in normal form we have to let $\psi \to \sqrt[4]{\phi} \psi$. Then we take the logarithmic derivative of $\psi$ in the new variable $w$
\be 
\Pi = - i\frac{d}{dw} \ln( \sqrt[4]{\phi} \psi)\,,
\ee 
and we get for it the Riccati equation
\be  \label{eqRiccatiGen}
\Pi(y)^2 - i \frac{1}{\sqrt{\phi}}\frac{d}{dy}\Pi(y) =e^{2\th}-e^\th V(y) -U(y) \,,
\ee
with 
\be 
\ba
V(y) &= \begin{cases}
-\frac{ 2 q   e^y}{e^{-y}+e^{2 y}} \qquad &N_f=1 \\
-\frac{q_1 e^y +q_2 e^{-y} }{\cosh(2y)}\qquad &N_f=2\,,
\end{cases}\\
U(y) &= \begin{cases}
-\frac{p^2}{e^{-y}+e^{2 y}}+\frac{e^y-40 e^{4 y}+4 e^{7 y}}{16 \left(e^{3 y}+1\right)^3} \qquad &N_f=1\\
\frac{1}{2\cosh(2y)}\left[ -p^2-1+\frac{5}{4} \tanh ^2(2 y)\right]\qquad &N_f=2\,.
\end{cases}
\ea
\ee  
The first asymptotic expansion we make is the one for $y \to \pm\infty$, in the formal parameter $e^{\mp y}$. The Riccati equation gets approximated, at the leading and subleading order as
\be 
\Pi(y)^2 - i \frac{1}{\sqrt{\phi}}\frac{d}{dy}\Pi(y)  \simeq \begin{cases}
e^{2\th}+2 e^\th \delta_+ q e^{-y} \qquad &N_f=1\\
e^{2\th}+2e^\th q_{1,2} e^{\mp y} \qquad &N_f=2 
\end{cases} \qquad y \to \pm \infty \,,
\ee 
where for $N_f=1$ $\delta_+=1$ for $y \to +\infty$, $\delta_+=0$ for $y \to -\infty$.
Then the solution is asymptotic to
\be 
\Pi(y)\simeq \begin{cases}
 e^\th + \delta_+ q e^{-y} \qquad &N_f=1\\
e^\th +  q_{1,2} e^{\mp y} \qquad &N_f=2 
\end{cases} \qquad y \to \pm \infty\,.
\ee
This leading expansion for $y \to \pm \infty$ allows us to fix the regularization in the integrals formulas we now write for the (logarithm) of $\psi_{-,0}$, for $N_f=1$
\be 
\ba \label{psi-0exact1}
\psi_{-,0}(y)&=\frac{2^{-\frac{1}{2}}e^{-\frac{1}{2}\th}}{\sqrt[4]{e^{2y}+e^{-y}}}\exp \left \{- e^\th (2e^{-y/2}-e^y)+2q  \ln (1+e^{y/2}) \right \}\times
 \\
 &\exp\left \{ \int_{-\infty}^y d y'\, \left[ \sqrt{e^{2y'}+e^{-y'}} \Pi(y',\th,p,q)-e^\th \left(e^{y'}+e^{-y'/2}\right)-q  \frac{1}{1+e^{-y'/2}} \right] \right\}\,,
\ea
\ee 
and for $N_f=2$ 
\be 
\ba 
\psi_{-,0}(y)&=\frac{2^{-\frac{1}{2}-q_2}e^{-(\frac{1}{2}+q_2)\th}}{\sqrt[4]{e^{2y}+e^{-2y}}}\exp \left \{-  e^\th(e^{-y}-e^y)+2q_1  \ln (1+e^{y/2})-2q_2   \ln (1+e^{-y/2})] \right \}\times
 \\
 &\exp\left \{ \int_{-\infty}^y d y'\, \left[ \sqrt{e^{2y'}+e^{-2y'}} \Pi(y',\th,p,q_1,q_2)- e^\th (e^{y'}+e^{-y'})-q_1 \frac{1}{1+e^{-y'/2}}-q_2 \frac{1}{1+e^{y'/2}}  \right] \right\}\,.
\ea
\ee
Then from the limit formula for $Q$ we get also an integral expression for it, for $N_f=1$
\be 
\ba \label{lnQPi}
\ln Q_+(\th) &=  \int_{-\infty}^\infty  d y\,\left[\sqrt{e^{2y}+e^{-y} }\Pi(y,\th,q,p)-e^\th e^y - e^\th e^{-y/2} -q  \frac{1}{1+e^{-y/2}} \right]  -\left(\th + \ln 2\right)q\,,
\ea 
\ee 
and for $N_f=2$
\begin{align}  \label{Q2++}\begin{split}
\ln Q_{+,+}(\th) &=\int_{-\infty}^\infty d y\biggl[\sqrt{2 \cosh (2y) } \Pi(y,\th,q_1,q_2,p)- 2e^{\th}\cosh y -  \left(\frac{q_1}{1+e^{-y/2}}+\frac{q_2}{1+e^{y/2}} \right) \biggr]\\
&-\left(\th+ \ln 2 \right) (q_1+q_2)\,.
\end{split}
\end{align}
To get the vacuum eigenvalues of the local integrals of motion (LIMs), we make instead the $\th\to +\infty$ asymptotic expansion (denoted by $\doteq$), at all orders
\be 
\Pi(y,\th) \doteq e^\th+\sum_{n=0}^\infty \Pi_n(y) e^{-n \th} \qquad \th \to +\infty\,.
\ee 
Its coefficients $\Pi_n$ satisfy the recursion relation
\be  \label{WKBRecInt}
\Pi_{n+1} = \frac{1}{2} \left(\frac{i}{\sqrt{\phi}} \frac{d}{dy}\Pi_n -\sum_{m=0}^n \Pi_m \Pi_{n-m} \right) \qquad n \geq 1\,,
\ee 
with initial conditions
\be 
\ba 
\Pi_0 &=-\frac{1}{2} V\\
\Pi_1 &= \frac{1}{2} \left(\frac{i}{\sqrt{\phi}} \frac{d}{dy}\Pi_0 -\Pi_0^2-U\right) \,.
\ea 
\ee
The expansion of $\ln Q$ in terms of the LIMs is, for $N_f=1$
\be \label{asyQLIM1}
\ln Q_+(\th) \doteq -\frac{4   \sqrt{3\pi^{3}}}{\Gamma \left(\frac{1}{6}\right) \Gamma \left(\frac{1}{3}\right)} e^\th-( \th + \frac{1}{3} \ln 2) q - \sum_{n=1}^\infty e^{-n \th}c_n \mathbb{I}_n  \qquad \th \to +\infty\,,
\ee 
and for $N_f=2$
\be \label{asyQLIM2}
\ln Q_{+,+}(\th) \doteq - \frac{4\sqrt{\pi^3 }  }{\Gamma \left(\frac{1}{4}\right)^2} e^\th-( \th + \frac{1}{2} \ln 2) (q_1+q_2) -\sum_{n=1}^\infty e^{-n \th} c_n \mathbb{I}_n \qquad \th \to +\infty\,,
\ee 
with the local integrals of motion $\mathbb{I}_n$ times some normalization constants $c_n$ given by the integrals
\be 
c_n \mathbb{I}_n(p,q) =- i \int_{-\infty}^\infty dy\,\sqrt{\phi(y)} \Pi_n(y,p,q)  \qquad n \geq 1\,.
\ee 
$\mathbb{I}_n(p,q)$ are in general polynomials  in $p,q$, where $q$ of course here stands for either $q$ for $N_f=1$ or $(q_1,q_2)$ for $N_f=2$. We checked the first ones for $N_f=1$ match those of IPHM~\cite{FateevLukyanov:2005}:
\be 
\ba 
\mathbb{I}_1(p,q)&=\frac{1}{12}  \left(4 q^2-12 p^2-1\right)\\
	\mathbb{I}_2(p,q)&= \frac{1}{6\sqrt{3}}q\left(\frac{20}{3} q^2-12 p^2-3\right)\,.
\ea
\ee

\subsection{Integrability TBAs} \label{subsecintTBA}

Let us define as usual the pseudoenergy $\ve(\th)=-\ln Y(\th)$ and $L = \ln [1+\exp (-\ve)]$ (with suitable subscripts omitted of course). 
Using the analytic properties of pseudoenergy $\epsilon$, we can transform the $Y$ system (\ref{int-Ysys1}) into the following ``integrability TBAs". For $N_f=1$~\cite{FateevLukyanov:2005}
\begin{align} \label{int-TBA}
\begin{split}
\ve_+(\th) &= \frac{12\sqrt{ \pi ^{3}}}{\Gamma \left(\frac{1}{6}\right) \Gamma \left(\frac{1}{3}\right)}e^\theta - \frac{4}{3} i \pi q -  (\vp_{+ +}\ast L_+)(\theta) - (\vp_{+ -} \ast L_-)(\theta)\\
\ve_-(\th) &= \frac{12\sqrt{\pi ^{3}}}{\Gamma \left(\frac{1}{6}\right) \Gamma \left(\frac{1}{3}\right)}e^\theta + \frac{4}{3} i \pi q -  (\vp_{+ +}\ast L_-)(\theta) - (\vp_{+ -} \ast L_+)(\theta)\,,
\end{split}
\end{align}
and for $N_f=2$
\be
\ba 
\varepsilon_{+,+}(\th) &= \frac{8 \sqrt{\pi^3 }  }{\Gamma \left(\frac{1}{4}\right)^2}e^{\th}-i \pi (q_1-q_2)- \varphi \ast (L_{+ -}+ L_{- +}) \\
\varepsilon_{+,-}(\th) &= \frac{8 \sqrt{\pi^3 }  }{\Gamma \left(\frac{1}{4}\right)^2}e^{\th}-i \pi (q_1+q_2)- \varphi \ast (L_{+ +}+ L_{- -}) \\
\varepsilon_{-,+}(\th) &=\frac{8 \sqrt{\pi^3 }  }{\Gamma \left(\frac{1}{4}\right)^2}e^{\th}+i \pi (q_1+q_2)- \varphi \ast (L_{- -}+ L_{+ +}) \\
\varepsilon_{-,-}(\th) &= \frac{8 \sqrt{\pi^3 }  }{\Gamma \left(\frac{1}{4}\right)^2}e^{\th}+i \pi (q_1-q_2)- \varphi \ast (L_{ -+}+ L_{+- }) \,.\\
\ea  \label{int-TBA2}
\ee 
The leading (or driving) term follows directly from the expansions~\eqref{asyQLIM1}-\eqref{asyQLIM2} under the definitions for $Y= \exp( - \varepsilon)$ \eqref{intYdef1}-\eqref{intYdef2}. The symbol $\ast$ stands for the $(-\infty,+\infty)$ convolution, which for general functions $f,g$ is
\be 
(f \ast g)(\th) = \int_{-\infty}^\infty  \frac{d\th'}{2\pi}\,f(\th-\th') g(\th')\,.
\ee
The kernel for $N_f=2$ is the simple usual hyperbolic secant~\cite{Zamolodchikov:1991}
\be  \label{kernelNf2}
\varphi(\th) =   \frac{1}{\cosh \th}\,,
\ee
while the one for $N_f=1$ is slightly more involved (as a consequence of the shifts in $\th$ also on the RHS of the $Y$ system~\eqref{int-Ysys1}) but can be obtained by taking Fourier transform as explained in~\cite{FabbriFioravantiPiscagliaTateo:2013}
\be 
\vp_{+ \pm}(\th) =  \frac{\sqrt{3}}{2 \cosh \th\pm 1} \,. 
\ee 

We notice that $q,q_1,q_2$ enter the integrability TBAs as chemical potentials~\cite{KlassenMelzer2:1990}. 
In these TBAs the parameter $p$ does not appear, but it enters in the boundary condition for the solution $\varepsilon$ at $\th \to -\infty$, for $N_f=1$ 
\be  \label{bdyint1}
\ve_+(\th,p)\simeq 6 p \th - i \pi q-2 C_1(p,q) \qquad \th \to -\infty \,,
\ee 
and for $N_f=2$
\be   \label{bdyint2}
 \varepsilon_{+,+}(\th,p) \simeq 4 p \th -i \pi (q_1-q_2) -2 C_2(p,q_1,q_2) \qquad \th \to -\infty\,,
 \ee 
with 
\begin{align}
 C_1(p,q) &=\ln \left[\frac{2^{ -p}\Gamma(2  p) \Gamma(1 + 2  p)}{\sqrt{2\pi}\sqrt{\Gamma(\frac{1}{2}+ p + q)\Gamma(\frac{1}{2}+ p -q)}}\right]   \label{Cint1}\\
C_2(p,q_1,q_2) &=\ln\left[\frac{2^{1-2 p} p \,\Gamma (2 p)^2}{\sqrt{\Gamma \left(p+\frac{1}{2}-q_1\right)\Gamma \left(p+\frac{1}{2}+q_1\right)\Gamma \left(p+\frac{1}{2}-q_2\right)\Gamma \left(p+\frac{1}{2}+q_2\right)}}\right]  \label{Cint2} \,.
\end{align}
The derivation of these boundary conditions from the $\theta \to - \infty$ perturbative solution of ODEs \eqref{ODEint1}, \eqref{ODEint2} is explained in appendix \ref{appForcing}. So, to fix $p$ and reproduce the boundary conditions~\eqref{bdyint1}-\eqref{bdyint2} as $\th \to - \infty$, we have to add and subtract outside and inside the convolutions suitable auxiliary functions $f_{N_f}(\theta)$, $F_{N_f}(\theta)$ defined for $\theta \in \mathbb{R}$\footnote{It turns out that the constant term $i \pi q$ in the boundary condition~\eqref{bdyint1} needs not this kind of engineering, as it is automatically produced by the complex valued convolution.}. In details, for $N_f=1$ the fully explicit integrability TBA reads
\be
\ba \label{int-TBA-full-1}
\varepsilon_+(\theta ) &=\frac{12 \sqrt{\pi ^{3}}}{\Gamma \left(\frac{1}{6}\right) \Gamma \left(\frac{1}{3}\right)} e^\theta - \frac{4}{3} i \pi q  + f_1(\th)- (\vp_{++} \ast (L_+ +   F_1))(\th)- (\vp_{+-} \ast (L_{-} +   F_1))(\th)  \\
\varepsilon_-(\theta ) &=\frac{12\sqrt{\pi ^{3}}}{\Gamma \left(\frac{1}{6}\right) \Gamma \left(\frac{1}{3}\right)} e^\theta + \frac{4}{3} i \pi q + f_1(\th)- (\vp_{++} \ast (L_-  +F_1))(\th)- (\vp_{+-} \ast (L_+  + F_1))(\th)\,,
\ea
\ee
where the new terms are obtained as just explained and explicitly read
\be 
\ba \label{int-TBA-full-2}
F_1&(\th) =- 3 p  \ln \left[1+e^{-2 \th} \right] -C_1(1- \tanh \th)\,,\\
f_1& (\th)= (\varphi_{++}+\varphi_{+-} ) \ast F_1 \\
=&- 3 p \left \{\ln \left[1+e^{-(\th+i  \pi/6)}\right]  +\ln \left[1+e^{-(\th-i  \pi/6)}\right]   \right \}-C_1 \left[1- \frac{1}{2}\tanh \left( \frac{\th}{2} + \frac{i \pi }{12} \right)-  \frac{1}{2}\tanh \left( \frac{\th}{2} - \frac{i \pi }{12} \right)\right]\,.
\ea 
\ee 
\begin{figure}
\centering
\includegraphics[width=0.48\textwidth]{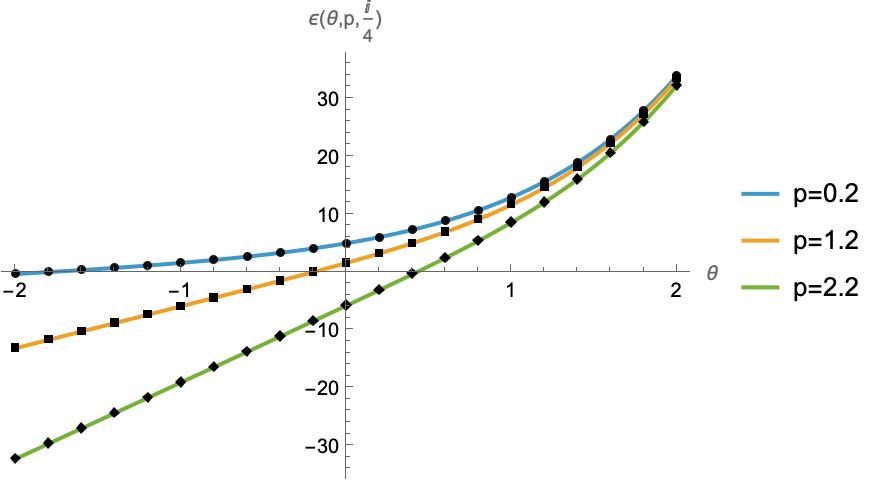}\quad\includegraphics[width=0.48\textwidth]{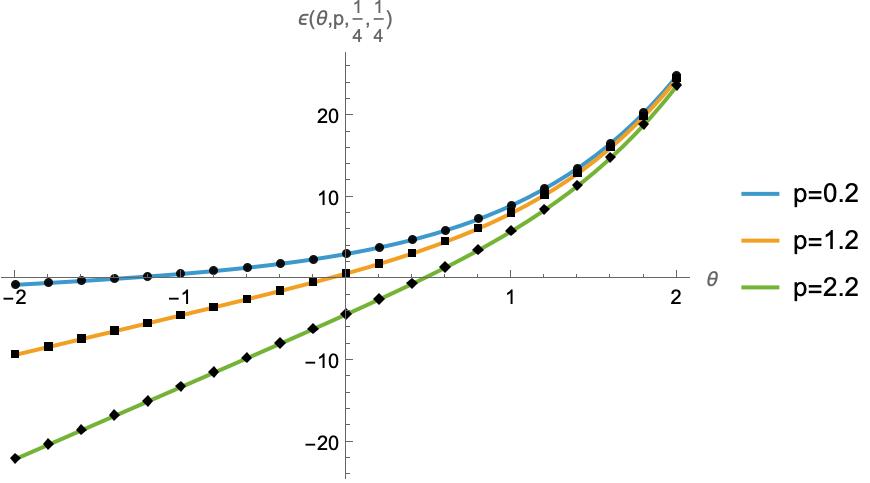}
\caption{Plot of the solution $\varepsilon(\theta) $ of the integrability TBAs \eqref{int-TBA-full-1}, \eqref{int-TBA-full-2}, (the colored continuous curves) vs the Riccati numeric solution as in \eqref{lnQPi}, \eqref{Q2++} (the black dots).}
\label{plotRiccatiTBAHp}
\end{figure}
We notice that boundary condition~\eqref{bdyint1} requires strictly $p>0$, which in gauge theory corresponds to $u/\Lambda_{1,2}^2>0$ by~\eqref{DictGau1}. However, we shall see that we can solve the TBA in gauge variables for $u/\Lambda_{1,2}^2 \in \mathbb{C}$ (small), thus providing an analytic continuation of the integrability TBA.
Similarly for $N_f=2$ the corresponding auxiliary functions are
\be 
\ba
F_2(\th) &= -2 p  \ln \left[1+e^{-2 \th} \right]  -C_2(1- \tanh \th)\,,  \\
f_2(\th)&=2\varphi  \ast F_2 = -4 p  \ln \left[1+e^{-\th}\right]   -C_2 \left[1-  \tanh \left( \frac{\th}{2}  \right)\right]\,,
\ea 
\ee 
and the fully explicit integrability TBA reads
\be
\ba
\varepsilon_{\pm,\pm}(\th) &= \frac{8 \sqrt{\pi^3 }  }{\Gamma \left(\frac{1}{4}\right)^2}e^{\th}+f_2(\theta)\mp i \pi (q_1-q_2)- (\varphi \ast (L_{+ -}+ L_{- +}+2 F_2))(\theta) \\
\varepsilon_{\pm,\mp}(\th) &= \frac{8 \sqrt{\pi^3 }  }{\Gamma \left(\frac{1}{4}\right)^2}e^{\th}+f_2(\theta)\mp i \pi (q_1+q_2)- (\varphi \ast (L_{+ +}+ L_{- -}+2 F_2) )(\theta)\,.
\ea
\ee
We notice that~\eqref{int-TBA2} generalizes the TBA found in  \cite{FateevLukyanov:2005} for the Perturbed Hairpin IM and therefore we shall call the corresponding IM as \emph{Generalized Perturbed Hairpin}.

\begin{figure}
\centering
\includegraphics[width=0.48\textwidth]{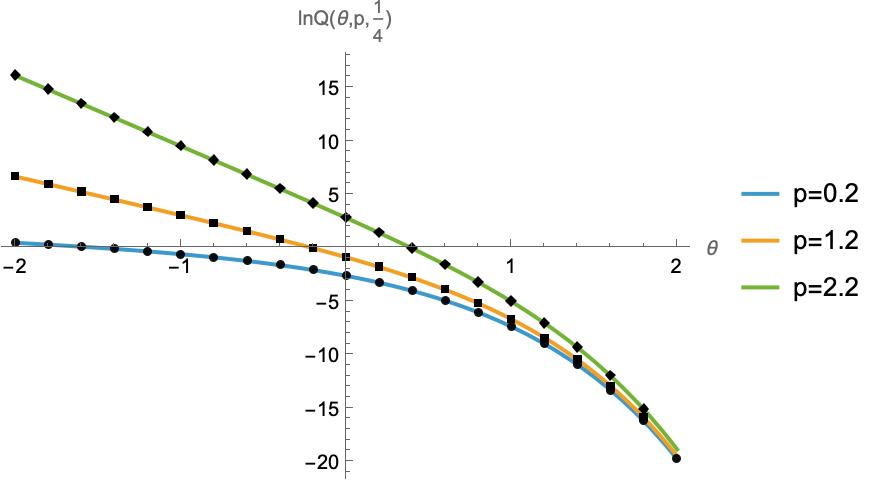}\quad\includegraphics[width=0.48\textwidth]{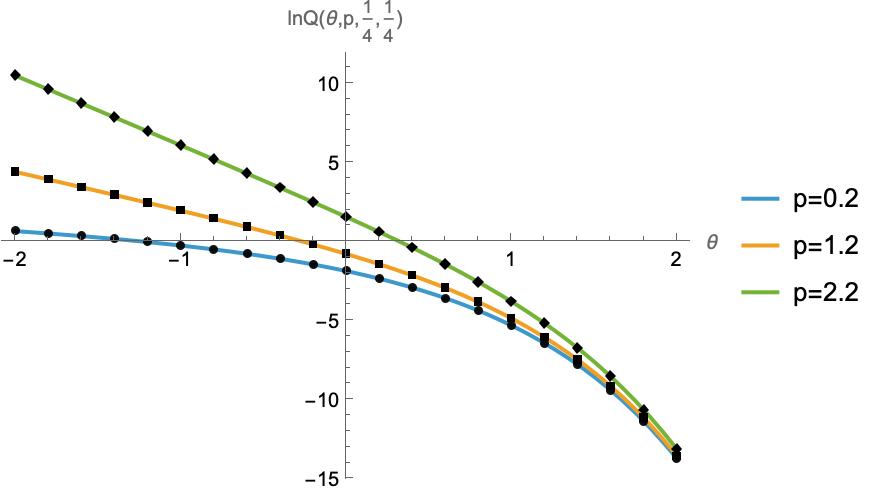}
\caption{Plot of $\ln Q(\theta)$, using the TBA solution as in \eqref{TBAQ1}, \eqref{TBAQ2} (the colored continuous curves) vs the Riccati numeric solution as in \eqref{lnQPi}, \eqref{Q2++} (the black dots).}
\label{plotRiccatiTBAGhp}
\end{figure}

Now from the TBA solution we can obtain also $Q$ as follows. Taking products and ratios of the $QQ$ system~\eqref{QQY1}
\be 
\ba 
[Q_+(\th+i \pi/2)Q_-(\th+i\pi/2)][Q_+(\th-i \pi/2)Q_-(\th-i\pi/2)] &=[1+Y_+(\th)][1+Y_-(\th)]\\
\left[\frac{Q_+(\th+i \pi/2)}{Q_-(\th+i\pi/2)} \right]\left[\frac{Q_+(\th-i \pi/2)}{Q_-(\th-i\pi/2)} \right]^{-1} &=e^{-2\pi i q}\frac{1+Y_+(\th)}{1+Y_-(\th)}\,,
\ea
\ee
we easily deduce for $N_f=1$ the following integral expression for $Q$  
\be
\ba \label{TBAQ1}
\ln Q_\pm(\th) &= -\frac{4   \sqrt{3\pi^{3}}}{\Gamma \left(\frac{1}{6}\right) \Gamma \left(\frac{1}{3}\right)} e^\th\mp( \th + \frac{1}{3} \ln 2) q \\
&+\frac{1}{2} \int_{-\infty}^\infty \frac{d\th'}{2\pi}\left \{ \frac{\ln[1+\exp \{-\varepsilon_+(\th')   \} ][1+\exp \{-\varepsilon_-(\th')  \} ]}{\cosh(\th-\th')}\mp i \frac{e^{\th'-\th}}{\cosh(\th-\th')} \ln\left[\frac{1+\exp \{-\varepsilon_-(\th')  \} }{1+\exp \{-\varepsilon_+(\th') \} }\right]\right \} \,.
\ea
\ee
A similar derivation for $N_f=2$ leads to
\be
\ba \label{TBAQ2}
&\ln Q_{\pm,\pm}(\th) = -\frac{4   \sqrt{\pi^{3}}}{\Gamma \left(\frac{1}{4}\right)^2} e^\th\mp( \th + \frac{1}{2} \ln 2) (q_1+q_2) \\
&+\frac{1}{2} \int_{-\infty}^\infty \frac{d\th'}{2\pi}\left \{ \frac{\ln[1+\exp \{-\varepsilon_{+,-}(\th')   \} ][1+\exp \{-\varepsilon_{-,+}(\th')  \} ]}{\cosh(\th-\th')}\mp i \frac{e^{\th'-\th}}{\cosh(\th-\th')} \ln\left[\frac{1+\exp \{-\varepsilon_{+,-}(\th')  \} }{1+\exp \{-\varepsilon_{-,+}(\th') \} }\right]\right \} 
\,.
\ea
\ee
We can check the TBA solution $\varepsilon(\theta)$ and these formulae for $\ln Q(\theta)$ by comparing  them with the integration of exact Riccati numeric solution, via \eqref{lnQPi} and \eqref{Q2++}, as shown in figures \ref{plotRiccatiTBAHp} and \ref{plotRiccatiTBAGhp}.

\section{Integrability $Y$ function and gauge periods} \label{secY}

\subsection{Gauge TBAs}

To establish a connection between integrability and gauge theory, first of all we need to express all integrability definitions and relations in terms of gauge variables through the parameter dictionaries~\eqref{DictGau1}-\eqref{DictGau2}. For example, for $N_f=1$ the explicit relation between $Q$ and $Y$~\eqref{intYdef1} becomes
\begin{align} 
Y_{\pm,0}(\th)&= e^{\mp 2 \pi \frac{m}{\Lambda_1}e^\th}Q_{\pm,2}(\th-i\pi/6)Q_{\pm,1}(\th+i\pi/6)\\
Y_{\pm,1}(\th)&= e^{\mp 2 \pi \frac{e^{-2 \pi i/3}m}{\Lambda_1}e^\th}Q_{\pm,0}(\th-i\pi/6)Q_{\pm,2}(\th+i\pi/6)\\
Y_{\pm,2}(\th)&= e^{\mp 2 \pi \frac{e^{2 \pi i /3}m}{\Lambda_1}e^\th}Q_{\pm,1}(\th-i\pi/6)Q_{\pm,0}(\th+i\pi/6)\,,
\end{align}
where we defined $6$ new gauge $Q$ and $Y$ functions, as (with $k=0,1,2$)
\be  \label{QYk}
Q_{\pm,k}(\th)=Q(\th,-u_k, \pm m_k,\Lambda_1)\,, \quad Y_{\pm,k}(\th)=Y(\th,u_k,\pm i m_k,\Lambda_1)\,,
\ee 
and we denote
\be 
u_k = e^{2\pi i k/3} u \qquad m_k = e^{-2 \pi ik/3} m  \qquad k=0,1,2\,.
\ee 
Then, the gauge Y system for $N_f=1$~\eqref{int-Ysys1} becomes: 
\be \label{gauYsys1}
\ba 
Y_{\pm,0}(\theta+i\pi/2)Y_{\pm,0}(\theta-i\pi/2)&=\left[1+Y_{\pm,1}(\theta+i\pi/6)\right]\left[1+Y_{\pm,2}(\theta-i\pi/6)\right]\\
Y_{\pm,1}(\theta+i\pi/2)Y_{\pm,1}(\theta-i\pi/2)&=\left[1+Y_{\pm,2}(\theta+i\pi/6)\right]\left[1+Y_{\pm,0}(\theta-i\pi/6)\right]\\
Y_{\pm,2}(\theta+i\pi/2)Y_{\pm,2}(\theta-i\pi/2)&=\left[1+Y_{\pm,0}(\theta+i\pi/6)\right]\left[1+Y_{\pm,1}(\theta-i\pi/6)\right]\,.
\ea 
\ee
Instead, for $N_f=2$ we have $8$ new $Y$ functions
\be 
Y_{\pm,\pm} (\th) =Y(\th,u,\pm m_1,\pm m_2 ,\Lambda_2)\qquad \bar{Y}_{\pm,\pm} (\th) =Y(\th,-u,\mp i m_1,\pm i m_2 ,\Lambda_2)\,,
\ee
and the $Y$ system \eqref{Ysyst2} becomes:  
\be \label{gauYsys2}
\ba 
\bar{Y}_{\pm,\pm} (\th+i \pi/2)\bar{Y}_{\pm,\pm} (\th-i \pi/2)&=[1+Y_{\pm,\pm}(\th)][1+Y_{\mp,\mp}(\th)] \\
Y_{\pm,\pm} (\th+i \pi/2)Y_{\pm,\pm} (\th-i \pi/2)&=[1+\bar{Y}_{\pm,\pm}(\th)][1+\bar{Y}_{\mp,\mp}(\th)]\,.
\ea 
\ee
We notice that, in terms of gauge variables, the $Y$ systems simplify their dependence on the (no longer flipped) masses, but the number of $Y$ functions increases\footnote{This also happens for the $SU(2)$ $N_f=0$ theory (where it doubles)~\cite{FioravantiGregori:2019}.}.

Again, as explained in~\cite{FabbriFioravantiPiscagliaTateo:2013}, it straightforward to invert the Y-systems \eqref{gauYsys1} and \eqref{gauYsys2} into the following ``gauge TBAs", for $N_f=1$:
\be 
\ba 
\label{ga-TBA}
\ve_{\pm,0}(\th)&=\ve^{(0)}_{\pm,0}e^\theta - \left(\varphi_+ \ast L_{\pm,1}\right)(\th)- \left(\varphi_- \ast L_{\pm,2}\right)(\th) \\
\ve_{\pm,1}(\th)&=\ve^{(0)}_{\pm,1} e^\theta - \left(\varphi_+ \ast L_{\pm,2}\right)(\th)- \left(\varphi_- \ast L_{\pm,0}\right)(\th)\\
\ve_{\pm,2}(\th)&=\ve^{(0)}_{\pm,2} e^\theta - \left(\varphi_+ \ast L_{\pm,0}\right)(\th)- \left(\varphi_- \ast L_{\pm,1}\right)(\th)\,,
\ea 
\ee 
and for $N_f=2$:
\be \label{ga-TBA2}
\ba 
\varepsilon_{\pm,\pm}(\th)&= \varepsilon_{\pm,\pm}^{(0)} e^{\th}- \varphi \ast (\bar{L}_{\pm \pm}+ \bar{L}_{\mp \mp})(\th) \\
\bar{\varepsilon}_{\pm,\pm}(\th)&= \bar{\varepsilon}_{\pm,\pm}^{(0)} e^{\th}- \varphi \ast (L_{\pm \pm}+ L_{\mp \mp})(\th) \,.\\
\ea 
\ee 
The TBA for $N_f=2$ still involves the kernel \eqref{kernelNf2}, while new kernels for $N_f=1$ are defined as
\be \label{kern1}
\varphi_\pm(\th) = \frac{1}{\cosh (\th  \pm i \pi/6)}\,.
\ee 
The forcing terms for $N_f=1$ write explicitly, for $k=0,1,2$:
\be \label{eps0gau}
\ve^{(0)}_{\pm,k}=-e^{-i\pi/6}\ln Q^{(0)}(-e^{-2\pi i/3}u_k,\pm e^{2\pi i/3}m_k,\Lambda_1)-e^{i\pi/6}\ln Q^{(0)}(-e^{2\pi i/3} u_k,\pm e^{-2\pi i/3}m_k,\Lambda_1)\pm\frac{8}{3} \pi  \frac{m_k}{\Lambda_1} \,,
\ee
in terms of the leading order coefficient of $\ln Q(\theta,u,m,\Lambda_1) \simeq e^\theta \ln Q^{(0)}(u,m,\Lambda_1)\simeq \frac{\Lambda_1}{2\hbar} \ln Q^{(0)}(u,m,\Lambda_1)$, as $\theta \to + \infty$ or $\hbar \to 0$, given by
\be \label{lnQ0gauint}
\ba 
\ln Q^{(0)}(u,m,\Lambda_1) = \int_{-\infty}^\infty \left[\sqrt{e^{2y}+e^{-y}+\frac{4 m }{\Lambda_1}e^y+\frac{4u}{\Lambda_1^2}}-e^y-e^{-y/2}-2\frac{m}{\Lambda_1}\frac{1}{1+e^{-y/2}}\right]\,d y\,.
\ea 
\ee 
Similarly for $N_f=2$ the forcing terms are expressed as
\be 
\ba 
 \varepsilon_{\pm,\pm}^{(0)} &=-\ln Q^{(0)}(u, m_1, m_2,\Lambda_2) -\ln Q^{(0)}(u, -m_1, -m_2,\Lambda_2)  \mp \frac{4 \pi i}{\Lambda_2}(m_1-m_2) \\
 \bar{\varepsilon}_{\pm,\pm}^{(0)} &=-\ln Q^{(0)}(-u, -i m_1, i m_2,\Lambda_2) -\ln Q^{(0)}(-u, i m_1, - i m_2,\Lambda_2)  \mp \frac{4 \pi }{\Lambda_2}(m_1+m_2)\,,
\ea
\ee
in terms of the leading order coefficient of $\ln Q(\theta,u,m_1,m_2,\Lambda_2) \simeq e^\theta \ln Q^{(0)}(u,m_1,m_2,\Lambda_2)$, as $\theta \to + \infty$ or $\hbar \to 0$, that is
\begin{align} \label{lnQ0gauint2}
&\ln Q^{(0)}(u,m_1,m_2,\Lambda_2) \nonumber \\
&= \int_{-\infty}^\infty\left[\sqrt{2 \cosh(2 y) +\frac{8 m_1}{ \Lambda_2}e^y+ \frac{8 m_2}{\Lambda_2}e^{-y}+\frac{16 u}{\Lambda_2^2}} -2\cosh y -\frac{4 m_1}{\Lambda_2}\frac{1}{1+e^{-y/2}}- \frac{4 m_2}{\Lambda_2}\frac{1}{1+e^{y/2}}\right]\, dy \,.
\end{align} 
The analytic computation of the integrals \eqref{lnQ0gauint} and \eqref{lnQ0gauint2} is explained in appendix \ref{appForcing}.

\begin{figure} \label{fig:TBARiccati}
\centering
\includegraphics[width=0.5\textwidth]{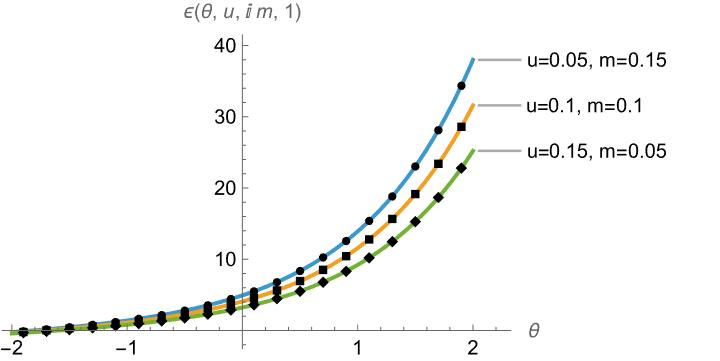}
\includegraphics[width=0.45\textwidth]{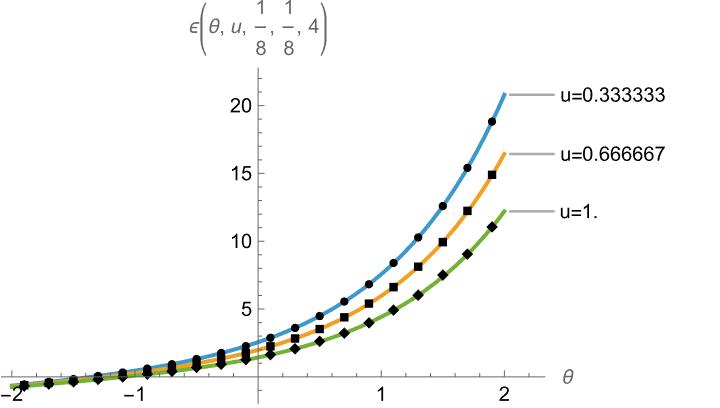}
\caption{Comparison of the solution of the gauge TBA \eqref{ga-TBA}, \eqref{ga-TBA2} (the colored continuous curves) vs. Riccati ODE numeric integration (the black dots), for $N_f=1$ and $N_f=2$ on the left and right respectively.} 
\end{figure}

We can also use a boundary condition at $\theta \to - \infty$, which now is not strictly necessary but just improves the numerical precision\footnote{We recall that for the integrability TBAs~\eqref{int-TBA}, \eqref{int-TBA2}, the boundary condition is strictly necessary to fix $p$, which does not enter the forcing term at $\theta \to + \infty$.}. As explained in appendix \ref{appForcing}, it is easy to find it to be, for $N_f=1$
\be\label{ga-bdy-cond}
\ve_{\pm,k}(\theta)\simeq -2 \ln \left(-\frac{2}{\pi}\th \right)\simeq \tilde{f}_1(\theta),\quad \theta \to -\infty\,,
\ee
where we define also the auxiliary function
\be
\hat{f}_1(\th)=-\ln\Big(1+\frac{2}{\pi}\ln\big(1+e^{-\theta-\frac{\pi i}{6}}\big)\Big)-\ln\Big(1+\frac{2}{\pi}\ln\big(1+e^{-\theta+\frac{\pi i}{6}}\big)\Big)\,,
\ee
to be inserted into the TBA as follows
\be \label{gauTBAbdy}
\ba
\ve_{\pm,k}(\th)&=\ve_{\pm,k}^{(0)}e^{\theta}+\hat{f}_1(\th)-\left(\varphi_{+}\ast\big(L_{\pm,(k+1)\:{\rm mod}\,3}+\hat{F}_1\big)\right)(\th)-\left(\varphi_{-}\ast\big(L_{\pm,(k+2)\,{\rm mod}\,3}+\hat{F}_1\big)\right)(\th),
\ea
\ee
where $\hat{F}_1$ is fixed by $\hat{f}_1=(\varphi_{+}+\varphi_{-})\ast\hat{F}_1$ as
\be
\hat F_1(\theta) = -\log \left(1+\frac{2}{\pi } \log \left(1+ e^{-2\theta}\right)\right)\,.
\ee
Similarly for $N_f=2$ we find the boundary condition at  $\th \to  -\infty$ to be 
\be \label{ga-bdy-cond-2}
\varepsilon_{\pm,\pm}(\th) \simeq -2\ln \left(-\frac{2\th}{\pi} \right)\simeq \hat{f}_2(\th) \qquad \th \to - \infty\,,
\ee
where the auxiliary function in the TBA is
\be
\hat{f}_2(\th)=-2\ln\Big(1+\frac{2}{\pi}\ln\big(1+e^{-\theta}\big)\Big)\,,
\ee
and correspondingly
\be
\hat F_2 (\theta)= -\log \left(1+\frac{2}{\pi } \log \left(1+ e^{-2\theta}\right)\right)\,.
\ee
Eventually the TBA for $N_f=2$ reads
\be \label{ga-TBA2bdy}
\ba 
\varepsilon_{\pm,\pm}(\th)&= \varepsilon_{\pm,\pm}^{(0)} e^{\th}+\hat{f}_2(\th)- \varphi \ast (\bar{L}_{\pm \pm}+ \bar{L}_{\mp \mp}+2\hat F_2)(\th) \\
\bar{\varepsilon}_{\pm,\pm}(\th)&= \bar{\varepsilon}_{\pm,\pm}^{(0)} e^{\th}+\hat{f}_2(\th)- \varphi \ast (L_{\pm \pm}+ L_{\mp \mp}+2\hat F_2)(\th) \,.\\
\ea 
\ee 

We remark we can also compute $\varepsilon$ through the numerical solution of the Riccati equation, as explained in section \ref{sec:QIntExact}. The result matches very well with the TBA solution and is shown in figures \ref{fig:TBARiccati}.

\subsection{Seiberg-Witten gauge-integrability identification}

In this subsection we prove the identification between the integrability pseudoenergy and the gauge SW periods, at the leading $\hbar\to 0$ ($\th\to +\infty$) order. In later subsections we will extend this at the $\hbar \neq 0$ exact level.\footnote{We recommend the reader to have a look also at appendix~\ref{proofNf=0} for the much simpler and illuminating proof for the $SU(2)$ $N_f=0$ gauge theory.} 

\subsubsection{Proof for the $N_f=1$ theory}

Let us consider first the $N_f=1$ gauge theory. We can explicitly write $\ve^{(0)}_{+,0}$ in~\eqref{eps0gau} as the sum of the following integral expressions
\begin{align} \label{lQ0+}
\begin{split}
&e^{-i\pi/6}\ln Q^{(0)}(-e^{-2\pi i/3}u,e^{2\pi i/3}m)=\\
&=  \int_{-\infty-2\pi i/3}^{\infty-2\pi i/3} \biggl[\sqrt{-e^{2 y} - \frac{4 m}{\Lambda_1}e^{+y} +\frac{4 u}{\Lambda_1^2}  - e^{-y}}-i e^y+ i e^{-y/2}-i 2\frac{m}{\Lambda_1}\frac{1}{1+e^{-y/2-\pi i/3}} \biggr]\,d y\,,
\end{split}
\end{align}
\begin{align}  \label{lQ0-}
\begin{split}
&e^{i\pi/6}\ln Q^{(0)}(-e^{2\pi i/3}u,e^{-2\pi i/3}m)=\\
&=  \int_{-\infty+2\pi i/3}^{\infty+2\pi i/3} \biggl[-\sqrt{-e^{2 y} - \frac{4 m}{\Lambda_1}e^{+y} +\frac{4 u}{\Lambda_1^2}  - e^{-y}}+i e^y- i e^{-y/2}+i 2\frac{m}{\Lambda_1}\frac{1}{1+e^{-y/2+\pi i/3}} \biggr]\,d y\,.
\end{split}
\end{align}
We notice that the integrands in \eqref{lQ0+} and \eqref{lQ0-} are the same except for the mass regularization term. So if we want to use only the former integrand we have to add the following term
\be \label{mregdiff}
2 \frac{ i m}{\Lambda_1}  \int_{-\infty+2\pi i/3}^{\infty+2\pi i/3} \left[\frac{1}{1+e^{-y/2+\pi i/3}} -\frac{1}{1+e^{-y/2-\pi i/3}}\right]\,d y=\frac{2}{\Lambda_1}\frac{4\pi m}{3}\,.
\ee 
Moreover we observe that the integrand of \eqref{lQ0+} can be manipulated as
\begin{align}
 \begin{split}  &e^{-i\pi/6} \ln Q^{(0)}(-e^{-2\pi i/3}u,e^{2 \pi i/3}m)\\ 
    &= 4 i \int_{-\infty-2\pi i/3}^{+\infty-2\pi i/3} d y\,\left[\frac{\frac{3}{8}e^{-y}+\frac{1}{2}\frac{m}{\Lambda_1} e^y-\frac{u}{\Lambda_1^2}}{\sqrt{e^{2y}+e^{-y}+\frac{4m}{\Lambda_1} e^y-\frac{4u}{\Lambda_1^2}}}+\frac{d}{d y}\sqrt{e^{2y}+e^{-y}+\frac{4m}{\Lambda_1} e^y-\frac{4u}{\Lambda_1^2}}-\mathrm{reg.}\right] \,,
   \end{split}
    \end{align}
     and eventually reduced, up to a total derivative and regularization, to the Seiberg-Witten differential $\lambda$, defined in the variable $x=- \frac{\Lambda_1^2}{4}e^{-y}$ as~\cite{BilalFerrari-massive:1997}
\begin{align}
 \begin{split}
  \lambda(x,-u) d x&= \frac{1}{2\pi  } \frac{-u-\frac{3}{2}x-\frac{\Lambda_1^3}{8}\frac{m}{x}}{\sqrt{x^3+u x^2+\frac{\Lambda_1^3 m}{4}x - \frac{\Lambda_1^6}{64}}} \,d x \\
  &=- \frac{i \Lambda_1}{\pi } \frac{\frac{3}{8} e^{-y}+\frac{1}{2}\frac{m}{\Lambda_1} e^y-\frac{u}{\Lambda_1^2}}{\sqrt{e^{2 y}+e^{-y}+\frac{4m}{\Lambda_1}e^y-\frac{4 u}{\Lambda_1^2}}}\, d y = \lambda(y,-u) \, d y\,.
  \end{split}
\end{align}
\begin{figure}[t]\label{figY1cycSW}
\centering
\includegraphics[width=0.9\textwidth]{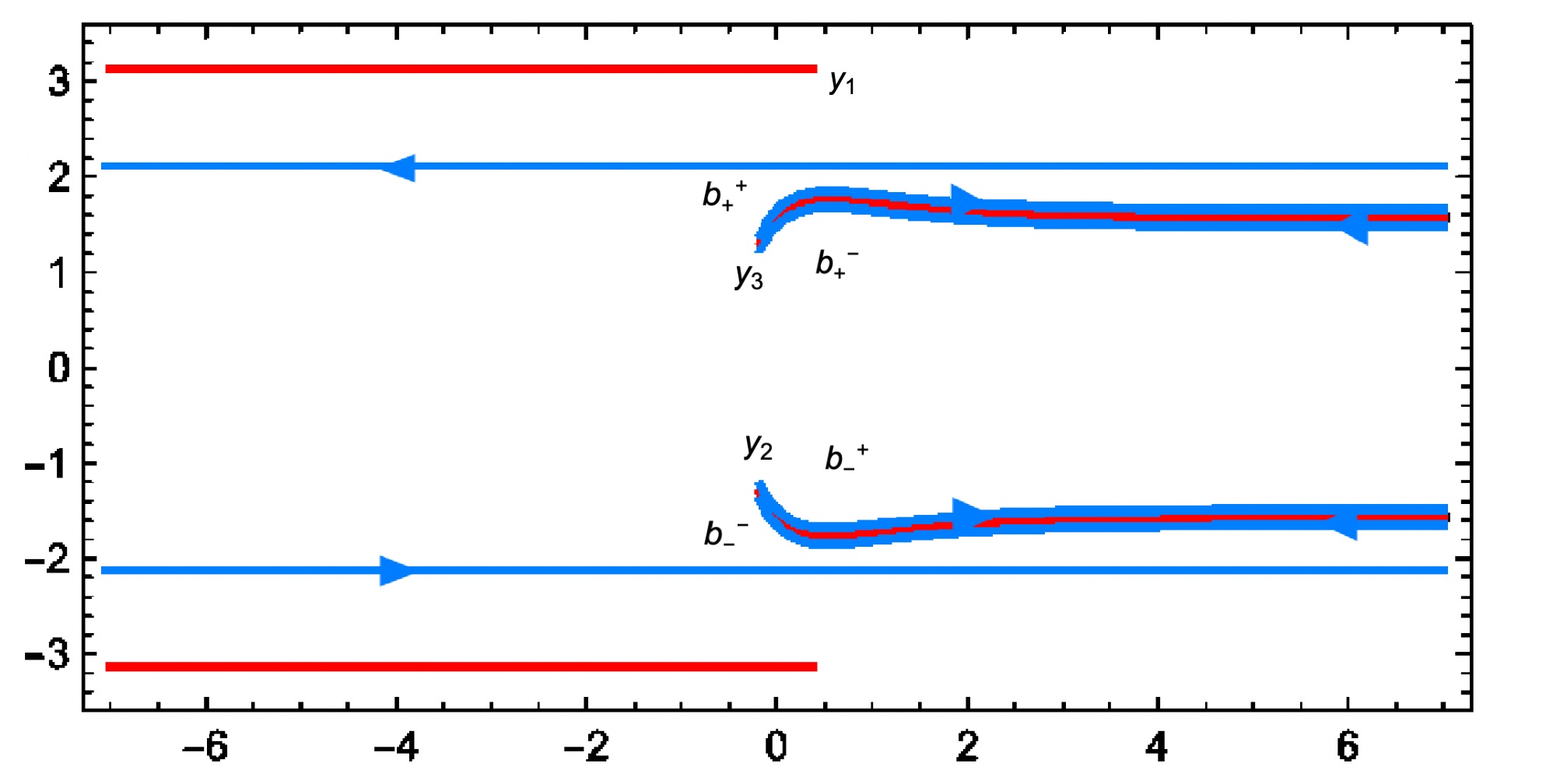}
\caption{A strip of the $y$ complex plane, where in blue we show the contour of integration of SW differential for the $SU(2)$ $N_f=1$ theory, and 
in red its branch cuts.}
\end{figure}
 Then we have
\be \label{proofNf1eps0lambda}
\varepsilon^{(0)}_+ =\varepsilon^{(0)}(u,i m)= \frac{4 \pi}{\Lambda_1}\left[ \int_{-\infty-2 \pi i/3}^{\infty- 2 \pi i/3} \tilde{ \lambda}(y,-u,m) \, dy +\int_{\infty+2 \pi i/3}^{\infty- 2 \pi i/3} \tilde{ \lambda}(y,-u,m) \, dy \right]\,,
\ee
with\footnote{The sign of the regularization depends on the sign chosen for the square root. We did not specify this before because it is relevant only when integrating along branch cuts to obtain the periods and not along the horizontal lines to obtain the $\ln Q$ function.}
     \be
    \tilde{ \lambda} (y,-u,m)= \lambda(y,-u,m) +\frac{d}{d y}\sqrt{e^{2y}+e^{-y}+\frac{4m}{\Lambda_1} e^y-\frac{4u}{\Lambda_1^2}}- \mathrm{reg.} \,
     \ee
  Now we consider for $-i \tilde\lambda(y)$ the countour of integration represented in figure~\ref{figY1cycSW}. We have horizontal branch cuts for $\Im y = \pm \pi$, $ \Re y < \Re y_1$ and, between them, other two curved branch cuts, whose upper and lower edges are denoted as $b_-^\pm$, $b_+^\pm$, respectively. These internal branch cuts start from the complex conjugates branch points $y_{2},y_3$ and have horizontal asymptotics $\Im y = \pm \frac{\pi}{2}$ for $\Re y \to + \infty$.\footnote{This can be shown easily by considering the asymptotics of $e^{2 y}+e^{-y}+\frac{4 m}{\Lambda_1}-\frac{4 u}{\Lambda_1^2}$ at $\Re y \to \pm \infty$ and $\Im y = \pm \frac{\pi}{2}, \pm \pi$, which are negative real.} Then, the cycle integral from $y_2$ to $y_3$ along the branch cuts defines the gauge period $a_1^{(0)}$ (\textit{cf.} appendix~\ref{appSWper})\footnote{Notice that doing a closed cycle integration the addition of a total derivative and regularization to $ \lambda$ is allowed. The regularization is also necessary since the branch cuts extend to infinity.}
\begin{align}
   a_1^{(0)}(-u,m,\Lambda_1) &=\oint \tilde{\lambda}(y,-u,m,\Lambda_1) \, dy=\oint \lambda(y,-u,m,\Lambda_1)\,.
\end{align}
We notice the following symmetry properties of $\tilde{\lambda}(y)$. Since for $y \in \mathbb{R}$ and $m, \Lambda > 0$ and $u>0$ not large we have
$    i \tilde{\lambda}(y) \in \mathbb{R} $, by Schwarz reflection principle the analytic continuation of $i \tilde{\lambda}(y)$ for $y \in \mathbb{C}$ satisfies
\begin{equation}
    i \tilde{\lambda}(y^*) = ( i \tilde{\lambda}(y))^* \qquad y \in \mathbb{C}\,.
\end{equation}
From this it also follows that along the upper $b_\pm^+$ and lower $b_\pm^-$ edge of the curved branch cuts, where $i \tilde\lambda \in i \mathbb{R}$, we have the sign properties
\begin{equation}\label{lambdasym}
    i \tilde\lambda(y) \Bigr |_{b_+^+}=- i \tilde\lambda(y) \Bigr |_{b_-^-}=- i \tilde\lambda(y) \Bigr |_{b_+^-}=+ i \tilde\lambda(y) \Bigr |_{b_-^+}\,,
\end{equation}
so that we can express the gauge period also as the following integral along $b_+^-$
\begin{equation}\label{proofNf1a1intb}
a_1^{(0)}=  4     \int_{b_+^-} \tilde{\lambda}(y) \, d y\,.
\end{equation}
On the other hand, by considering the integration contour $C_1=(-\infty+2\pi i/3,+\infty+2\pi i/3)\cup b_+^+\cup b_+^-\cup b_-^+ \cup b_-^-\cup(\infty-2\pi i/3,-\infty-2\pi i/3)$ (closed also at infinity) we obtain
\begin{align}
   0=  \oint_{C_1} i \tilde\lambda(y) \, d y =\int_{-\infty-2 \pi i/3}^{\infty- 2 \pi i/3} \tilde\lambda(y) \, dy +\int_{\infty+2 \pi i/3}^{\infty- 2 \pi i/3} \tilde\lambda(y) \, dy -4 \int_{b_+^-} i \tilde\lambda(y) \, d y\, ,\end{align}
and therefore by \eqref{proofNf1eps0lambda} and \eqref{proofNf1a1intb} the following relation between the SW gauge period and leading pseudoenergy ensues
\begin{equation}
\varepsilon^{(0)}(u,i m,\Lambda_1)=\frac{4  \pi }{\Lambda_1}a_1^{(0)}(-u,m,\Lambda_1)\,.
\end{equation}
Finally, by the change of basis of the gauge periods, for $u>0$ (\textit{cf.} appendix~\ref{appSWper})
\be 
\ba 
a^{(0)}(-u,m) &= -a_1^{(0)}(-u,m) + a_2^{(0)}(-u,m) +  m  \\
a^{(0)}_D(-u,m) &= -2a_1^{(0)}(-u,m) + a_2^{(0)}(-u,m) + \frac{3}{2} m \,,
\ea 
\ee 
we can write the following relations for all three forcing terms of the TBA:
\small
\be\label{ep0-in-aaD-nZ3}
\ba
\ve^{(0)}(u, i m)  &= \frac{4\pi}{\Lambda_1}  \left[a^{(0)}(-u,m)- a^{(0)}_D(-u,m)+\frac{1}{2}m \right]\\
\ve^{(0)}(e^{2\pi i/3}u, i e^{-2\pi i/3} m) &=\frac{4\pi}{\Lambda_1}  \left[a^{(0)}( -e^{2 \pi i/3}u,e^{-2 \pi i/3}  m)-a^{(0)}_D( -e^{2 \pi i/3}u,e^{-2 \pi i/3}  m)+  e^{-2 \pi i/3}  m \right]\\
\ve^{(0)}(e^{-2\pi i/3}u, i e^{2\pi i/3}m)  &=\frac{4\pi}{\Lambda_1}\left[-2a^{(0)}(-e^{-2 \pi i/3}u,e^{2 \pi i/3} m)+a^{(0)}_D(-e^{-2 \pi i/3}u,e^{2 \pi i/3} m)+\frac{1}{2}  e^{2 \pi i/3} m \right]\,.
\ea
\ee
\normalsize
In table \ref{tab:periods1WKB} we also check these expressions numerically, through the use of elliptic integrals expressions for the periods (\textit{cf.} appendix~\ref{app:Nf1SWper}) and the analytic series~\eqref{lnQ0gau} for $\ve^{(0)}_{\pm,k}$.

We notice that for all three pseudoenergies, the forcing term has the form of a central charge for the SW theory for $SU(2)$ with $N_f=1$~\cite{SeibergWitten:1994QCD}: 
\be \label{ZNf1}
Z =  n_m a_D^{(0)}-n_e a^{(0)} + s m\,,
\ee 
where $n_m$ and $ n_e$ are integers, while $s$ half-integer. Then, the mass of the BPS state is $M_{BPS}=|Z|$~\cite{BilalFerrari-massive:1997}. 
We find a perfect match between the expected electric and magnetic charges $n_e,n_m$ which multiply the periods $a^{(0)}$ and $a_D^{(0)}$ respectively (precisely, $(-1,0)$, $(1,-1)$ and $(0,1)$~\cite{BilalFerrariQCD:1996})\footnote{The mass constant term (physical flavour charge~\cite{Ferrari:1996de}) is ambiguous, but that it is just because the periods themselves are defined up to the well-known SW monodromy of exactly an integer multiple of $\frac{1}{2}m$~\cite{SeibergWitten:1994QCD,AlvarezGaumeMarinoZamora:1997,BilalFerrari-massive:1997}. We emphasize that that the central charge and mass of BPS states have no ambiguity.  We notice also that, in integrability, there is no ambiguity since the wave functions and therefore the $Q$ function in~\eqref{Qdefpsi} cannot change. In other words, we are fixing through integrability what is in gauge theory is in general ambiguous.} \footnote{The periods ($a^{(0)}$, $a_D^{(0)}$) are discontinuous on the moduli space, due to the singularities,
and can be analytically continued to $(\tilde{a}^{(0)}, \tilde{a}_D^{(0)})$ by using the monodromy matrix around the singularity on the moduli space. Correspondingly, the charges $(n_e, n_m)$ will also be transformed by the inverse of the monodromy matrix, because one needs to keep the physical mass and central charge invariant. Since the driving terms of TBA equations are given by the central charge, more precisely the BPS mass, the TBA equations are invariant under the monodromy transformation.}.

\begin{figure}[t]
\centering
\includegraphics[width=0.9\textwidth]{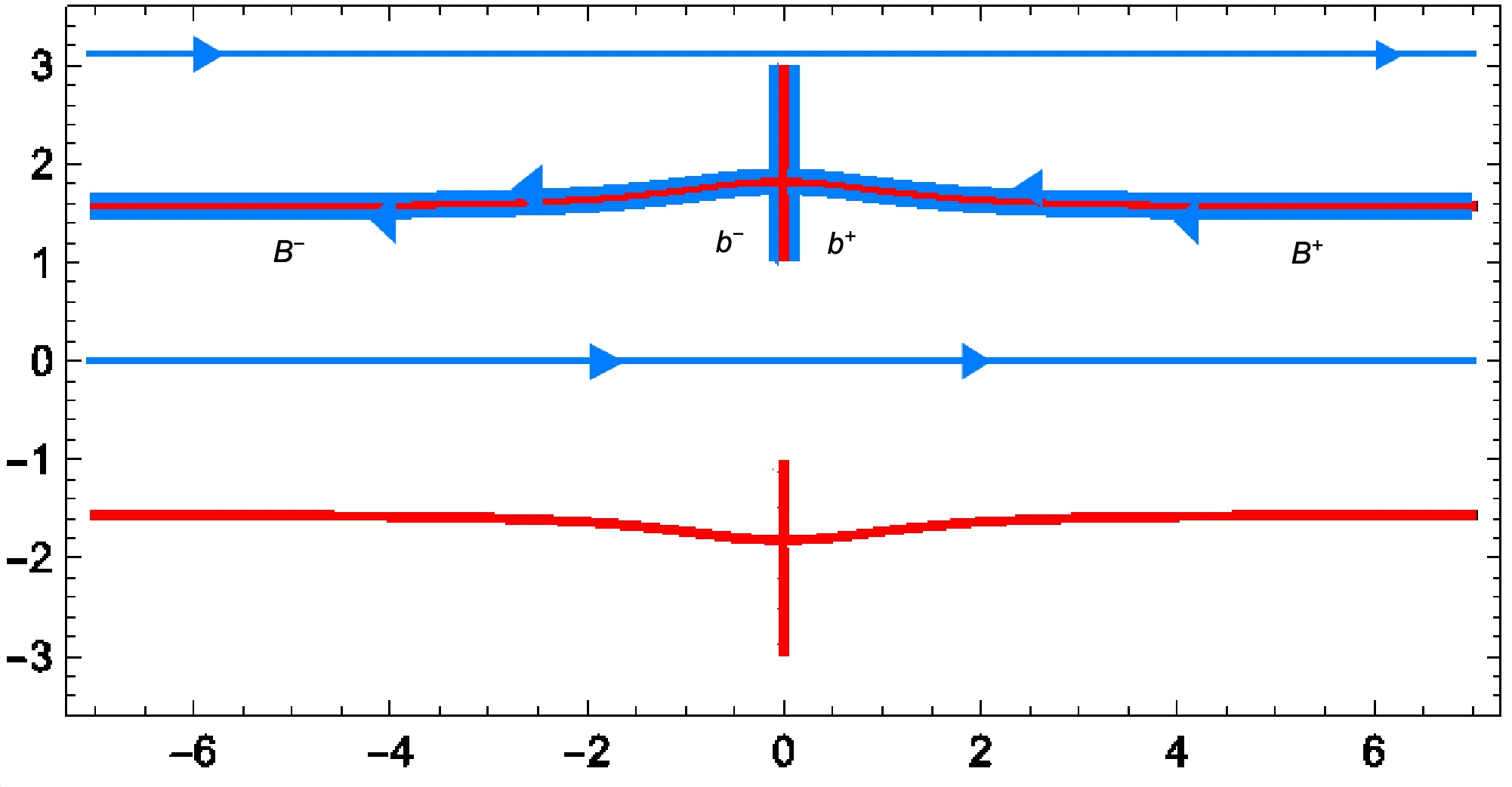}
\caption{A strip of the $y$ complex plane, where in blue we show the contour of integration of SW differential for the $SU(2)$ $N_f=2$ theory, and in red its branch cuts.}
\label{figY2cycSW}
\end{figure}

\subsubsection{Proof for the $N_f=2$ theory}

An analogue derivation can be produced for the $N_f=2$ gauge theory. Here we will report the details for the main intermediate results. 

The leading order of $\varepsilon$ as $\hbar \to 0$ (that is, $\theta\to \infty$) is
\be 
\ba
\varepsilon(\th,u,m_1,m_2,\Lambda_2) &\simeq e^\th \varepsilon^{(0)}(u,m_1,m_2,\Lambda_2) \\
&= e^\th  \left[-\ln Q^{(0)}(u,m_1,m_2,\Lambda_2) -\ln Q^{(0)}(u,-m_1,-m_2,\Lambda_2) +\frac{4  \pi i}{\Lambda_2} (m_1-m_2)\right]\,,
\ea
\ee 
with
\be 
\ba
\ln Q^{(0)}(u,m_1,m_2,\Lambda_2)= \int_{-\infty}^\infty \Bigl[&\sqrt{e^{2y}+e^{-2y}+ \frac{8m_1}{\Lambda_2} e^y+\frac{8m_2}{\Lambda_2}e^{-y} + \frac{16 u}{\Lambda^2}} \\ 
&-2\cosh y- \frac{4m_1}{\Lambda_2}\frac{1}{1+e^{-y/2}} - \frac{4m_2}{\Lambda_2}\frac{1}{1+e^{y/2}} \Bigr]\, d y\,,
\ea
\ee
\be
\ba
\ln Q^{(0)}(u,-m_1,-m_2,\Lambda_2)= \int_{-\infty}^\infty \Bigl[&\sqrt{e^{2y}+e^{-2y}- \frac{8m_1}{\Lambda_2} e^y-\frac{8m_2}{\Lambda_2}e^{-y} + \frac{16 u}{\Lambda^2}} \\ 
&-2\cosh y+ \frac{4m_1}{\Lambda_2}\frac{1}{1+e^{-y/2}} + \frac{4m_2}{\Lambda_2}\frac{1}{1+e^{y/2}} \Bigr]\, d y \,.
\ea
\ee 
Notice that in the latter expression we can trade the opposite sign of the masses as the shift $y \to y+i \pi$
\be 
\ba
\ln Q^{(0)}(u,-m_1,-m_2,\Lambda_2)= \int_{-\infty+i\pi}^{\infty+i\pi} \Bigl[&\sqrt{e^{2y}+e^{-2y}+ \frac{8m_1}{\Lambda_2} e^y+\frac{8m_2}{\Lambda_2}e^{-y} + \frac{16 u}{\Lambda^2}} \\ 
&+2\cosh y+ \frac{4m_1}{\Lambda_2}\frac{1}{1+i e^{-y/2}} + \frac{4m_2}{\Lambda_2}\frac{1}{1-i e^{y/2}} \Bigr]\, d y \,.
\ea
\ee
Then, we can integrate only the former integrand along the contour of figure~\ref{figY2cycSW}, while adding the following term which accounts for the difference in the regularizations
\be 
\int_{-\infty}^{\infty}\left[\frac{4 \left(m_1e^{y/2}+m_2\right)}{\Lambda_2 \left(e^{y/2}+1\right)}-\frac{4 \left(m_1e^{y/2}+i m_2\right)}{\Lambda_2 \left(e^{y/2}+i\right)}\right] \, dy=\frac{4 i \pi  (m_1-m_2)}{\Lambda_2}\,.
\ee 
Again we can reduce the integrand to the SW differential. As for the proof for $N_f=1$, we could start from the original cubic SW differential defined in appendix \ref{appSWper}, using some standard change of variable to transform it into quartic form. However, for simplicity we assume the quantum SW curve \eqref{sec:qSWNf1}, already in quartic form, to correctly reproduce it at the leading $\hbar \to 0$ order, as studied in previous literature \cite{ItoKannoOkubo:2017}. Then
\be 
\ba
\tilde\lambda(y,-u,-i m_1,i m_2,\Lambda_2) \, dx &= \lambda(y,-u,-i m_1,i m_2,\Lambda_2)+\frac{d}{dy}\left( ...\right)- \mathrm{reg.}\\
&=\sqrt{e^{2y}+e^{-2y}+ \frac{8m_1}{\Lambda_2} e^y+\frac{8m_2}{\Lambda_2}e^{-y} + \frac{16 u}{\Lambda^2}} - \mathrm{reg.} 
\ea
\ee 
So we can express the leading order pseudonergy as the following the closed integral
\be 
\ba 
\varepsilon^{(0)}(u,m_1,m_2,\Lambda_2) &= \oint \tilde\lambda(y)\, dy  - \frac{4 \pi i}{\Lambda_2}(m_1-m_2) \,.
\ea 
\ee 
We notice that for $y=t+i s$, with $t \in \mathbb R$ and $s>0$, along the horizontal branch cuts $B^\pm$ we have the properties
\be 
\ba 
\Re \mathcal{P}^{(0)}(y) &=0\\
\Im  \mathcal{P}^{(0)}(t+is)&=-\Im  \mathcal{P}^{(0)}(-t+is)\,,
\ea 
\ee 
so that the only contribution is from the vertical branch cuts $b^\pm$, which define the gauge period $a_2^{(0)}$. Eventually we obtain
\be 
\ba 
\varepsilon^{(0)}(u,m_1,m_2,\Lambda_2) 
&=\frac{8 \pi}{\Lambda_2}a_2^{(0)}(-u,-i m_1,i m_2,\Lambda_2) -\frac{4\pi i}{\Lambda_2} (m_1-m_2)\,.
\ea 
\ee 
Then, since for $u\to \infty$, $\Lambda_2>0,m_1>0,m_2 >0$ we have 
\be
a_2^{(0)}(-u,m_1,m_2,\Lambda_2)\simeq a_D(-u,m_1,m_2,\Lambda_2)\simeq \frac{i}{ \pi} \sqrt{ u} \ln \frac{u}{\Lambda_2^2}
\,,
\ee
the following relations between the TBA forcing terms and SW gauge periods ensue
\be \label{epsaDNf=2} 
\ba
\varepsilon^{(0)}(u,m_1,m_2,\Lambda_2) &=\frac{8  \pi}{\Lambda_2}a_D^{(0)}(-u,-i m_1,i m_2,\Lambda_2) -\frac{4\pi i}{\Lambda_2} (m_1-m_2) \\
\varepsilon^{(0)}(u,-m_1,-m_2,\Lambda_2) &=\frac{8 \pi}{\Lambda_2}a_D^{(0)}(-u,-i m_1,i m_2,\Lambda_2) +\frac{4\pi i}{\Lambda_2} (m_1-m_2)\\
\varepsilon^{(0)}(-u,-im_1,im_2,\Lambda_2) &=\frac{8 \pi}{\Lambda_2}a_D^{(0)}(u,m_1,m_2,\Lambda_2) \\
\varepsilon^{(0)}(-u,im_1,-im_2,\Lambda_2) &=\frac{8  \pi}{\Lambda_2}a_D^{(0)}(u,m_1,m_2,\Lambda_2)+\frac{8\pi}{\Lambda_2} (m_1+m_2) 
\,.
\ea
\ee
In table \ref{tab:periods11WKB} we also check these expressions numerically. Again the $\hbar \to 0$ leading pseudoenergies have the form of a SW central charge. In this way the TBA~\eqref{ga-TBA2} constitutes a generalization of that found in~\cite{Imaizumi:2021} $N_f=2$ gauge theory with equal masses $m_1 = m_2$ respectively (see a numerical test for different masses below in table~\ref{tab:periods11WKB}).

\subsection{Exact quantum gauge-integrability identification for $Y$} \label{subsecYex}

\begin{table}[t]
\centering
\begin{tabular}{c|c|c|c|c}
$n$ & WKB $\epsilon_{+,0}^{(n)}$& TBA $\epsilon_{+,0}^{(n)}$&WKB $\epsilon_{+,1}^{(n)}$&TBA $\epsilon_{+,1}^{(n)}$ \\
\hline
 $1$ &$ -0.203514 $&$ -0.203510$ & $-0.198772+0.00273796 i $& $-0.198768+0.00273793 i $\\
$ 2$ & $0.00807461 $& $0.00807458$ & $-0.0040373+0.00880084 i $& $-0.00403729+0.00880082 i$ \\
 $3 $&$ -0.000137236 $& $-0.000137240 $& $0.0127132\, -0.00741921 i$ & $0.0127132\, -0.00741919 i $\\
 $4$ &$ -0.00399816$ & $-0.00399815$ & $-0.0181678-0.00818084 i $& $-0.0181678-0.00818082 i $\\
$ 5 $&$ 0.0152519 $& $0.0152519$ & $-0.00762595+0.0907242 i$ & $-0.00762594+0.0907241 i$ \\
$ 6 $& $-0.059650$ & $-0.0596499 $&$ 0.454483\, -0.296835 i $&$ 0.454482\, -0.296834 i $\\
 $7$ &$ 0.234108 $&$ 0.234107$ & $-4.06371-2.48134 i $&$ -4.06370-2.48134 i $\\
 $8 $&$ -0.511981 $& $-0.511981$ &$ 0.25599\, +62.3103 i$ & $0.256243\, +62.3092 i $\\
 $9 $&$ -8.66103$ & $-8.65802$ & $866.785\, -505.439 i$ & $866.737\, -505.410 i$ \\
\end{tabular}
\caption{Comparison of WKB with TBA modes \eqref{tbaModesNf1} for $N_f=1$, with $u=\frac{1}{10}$, $m=\frac{1}{20\sqrt{2}}$, $\Lambda_1=1$.}\label{tab:higherWKBNf1}

\begin{tabular}{c|c|c|c|c}
$n$ & WKB $\varepsilon_{+,+}^{(n)}$ & TBA  $\varepsilon_{+,+}^{(n)}$ & WKB  $\bar\varepsilon_{+,+}^{(n)} $& TBA  $\bar\varepsilon_{+,+}^{(n)}$ \\
\hline
$ 1 $&$ -0.239513 $&$ -0.239514$ & $-0.502484 $& $-0.502486$ \\
 $2$ & $0.0158874$ & $0.0158875$ &$ 0.312000$  & $0.312001$ \\
$ 3 $& $-0.00707555 $& $-0.00707556$ &$ -1.32897 $& $-1.32897$ \\
 $4$ &$ 0.00852253 $& $0.00852254$ &$ 14.2777$  & $14.2777$ \\
$ 5$ & $-0.020378 $& $-0.020378$ & $-287.827$ & $-287.828$\\
 $6$ &$ 0.0815612 $& $0.0815613$ & $9362.12 $  & $9362.11$ \\
 $7$ & $-0.491013 $&$ -0.491012$ & $-447819$ & $-447818$ \\
 $8$ & $ 4.13913 $& $4.13846$ & $2.95855\times 10^7  $& $2.95848\times 10^7$ \\
$ 9 $&$ -46.4544$ &$ -46.4468$ & $-2.58042\times 10^9$ & $-2.58006\times 10^9$ \\
 \end{tabular}
\caption{Comparison of WKB with TBA modes \eqref{tbaModesNf2} for $N_f=2$, with $u=1$, $m_1=m_2=1/8$, $\Lambda_2=4$.}\label{tab:higherWKBNf2}
\end{table}

Let us consider higher $\hbar\to 0$ ($\theta\to + \infty$) asymptotic expansions coefficients for $\ln Q$ and $\varepsilon$:
\be 
\ba 
\ln Q (\th,u,\mathbf m,\Lambda_{N_f}) &\doteq \sum_{n=0}^\infty e^{\th(1-2n)} \ln Q^{(n)}(u,\mathbf{m},\Lambda_{N_f}) \qquad \th \to +\infty \\
\varepsilon (\th,u,\mathbf m,\Lambda_{N_f}) &\doteq \sum_{n=0}^\infty e^{\th(1-2n)} \varepsilon^{(n)}(u,\mathbf{m},\Lambda_{N_f}) \qquad \th \to +\infty \,.
\ea 
\ee 
They are related as follows: for $N_f=1$
\be
\varepsilon^{(n)}(u,m,\Lambda_1)= -e^{i\pi(2n-1)/6} \ln Q^{(n)}(-e^{-2\pi i/3}u_k,e^{2\pi i/3}m_k,\Lambda_1) -e^{-i\pi(2n-1)/6} \ln Q^{(n)}(-e^{2\pi i/3}u_k,e^{-2\pi i/3}m_k,\Lambda_1)  
\ee
and for $N_f=2$
\be
\varepsilon^{(n)}(u,m_1,m_2,\Lambda_2) = - \ln Q^{(n)}(u,m_1,m_2,\Lambda_2) - \ln Q^{(n)}(u,-m_1,-m_2,\Lambda_2)
\ee
These asymptotic coefficients can be computed at all orders for example through a numeric WKB recursion (by transforming \eqref{WKBRecInt} in gauge variables). Alternatively, they are directly and simply provided by the $\theta \to + \infty$ expansion of the TBA equations, through the following relations: for $N_f=1$
\be \label{tbaModesNf1}
\ve^{(n)}_{+,k}=\frac{(-1)^n}{\pi} \int_{-\infty}^\infty d \th \,e^{\th(2n+1)} \Biggl \{e^{-i\pi(2n-1)/6} L_{+,(k+1)\:{\rm mod}\,3}(\th)+e^{i\pi(2n-1)/6}L_{+,(k+2)\:{\rm mod}\,3}(\th) \Biggr \}
\ee
and for $N_f=2$
\be 
\ba \label{tbaModesNf2}
\varepsilon_{+,+}^{(n)} &=-\frac{1}{\pi} \int_{-\infty}^\infty d \th\, e^{\th(2n-1)} \left[ \bar{L}_{+,+}(\th)+\bar{L}_{-,-}(\th)\right]\\
\bar{\varepsilon}_{+,+}^{(n)} &=-\frac{1}{\pi} \int_{-\infty}^\infty d \th\, e^{\th(2n-1)} \left[ L_{+,+}(\th)+L_{-,-}(\th)\right]\,.
\ea
\ee
We emphasize that a single solution to the TBA equations allows to compute directly all infinite higher orders (albeit numerical precision might require special care). A comparison between these higher TBA modes and numeric WKB on $\ln Q$ is shown in tables \ref{tab:higherWKBNf1} and \ref{tab:higherWKBNf2}.

Now we want to connect the $\theta \to +\infty$ asymptotic expansion of the pseudoenergy $\varepsilon$ to that of the gauge periods, defined as
\be
a_k (\th,u,\mathbf m,\Lambda_{N_f}) \doteq \sum_{n=0}^\infty e^{-2n\th } a_k^{(n)}(u,\mathbf m,\Lambda_{N_f}) \qquad \th \to +\infty \,.
\ee
Interestingly, in the gauge theory literature we find analytic expressions for the first orders $a_k^{(n)}$, as differential operators acting on the leading order $a_k^{(0)}$. For the first 2 orders, they are the following: for $N_f=1$
\be \label{oper-higher}
\ba
a_k^{(1)}(u,m,\Lambda_1)=\left(\frac{\Lambda_1}{2 }\right)^2\frac{1}{12} &\left[\frac{\partial}{\partial u}+2 m\frac{\partial}{\partial m}\frac{\partial}{\partial u} +2 u \frac{\partial^2}{\partial u^2} \right]a_k^{(0)}(u,m,\Lambda_1)\\
a_k^{(2)}(u,m,\Lambda_1) =\left(\frac{\Lambda_1}{2 }\right)^4\frac{1}{1440}&\Biggl[28 m^2 \frac{\partial^2}{\partial u^2}\frac{\partial^2}{\partial m^2}+28 u^2 \frac{\partial^4}{\partial u^4}+132 m \frac{\partial^2}{\partial u^2}\frac{\partial}{\partial m}+56 m u \frac{\partial^3}{\partial u^3}\frac{\partial}{\partial m}\\&+81\frac{\partial^2}{\partial u^2} +124 u \frac{\partial^3}{\partial u^3}\Biggr]a_k^{(0)}(u,m,\Lambda_1)\,,
\ea
\ee
and for $N_f=2$~\cite{ItoKannoOkubo:2017}
\be 
\ba  \label{diffop11}
a_k^{(1)}(u,m_1,m_2,\Lambda_2) &= \left(\frac{\Lambda_2}{4 }\right)^2\frac{1}{6} \left[2 u \frac{\partial^2}{\partial u^2} +\frac{3}{2}\left(m_1 \frac{\partial }{\partial m_1}\frac{\partial }{\partial u}+ m_2 \frac{\partial }{\partial m_2} \frac{\partial }{\partial u} \right) +\frac{\partial }{\partial u} \right] a_k^{(0)}(u,m_1,m_2,\Lambda_2) \\
a_k^{(2)}(u,m_1,m_2,\Lambda_2)&=\left(\frac{\Lambda_2}{4 }\right)^4 \frac{1}{360} \Biggl [ 28 u^2 \frac{\partial^4}{\partial u^4} +120 u\frac{\partial^3}{\partial u^3} + 75\frac{\partial^2}{\partial u^2}  +42 \left(u  m_1 \frac{\partial}{\partial m_1}\frac{\partial^3}{\partial u^3} +u m_2\frac{\partial}{\partial m_2}\frac{\partial^3}{\partial u^3} \right)\\
& + \frac{345}{4} \left( m_1 \frac{\partial}{\partial m_1} \frac{\partial^2}{\partial u^2} +m_2 \frac{\partial}{\partial m_2} \frac{\partial^2}{\partial u^2}\right) +\frac{63}{4} \left(m_1^2 \frac{\partial^2}{\partial m_1^2} \frac{\partial^2}{\partial u^2}+m_2^2 \frac{\partial^2}{\partial m_2^2} \frac{\partial^2}{\partial u^2} \right)  \\
&+ \frac{126}{4} m_1 m_2  \frac{\partial}{\partial m_1} \frac{\partial}{\partial m_2} \frac{\partial^2}{\partial u^2} \Biggr ] a_k^{(0)}(u,m_1,m_2,\Lambda_2)\,.
\ea 
\ee
As we explained in~\cite{FioravantiGregori:2019}, the same operators can be used also to obtain the higher orders $\ln Q^{(n)}$, essentially because they can be derived from the WKB integrands, which are the same for both the periods $a_k$ and the $\ln Q$ function. Then, since we proved the identification between the leading orders $\varepsilon^{(0)}\propto a_k^{(0)}$, it follows that the same identification holds also for all higher asymptotic orders. The only difference is an alternating sign due to the fact that differential operators are odd under the inversion $u\to - u$, which the identification requires.  Therefore, we find the same higher WKB orders of $a_k$ to be given by the asymptotic expansion of the gauge TBAs~\eqref{ga-TBA} and~\eqref{ga-TBA2}, through the following identifications: for $N_f=1$
\be\label{highereps}
\ba
\ve^{(n)}_{+,k}
&=(-1)^n\frac{4\pi }{ \Lambda_1} \, a_1^{(n)}(-u_k,m_k,\Lambda_1)\,,
\ea
\ee
and similarly for $N_f=2$
\be 
\ba
\varepsilon_{+,+}^{(n)} 
&=(-1)^n\frac{8 \pi  }{\Lambda_2} a_2^{(n)}(-u,-i m_1, i m_2,\Lambda_2) \\
\bar{\varepsilon}_{+,+}^{(n)} 
&=(-1)^n\frac{8 \pi  }{\Lambda_2} a_2^{(n)}(u,m_1,  m_2,\Lambda_2)\,.
\ea
\ee
We show also a numerical proof of this in tables \ref{tab:periods1WKB} and \ref{tab:periods11WKB}. 
\begin{table}[t]
    \centering
  \small  \begin{tabular}{c|c|c|c|c|c}
$n$&    $u$& $m$  &  TBA $\ve^{(n)}(u,i m,1)$ &WKB $ \ve^{(n)}(u,i m,1)$ &WKB $ 4\pi (-1)^n a_1^{(n)}(-u,m,1) $\\
    \hline
$0$& $ 0.1$& $\frac{\sqrt 2}{40}$ &  $ 3.79071 $ &  $3.79071  $ &$ 3.79071 $\\
$1$& $ 0.1$& $\frac{\sqrt 2}{40}$ &  $-0.203510$ &  $-0.203514$ &$-0.203513$\\
$2$& $ 0.1$& $\frac{\sqrt 2}{40}$  & $0.00807458$ &  $0.00807461$ &$0.00807461$\\
\hline
$0$& $ 0.1 e^{2 \pi i/3}$& $\frac{\sqrt 2}{40} e^{-2 \pi i/3}$  & $4.80766\, -1.10016 i $ &  $4.80766\, -1.10016 i $ &$ 4.80766\, -1.10016 i$\\
$1$& $ 0.1 e^{2 \pi i/3}$& $\frac{\sqrt 2}{40} e^{-2 \pi i/3}$  & $-0.198768+0.00273793 i$ &  $-0.198772+0.00273796 i$ &$-0.198771+0.00273796 i$\\
$2$& $ 0.1 e^{2 \pi i/3}$& $\frac{\sqrt 2}{40} e^{-2 \pi i/3}$ &  $-0.0040373+0.00880082 i$ &  $-0.0040373+0.00880084 i$ &$-0.0040373+0.00880084 i$\\
\hline
$0$& $ 0.1 e^{-2 \pi i/3}$& $\frac{\sqrt 2}{40} e^{2 \pi i/3}$  & $ 4.80766\, +1.10016 i$ &  $4.80766\, +1.10016 i $ &$ 4.80766\, +1.10016 i$\\
$1$& $ 0.1 e^{-2 \pi i/3}$& $\frac{\sqrt 2}{40} e^{2 \pi i/3}$  & $-0.198768-0.00273793 i$ &  $-0.198772-0.00273796 i$ &$-0.198771-0.00273796 i$\\
$2$& $ 0.1 e^{-2 \pi i/3}$& $\frac{\sqrt 2}{40} e^{2 \pi i/3}$ &  $-0.0040373-0.00880082 i$ &  $-0.0040373-0.00880084 i$ &$-0.0040373-0.00880084 i$
    \end{tabular}
    \caption{Comparison of leading and first two higher $\hbar \to 0$ orders as computed from the $N_f=1$ TBA~\eqref{ga-TBA}, numeric WKB on $\ln Q$ \eqref{eps0gau} and the differential operators~\eqref{oper-higher} on the leading elliptic integrals for the periods~\eqref{perEllipticNf1}.  
      }
    \label{tab:periods1WKB}

\begin{tabular}{c|c|c|c|c|c|c}
$n$&$u$&$m_1$&$m_2$&$\mu_n$& TBA $\varepsilon^{(n)}(u,m_1,m_2,4)$ & WKB $ 2\pi (-1)^n  \left[a_D^{(n)}(-u,-im_1,i m_2 ,4)+\mu_n (m_1-m_2) \right]$\\
\hline
$0$&$1$&$1/8$&$1/8$&$-i/2$&$1.65590 $&$ 1.65590$\\$1$&$1$&$1/8$&$1/8$&$-i/2$&$-0.239514$&$-0.239513$\\
$2$&$1$&$1/8$&$1/8$&$-i/2$&$0.0158875$&$0.0158874$\\
\hline
$0$&$-1$&$-1/8i$&$1/8i$&$0$&$4.57036 $&$4.57036 $\\
$1$&$-1$&$-1/8i$&$1/8i$&$0$&$-0.502486$&$-0.502484$\\
$2$&$-1$&$-1/8i$&$1/8i$&$0$&$0.312001$&$0.312000$\\
\hline
$0$&$1$&$1/16$&$1/8$&$-i/2$&$1.64329+0.19635 i $&$ 1.64329+0.19635 i$\\
$1$&$1$&$1/16$&$1/8$&$-i/2$&$-0.237931$&$-0.237930$\\
$2$&$1$&$1/16$&$1/8$&$-i/2$&$0.0151357$&$0.0151358$\\
\hline
$0$&$-1$&$-1/16i$&$1/8i$&$0$&$ 4.77062$&$4.77062 $\\
$1$&$-1$&$-1/16i$&$1/8i$&$0$&$-0.500006$&$-0.500005$\\
$2$&$-1$&$-1/16i$&$1/8i$&$0$&$0.2941800$&$0.2941802$
\end{tabular}
\caption{Comparison of leading and higher asymptotic orders $\varepsilon^{(n)}$ as computed from the $N_f=2$ gauge TBA~\eqref{ga-TBA2} and $a_D^{(n)}$ from elliptic integrals (through differential operators~\eqref{diffop11} on~\eqref{perEllipticNf2}).}
\label{tab:periods11WKB} 
\end{table}
\normalsize
From this considerations it follows that we can write the asymptotic expansion for the pseudoenergy in terms fo the gauge periods, for $N_f=1$
\be 
\ba \label{NSYexp1}
\ve(\th,u,i m,\Lambda_1) &\doteq \sum_{n=0}^\infty e^{\th (1-2n)}\ve^{(n)}_{+,0}= \frac{4 \pi}{\Lambda_1}\sum_{n=0}^\infty e^{\th (1-2n)}(-1)^n a^{(n)}_1(-u, m,\Lambda_1)\,,\qquad \th \to +\infty\,,
\ea 
\ee
and for $N_f=2$
\be 
\ba 
&\ve(\th,u,  m_1,  m_2,\Lambda_2) \doteq \sum_{n=0}^\infty e^{\th (1-2n)}\ve^{(n)}_{+,+}\\
&= \frac{8 \pi}{\Lambda_2}\left[e^\th a^{(0)}_D(-u,-i m,+i m_2 , \Lambda_2)-\frac{1}{2 }(i m_1- i m_2) +\sum_{n=1}^\infty e^{\th (1-2n)}(-1)^n a^{(n)}_D(-u,-i m,+i m_2 , \Lambda_2) \right],\,\,\,\th \to +\infty\,.
\ea 
\ee

\begin{figure}[t]
    \centering
    \includegraphics[width=0.8\textwidth]{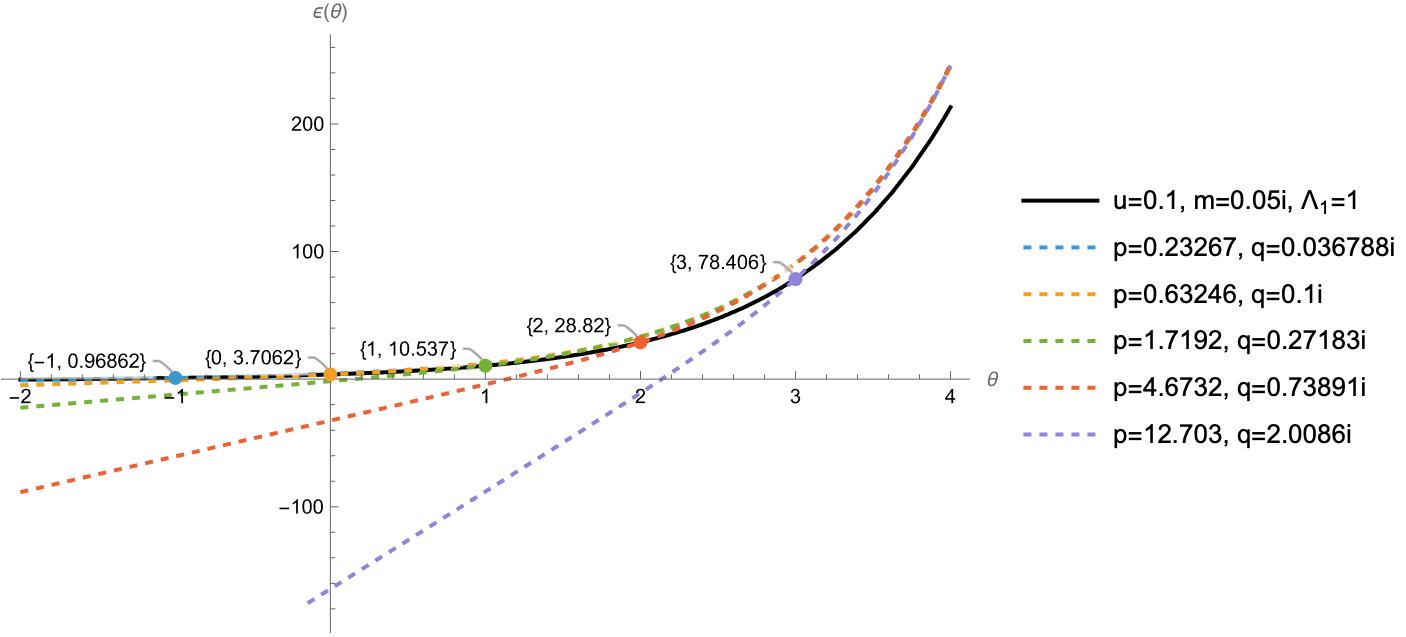}
    
\includegraphics[width=0.75\textwidth]{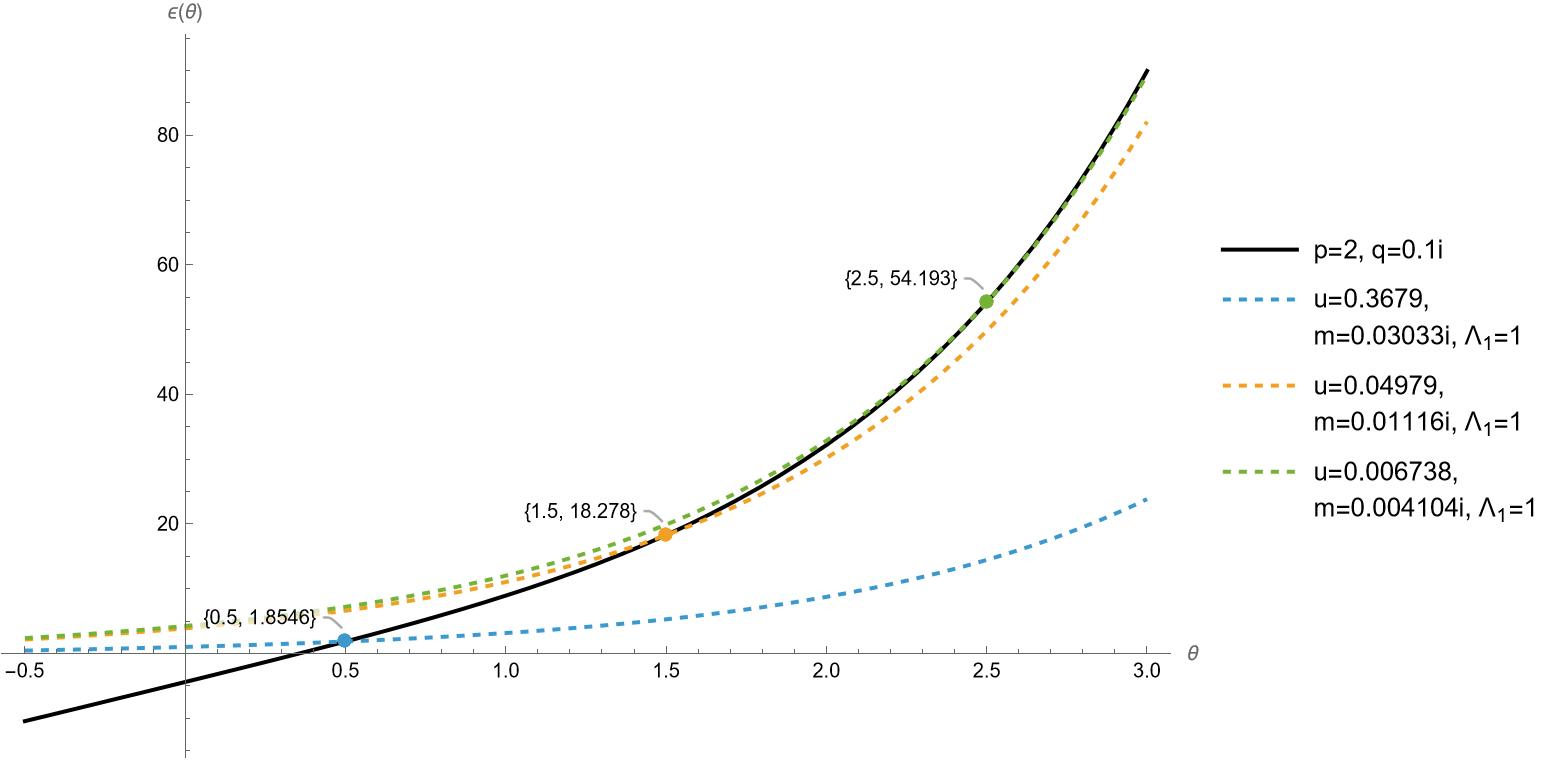}
    \caption{Plots showing the equivalence between the $N_f=1$ gauge and Perturbed Hairpin IM pseudoenergies $\ve(\th,u,m,\Lambda_1)$ and $\ve(\th,p,q)$, solutions of the TBAs \eqref{ga-TBA} and \eqref{int-TBA}. On the top a single gauge solution (in black) is matched by several integrability solutions (colored, dashed), while on the bottom the converse. The parameters match according to the dictionary \eqref{DictGau1}.}
    \label{fig:plotequivTBA1}
\end{figure}

\begin{figure}[t]
    \centering
    \includegraphics[width=0.8\textwidth]{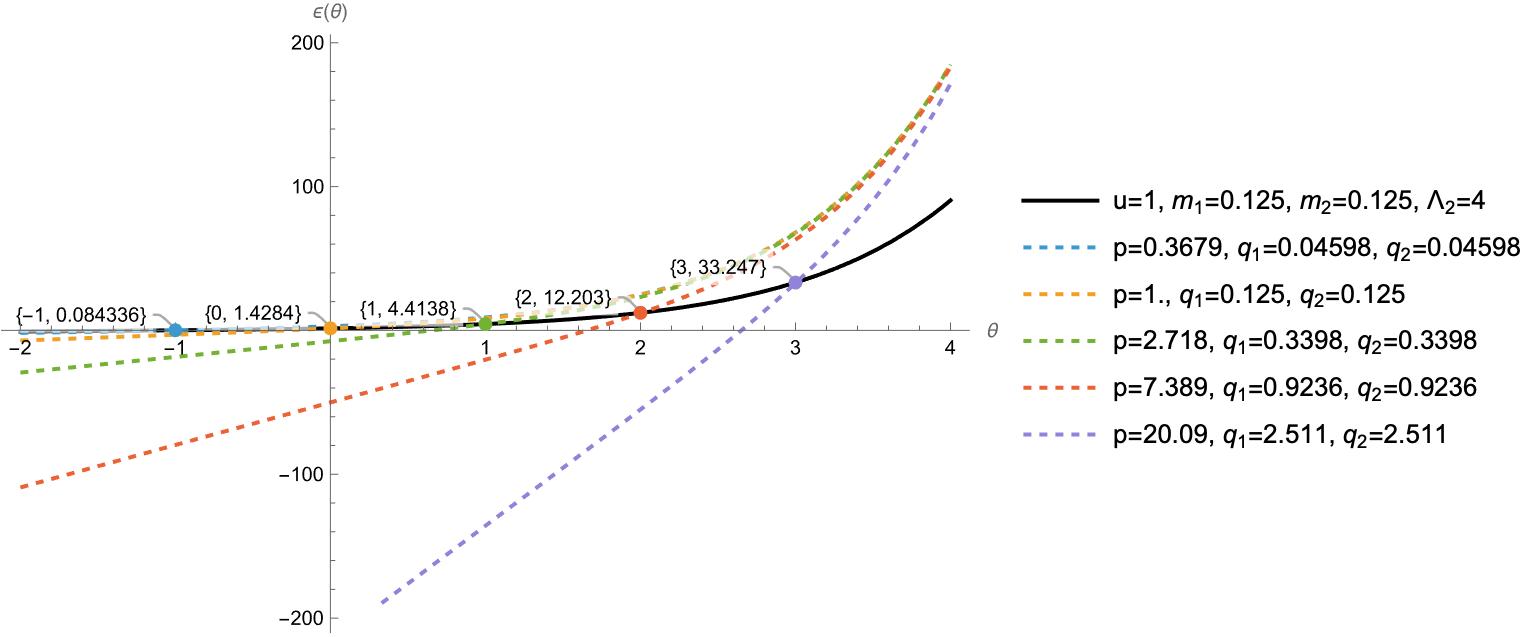}
    
\includegraphics[width=0.75\textwidth]{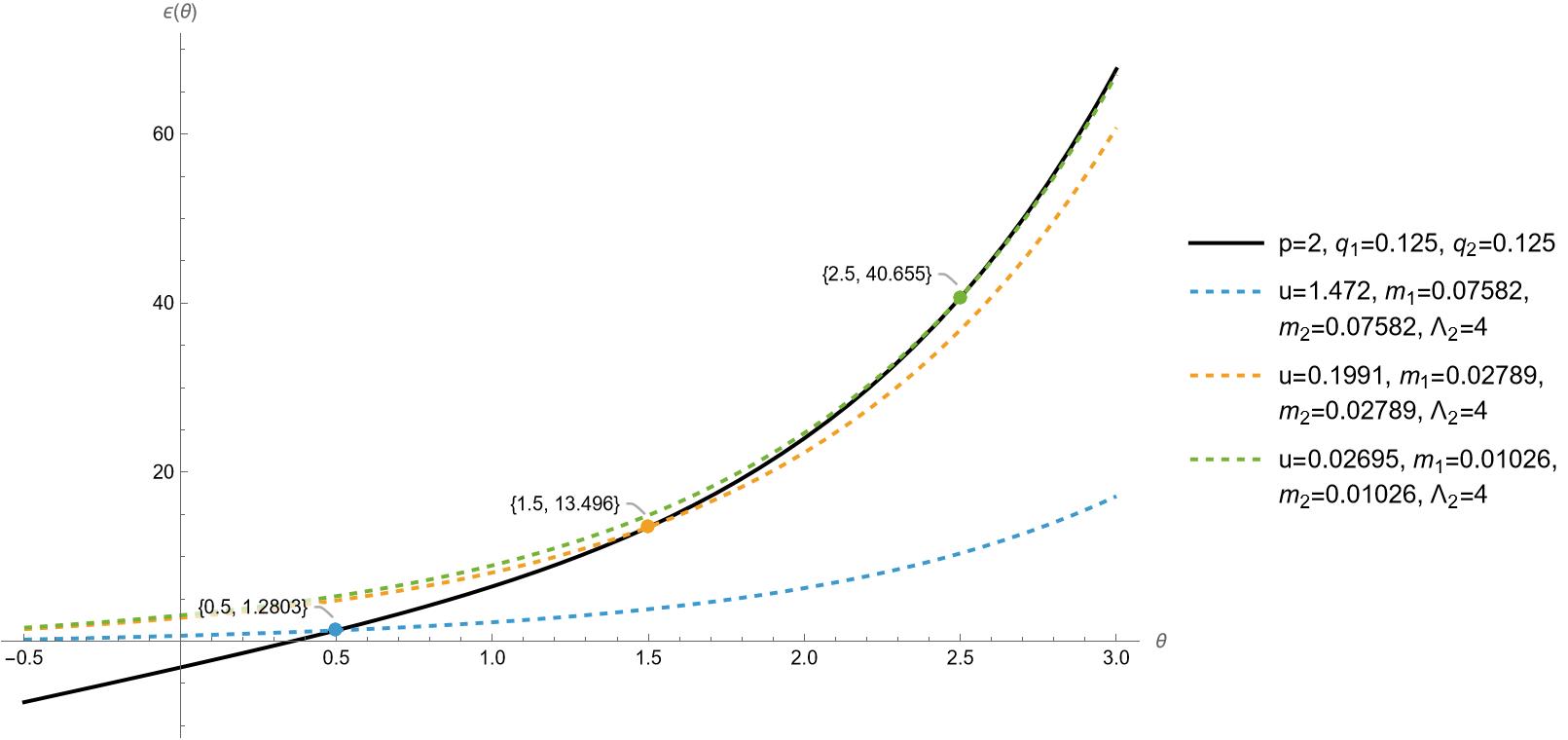}

    \caption{Plots showing the equivalence between the $N_f=2$ gauge and Generalized Perturbed Hairpin IM pseudoenergies $\ve(\th,u,m_1,m_2,\Lambda_2)$ and $\ve(\th,p,q_1,q_2)$, solutions of the TBAs \eqref{ga-TBA2} and \eqref{int-TBA2}. On the top a single gauge solution (in black) is matched by several integrability solutions (colored, dashed), while on the bottom the converse. The parameters match according to the dictionary \eqref{DictGau2}.}
    \label{fig:plotequivTBA2}
\end{figure}

Now, since the asymptotic expansion modes of the gauge TBA are the periods, we can think at the exact gauge pseudoenergy $\varepsilon$ as defining the exact periods $a_k$\footnote{The equivalence of this definition to the integral definition is argued in appendix~\ref{appProof} for the $N_f=0$ theory.}. Then we can prove the identification between the gauge periods and the integrability pseudoenergy, at the exact level, just by numerically checking that the gauge TBA and integrability TBA return the same values under the parameters maps \eqref{DictGau1}, \eqref{DictGau2}: 
\be
\ba  \label{numTBAeq}
\varepsilon(\th ,p, q)&=\varepsilon(\th ,u,m,\Lambda_1)\,,\qquad  &\frac{u}{\Lambda_1^2}=\frac{1}{4}p^2e^{-2\theta }\qquad &\frac{m}{\Lambda_1}=\frac{1}{2 }\,q\, e^{-\theta_0} \,,\\
\varepsilon(\th,p,q_1,q_2) &= \varepsilon(\th,u,m_1,m_2,\Lambda_2)\,, \qquad &\frac{u}{\Lambda_2^2} = \frac{1}{16} p^2 e^{-2\th} \, \quad &\frac{m_i}{\Lambda_2} = \frac{1}{4} q_i e^{-\theta}\,.
\ea
\ee
In other words, the integrability and gauge TBAs are equivalent.
The numerical check of this is shown in figures \ref{fig:plotequivTBA1} and \ref{fig:plotequivTBA2}.

However, further physical considerations are in order.
By the ODE/IM derivation of section \ref{sec:QIntExact}, the exact periods we are computing through TBA are integrals of $\CP(y)$, the solution of the Riccati equation, identified as the Seiberg-Witten quantum differential (see appendix~\ref{appProof}). However, more often in gauge theory the gauge periods they are defined from the perturbative $\Lambda_{N_f}\to 0$ Nekrasov-Shatashvili (NS) instanton series. Letting also $\hbar\to 0$, they are the following: for $N_f=1$~\cite{ItoKannoOkubo:2017}
\be
\ba \label{ainst}
a(\theta,u,m,\Lambda_1) &=\sqrt{u}-\frac{\Lambda_1^3 m \left(\frac{1}{u}\right)^{3/2}}{2^4 }+\frac{3\Lambda_1^6 \left( \frac{1}{u}\right)^{5/2}}{2^{10} }+... \\
&+\hbar(\theta)^2 \left( -\frac{\Lambda_1^3 m \left(\frac{1}{u}\right)^{5/2}}{2^6 }+\frac{15 \Lambda_1^6 \left( \frac{1}{u}\right)^{7/2}}{2^{12} }- \frac{35 \Lambda_1^6 m^2 \left(\frac{1}{u} \right)^{9/2}}{2^{11} }+...\right)\,, 
\ea 
\ee 
\be 
\ba \label{aDinst}
a_D(\th,u,m,\Lambda_1) &=\frac{i}{2 \pi}\Biggl[a(\theta,u,m,\Lambda_1) \left(i \pi - 3 \ln \frac{16 u}{\Lambda_1^2}\right)+\left(6\sqrt{u}+\frac{m^2}{\sqrt{u}}+\frac{\frac{m^4}{6}-\frac{1}{4}\Lambda_1^3 m}{u^{3/2}}+...\right)\\
&\quad+\hbar(\theta)^2 \left( -\frac{1}{4\sqrt{u}}-\frac{m^2}{12 u^{3/2}}+\frac{-\frac{9}{64}\Lambda_1^3 m -\frac{m^4}{12}}{u^{5/2}}+...\right)\Biggr]\,,
\ea 
\ee 
and similarly for $N_f=2$ (with equal masses for simplicity)
\be
\ba \label{ainst2}
a(\theta,u,m,m,\Lambda_2)&=\sqrt{u}-\frac{\Lambda _2^2 m^2}{16 u^{3/2}}+\frac{\Lambda _2^4 \left(-15 m^4+6 m^2 u-u^2\right)}{1024 u^{7/2}}\\
&+ \hbar^2(\theta) \left( -\frac{\Lambda _2^2 m^2}{64 u^{5/2}}+\frac{\Lambda _2^4 \left(-35 m^4+15 m^2 u-2 u^2\right)}{2048 u^{9/2}}\right)\,,
\ea
\ee
\be
\ba \label{aDinst2}
a_D(\theta,u,m,m,\Lambda_2)&=\frac{i}{2\pi } \Biggl [ 2 a(\theta,u,m,m,\Lambda_2) \left (i \pi - 2\ln \frac{8u}{\Lambda_2}-\ln(u-m^2) \right)  \\
&+\hbar(\theta) ^2\left(-\frac{1}{6 \sqrt{u}}+\frac{m^2}{12 u^{3/2}}+\frac{m^4}{4 u^{5/2}}+... \right)
\Biggr ]\,.
\ea
\ee
\begin{table}[t]
    \centering
    \begin{tabular}{c|c|c|c|c}
    $\theta$& $p$ &$q$ & $\varepsilon(\theta,p,q)$ TBA & $\frac{2 \pi}{\hbar} (a-a_D -\frac{i}{2}m) $ NS\\
    \hline
     $-5$&$5$&$0.1 i$&$-188.815$&$-188.815$\\
      $-2.5$&$5$&$0.1 i$&$-113.815$&$-113.815$\\
      $0$&$5$&$0.1 i$&$-38.8108$&$-38.8234$\\
      \hline
    $-5$&$10$&$0.1 i$&$-419.457$&$-419.456$\\
    $-2.5$&$10$&$0.1 i$&$-269.457$&$-269.456$\\
     $0$&$10$&$0.1 i$&$-119.457$&$-119.457$
 \bigskip
    \end{tabular}

 \begin{tabular}{c|c|c|c|c|c}
    $\theta$& $p$ &$q_1$&$q_2$ & $\varepsilon(\theta,p,q_1,q_2)$ TBA & $\frac{2 \pi}{\hbar} a_D $ NS\\
    \hline
    $-5$&$5$&$0.1 $&$0.1$&$-126.124$&$-126.122 $\\   $-2.5$&$5$&$0.1 $&$0.1$&$-76.1208$&$-76.1223 $\\   $0$&$5$&$0.1 $&$0.1$&$-26.1022$&$-26.1024 $\\
    \hline
 $-5$&$10$&$0.1 $&$0.1$&$-279.869$&$-279.865$\\
  $-2.5$&$10$&$0.1 $&$0.1$&$-179.866$&$-179.865 $\\ $0$&$10$&$0.1 $&$0.1$&$-79.8598$&$-79.8615 $   \end{tabular}
    \caption{Tables showing the match between the integrability pseudoenergies, solutions of the TBAs \eqref{int-TBA} and \eqref{int-TBA2}, in the $\theta \lesssim 0$ non-perturbative region and the NS instanton expansion for the gauge periods \eqref{ainst}-\eqref{aDinst2}, for $N_f=1,2$ in the top and bottom respectively.}
    \label{tab:inst}
\end{table}
By comparing these expansions with the exact pseudoenergy - as shown in table~\ref{tab:inst} - we effectively test our identications in the opposite regime $\theta \lesssim 0$. 
So we can finally state the following gauge-integrability \emph{exact} identifications. For the $N_f=1$ theory, for $u,m,\Lambda_1>0$:
\be \label{gauint1}
\ve(\th,p, q) = \frac{2\pi  i}{\h(\th-i\pi/2)}\left[a(\th-i\pi/2,-u,- i m,\Lambda_1)-a_D(\th-i\pi/2,-u,-i m,\Lambda_1) -\frac{i}{2}m  \right] \,,
\ee 
or more generally for $u,m \in \mathbb{C}$, with $\arg u = - \arg m$
\be 
\ba\label{gaupseudocycles}
\ve(\th,p,q) &=
\frac{2\pi  \,i}{\h(\th-i\pi/2)}\, a_1(\th-i\pi/2,-u,-i m)\,. 
\ea 
\ee 
Similarly for the $N_f=2$ theory, for $u,m,\L_2>0$:\footnote{We remark that other two pseudoenergies imply relations with imaginary $p$ parameter, which are not directly implemented in the integrability variables (since the integrability TBA does not converge). However, they will be implemented in the gravity variables in section~\ref{gravity} (since in~\eqref{quanteps2} precisely this range of parameters is involved).}
\be \label{gauint2}
\ba
\varepsilon(\th,p,q_1,q_2)&=\frac{2  \pi i}{\hbar(\theta-i \pi/2)}\left[a_D(\th-i \pi/2,-u,-i m_1,i m_2,\Lambda_2) -\frac{i}{2 } (m_1-m_2) \right] \,.
\ea
\ee
Relations~\eqref{gauint1}-\eqref{gauint2} show a new connection between the $SU(2)$ $N_f=1,2$ gauge periods and the $Y$ function (Generalized) Perturbed Hairpin integrable model. This generalizes to the case of massive hypermultiplets matter the integrability-gauge correspondence already developed for the $SU(2)$ $N_f=0$ and the self-dual Liouville model (\textit{cf.} the first~\eqref{QTcycles0}, with $Q=\sqrt{Y}$)~\cite{FioravantiGregori:2019}. 


We conclude this section with some observations. The RHS of \eqref{gauint1} and~\eqref{gauint2} are expressions for a $N_f=1,2$ SW exact central charge. As explained in appendix~\ref{appProof} by considering different particles in the spectrum, different relations could be found (like those for the $N_f=0$ and $N_f=1$ theory in~\cite{GrassiGuMarino,GrassiHaoNeitzke:2021}). Besides, we remark that these gauge-integrability identifications hold as they are written only in a restricted strip of of the complex $\th$ plane: $\Im \th<\pi/3$ and $\Im \th < \pi/2 $ for the $N_f=1$ and $N_f=2$ theory. Beyond such strips the gauge TBAs~\eqref{ga-TBA}~\eqref{ga-TBA2} requires analytic continuation (of its solution) since poles of the kernels are found on the $\th'$ integrating axis. A modification of TBA equations, as usually done in integrability by adding the residue, is possible, but then the $Y$s no longer identifies with the gauge periods: in fact the former are entire functions while the latter are not~\cite{Fioravanti-DDV-1996,DestriDeVega:1997,CecottiDelZotto:2014,ItoMarinoShu:2018}. A similar singular behaviour is found using the $y$ integral definition of gauge periods, as explained in appendix~\ref{appProof}. This is a manifestation of the so-called wall-crossing phenomenon, whereby the spectrum of SW theory changes and therefore a fundamental change in its relation to integrability is to be expected. We hope to investigate further on this issue in the future.

\section{Integrability $T$ function and gauge periods} \label{secT}

In the previous section, we established an identification between the $Y$ function and the gauge periods. In this subsection, we want to prove a similar identification for the $T$ functions. To do that, in the first subsection we will first establish a connection between (quadratic combinations of) the $T$ functions and the Floquet exponent $\nu$. Then, in the second subsection we will identify the latter with the gauge period $a$.

\subsection{$T$ function and Floquet exponent}

One says $\nu$ is a Floquet or characteristic exponent of the ODEs \eqref{ODEint1},~\eqref{ODEint2}, if and only if $e^{\pm 2 \pi i\nu}$ are eigenvalues of the periodicity operator $\Upsilon$, defined as
\be 
\Upsilon \psi(y) = \psi(y+2 \pi i)\,.
\ee
Now observe we can express $\Upsilon$ in terms of the $\Omega_\pm$ symmetry operators \eqref{OmegaSym1}, \eqref{OmegaSym2}, for $N_f=1$ as
\be 
\Upsilon  =  \Omega_+^2 \Omega_-^{-1}\,,
\ee 
and for $N_f=2$ as
\be 
\Upsilon  =  \Omega_+^2 \Omega_-^{-2}\,.
\ee 
So we rewrite the lateral connection relations \eqref{connLatNf1}, \eqref{connLatNf2}, for $N_f=1$
\be 
\ba
\psi_{+,-1}(y+2\pi i) =\psi_{+,1}&= -e^{2 \pi i q}\psi_{+,-1} +i e^{i \pi q} \tilde{T}_{+}(\th )\psi_{+,0} \\
\psi_{+,0}(y+2\pi i) =\psi_{+,2}&=- e^{i \pi q} \tilde{T}_{-}(\th+i \pi/3)\psi_{+,-1} +[-e^{-2\pi i q}+ \tilde{T}_{-}(\th+i \pi/3) \tilde{T}_{+}(\th )]\psi_{+,1}\,,
\ea
\ee
and for $N_f=2$
\be 
\ba
\psi_{+,-1}(y+2\pi i) =\psi_{+,1}&= -e^{2 \pi i q_1}\psi_{+,-1} +i e^{i \pi q_1} \tilde{T}_{+,+}(\th )\psi_{+,0} \\
\psi_{+,0}(y+2\pi i) =\psi_{+,2}&=- e^{i \pi q_1} \tilde{T}_{-,+}(\th+i \pi/2)\psi_{+,-1} +[-e^{-2\pi i q_1}+ \tilde{T}_{-,+}(\th+i \pi/2) \tilde{T}_{+,+}(\th )]\psi_{+,1}\,.
\ea
\ee
We can write these also in matrix form as
\be 
\Upsilon  \psi_+ = \mathcal{T}_+ \psi_+\,,
\ee
where we defined the vector $\psi=(\psi_{+,-1},\psi_{+,0})$ and the lateral connection matrices, for $N_f=1$
\be 
\mathcal{T}_+ = \begin{pmatrix}
-e^{2 \pi i q} & e^{i \pi q} \tilde{T}_{+,+}(\th ) \\
e^{i \pi q} \tilde{T}_{-}(\th+i\pi/3) & [-e^{-2\pi i q}+\tilde{T}_{-}(\th+i\pi/3)\tilde{T}_{+}(\th ) ]
\end{pmatrix}\,,
\ee
and for $N_f=2$
\be 
\mathcal{T}_+ = \begin{pmatrix}
-e^{2 \pi i q_1} & e^{i \pi q_1} \tilde{T}_{+,+}(\th ) \\
e^{i \pi q_1} \tilde{T}_{-,+}(\th+i\pi/2) & [-e^{-2\pi i q_1}+\tilde{T}_{-,+}(\th+i\pi/2)\tilde{T}_{+,+}(\th ) ]
\end{pmatrix}\,.
\ee
Since $e^{\pm 2 \pi i\nu}$ are eigenvalues of $\mathcal{T}_+ $, it then follows that $\nu$ is determined from the following relation:
\be 
2\cos 2 \pi \nu = \rm{tr} \, \mathcal{T}_+ \,.
\ee
This reads more explicitly, for $N_f=1$
\be  \label{Ttnu1}
2 \cos 2\pi \nu + 2 \cos 2 \pi q 
= \tilde{T}_{+}(\th )\tilde{T}_{-}(\th+i\frac{\pi}{3})= \tilde{T}_{+}(\th )\tilde{T}_{-}(\th-i\frac{\pi}{3})\,,
\ee 
and for $N_f=2$
\be  \label{Ttnu2}
2 \cos 2\pi \nu + 2 \cos 2 \pi q_1 
= \tilde{T}_{+,+}(\th )\tilde{T}_{-,+}\,.(\th+i\frac{\pi}{2})=\tilde{T}_{+,+}(\th )\tilde{T}_{-,-}\,.(\th)\,,
\ee 
where in the last equalities we used the periodicity properties \eqref{Tper1} and \eqref{Tper2}. 
Similarly we can prove relations for $T$, for $N_f=1$
\be  \label{Tnu1}
2 \cos 4\pi \nu  +2 
= T_{+}(\th)T_{+}(\th+i \frac{2\pi}{3})=T_{+}^2(\th)\,,
\ee 
(through $\Upsilon^2$) and for $N_f=2$
\be \label{TTnuNf=2}
2 \cos 2\pi \nu + 2 \cos 2 \pi q_2 
= T_{+,+}(\th)T_{+,-}(\th+i \frac{\pi}{2})=T_{+,+}(\th)T_{-,-}(\th)\,.
\ee 
\begin{table} 
\centering
$\begin{array}{c|c|c|c|c}
\th & p & q   &T(\th,p,q) \text{ TBA} & 2 \cos 2\pi \nu(\theta,p,q) 
\text{ Hill} \\
\hline
-8 &0.2&0.3&0.618034& 0.618034 \\
-6 &0.2&0.3 &0.618033 & 0.618033 \\
-4&0.2&0.3 &0.617763 & 0.617772 \\
-2&0.2&0.3 & 0.511941 &0.511946\\
-1&0.2&0.3 & -1.62208&-1.62205\\
0 &0.2&0.3&-81.7391&-81.7392\\
1&0.2&0.3 &63194.9 &63194.9 \\
2&0.2&0.3&-4.39743\cdot 10^{11} &-4.39743\cdot 10^{11} \\
3&0.2&0.3&-1.16022\cdot10^{33}&-1.16022\cdot10^{33}\\
4&0.2&0.3&-1.87911\cdot10^{92}
&-1.87912\cdot10^{92}
\end{array}$
\caption{Comparison of $T $, as computed from the TBA \eqref{int-TBA} and $TQ$ system \eqref{TQ1}, with \eqref{TBAQ1} and \eqref{TBAQCont1}, and $2 \cos 2 \pi \nu $, as computed from the Hill's determinant (\textit{cf.} appendix~\ref{appHill}). This confirms relation \eqref{TnuFL} for the $N_f=1$ theory.}
\label{tabFLconj}
\end{table}
We also notice that for $N_f=1$, from the $T$ periodicity $T_+(\th+i\pi/3) = T_-(\th)$ it follows the Floquet (anti)-periodicity
\be \label{nuper}
\nu(\th+i \frac{\pi}{3},-q) =\nu(\th,q)= \pm\nu(\th-i \frac{\pi}{3},-q)   \qquad \rm{mod}(n) \in \mathbb{Z}\,.
\ee
Thus we prove the following conjecture of~\cite{FateevLukyanov:2005}: 
\be  \label{TnuFL}
T(\th,p,q) =2 \cos \{2\pi \nu(\theta,p,q)\}= \exp \{-2 \pi i \nu(\th+i\pi/3,p,-q)\}+ \exp \{ 2\pi  i\nu(\th-i\pi/3,p,-q)\}\,,
\ee 
which follows immediately from~\eqref{Tnu1} and~\eqref{nuper}.

In practice, the Floquet exponent $\nu$ can be
computed through the Hill determinant method~\cite{whittaker_watson_1996}, as explained in appendix~\ref{appHill}. Instead, $T_+(\theta)$, with $\theta \in \mathbb R$, can be computed as in \eqref{TQ1}, 
in terms of $Q_+(\theta)$, which is given directly from the TBA solution as in \eqref{TBAQ1}, and of $Q_+(\theta \pm 2 \pi i/3)$, which is obtained by analytic continuation of the same formula, by adding the residue of the integral kernel as follows:
\be
\ba \label{TBAQCont1}
 \ln &Q_+(\theta \pm 2 \pi i/3) 
 = -\frac{4   \sqrt{3\pi^{3}}}{\Gamma \left(\frac{1}{6}\right) \Gamma \left(\frac{1}{3}\right)} e^{\th\pm 2 \pi i/3}-( \th \pm \frac{2 \pi i}{3}+ \frac{1}{3} \ln 2) q \\
&+\frac{1}{2} \int_{-\infty}^\infty \frac{d\th'}{2\pi}\left \{ \frac{\ln[1+\exp \{-\varepsilon_+(\th')   \} ][1+\exp \{-\varepsilon_-(\th')  \} ]}{\cosh(\th\pm 2 \pi i/3-\th')}- i \frac{e^{\th'-\th\mp 2 \pi i/3}}{\cosh(\th\pm 2 \pi i/3-\th')} \ln\left[\frac{1+\exp \{-\varepsilon_-(\th')  \} }{1+\exp \{-\varepsilon_+(\th') \} }\right]\right \} \\
&+\frac{1}{2} \left\{\ln[1+\exp \{-\varepsilon_+(\th\pm i \pi/6)   \} ][1+\exp \{-\varepsilon_-(\th\pm i \pi/6)  \} ]\mp\ln\left[\frac{1+\exp \{-\varepsilon_-(\th\pm i \pi/6)  \} }{1+\exp \{-\varepsilon_+(\th\pm i \pi/6) \} }\right] \right\}\,.
\ea
\ee
Similarly for $N_f=2$ $T_{+,+}(\theta)$, with $\theta \in \mathbb R$, can be computed as in \eqref{TQ2}, 
in terms of $Q_{+,+}(\theta)$, which is given by the TBA solution as in \eqref{TBAQ2}, and of $Q_{+,-}(\theta \pm i \pi /2)$, by adding half of the residue of the integral kernel as follows:
\be
\ba \label{TBAQCont2}
 \ln &Q_{+,+}(\theta \pm i \pi/2) 
 = -\frac{4   \sqrt{\pi^{3}}}{\Gamma \left(\frac{1}{4}\right)^2 } e^{\th\pm i \pi/2}-( \th \pm \frac{i \pi}{2}+ \frac{1}{2} \ln 2) (q_1+q_2) \\
&\mp\frac{i}{2} \mathrm{P}\int_{-\infty}^\infty \frac{d\th'}{2\pi}\left \{ \frac{\ln[1+\exp \{-\varepsilon_{+,-}(\th')   \} ][1+\exp \{-\varepsilon_{-,+}(\th')  \} ]}{\sinh(\th-\th')}- i \frac{e^{\th'-\th\mp i\pi/2}}{\sinh(\th-\th')} \ln\left[\frac{1+\exp \{-\varepsilon_{+,-}(\th')  \} }{1+\exp \{-\varepsilon_{-,+}(\th') \} }\right]\right \} \\
&+\frac{1}{4} \left\{\ln[1+\exp \{-\varepsilon_{+,-}(\th )   \} ][1+\exp \{-\varepsilon_{-,+}(\th )  \} ]\mp\ln\left[\frac{1+\exp \{-\varepsilon_{+,-}(\th )  \} }{1+\exp \{-\varepsilon_{-,+}(\th ) \} }\right] \right\}\,,
\ea
\ee
where the singular integral has to be evaluated as principal value. 
Proceeding in this way, we can show also a numerical proof of \eqref{TnuFL} and \eqref{TTnuNf=2} in tables~\ref{tabFLconj} and \ref{tab:TvsNuNf2}.

The above relations \eqref{Ttnu1}-\eqref{TTnuNf=2} between $T$ and $\nu$ generalize analogue ones previously found numerically for $SU(2)$ $N_f=0$ and $SU(3)$ $N_f=0$~\cite{ZamolodchikovMemorial,FioravantiGregori:2019,FioravantiPoghossian:2019}. Similar results were derived for the Doubly Confluent Heun equation in~\cite{Ronveaux:1995}.

\begin{table}
\centering
$\begin{array}{c|c|c|c|c|c}
\theta & p& q_1 & q_2 &  T_{+,+}(\theta)T_{-,-}(\theta) \text{ TBA}& 2 \cos 2 \pi \nu(\theta,p,q_1,q_2) + 2 \cos 2 \pi q_2  \,\text{Hill}\\
\hline
 -10 & 0.2 & 1/8 & 1/8 & 2.03225 & 2.03225 \\
 -6 & 0.2 & 1/8 & 1/8 & 2.03222 & 2.03222 \\
 -2 & 0.2 & 1/8 & 1/8 & 1.94329 & 1.94329 \\
 -1 & 0.2 & 1/8 & 1/8 & 1.02423 & 1.02427 \\
 0 & 0.2 & 1/8 & 1/8 & -3.84272 & -3.84291 \\
 1 & 0.2 & 1/8 & 1/8 & -5134.58 & -5134.58 \\
 2 & 0.2 & 1/8 & 1/8 & 2.44329\times 10^{11} & 2.44330\times 10^{11} \\
 3 & 0.2 & 1/8 & 1/8 & 8.74674\times 10^{29} & 8.74675\times 10^{29} \\
 4 & 0.2 & 1/8 & 1/8 & -1.08768\times 10^{80} & -1.08769\times 10^{80}\\
\end{array}$
\caption{Comparison of $T$, as computed from the TBA \eqref{int-TBA2} and $TQ$ system \eqref{TQ2}, with \eqref{TBAQ2} and \eqref{TBAQCont2}, and $ \nu $, as computed from the Hill's determinant (\textit{cf.} appendix~\ref{appHill}). This confirms relation \eqref{TTnuNf=2} for the $N_f=2$ theory.}\label{tab:TvsNuNf2}
\end{table}


\subsection{Exact quantum gauge-integrability identification for $T$} \label{subsec:Ta}

In this subsection, we are going to prove the following identification between the Floquet exponent and the gauge period
\be \label{anu}
\nu =  \frac{a}{\hbar}\qquad \rm{mod}( n)\,,\quad n \in \mathbb{Z}\,.
\ee

\begin{figure}
\centering
\includegraphics[width=0.49\textwidth]{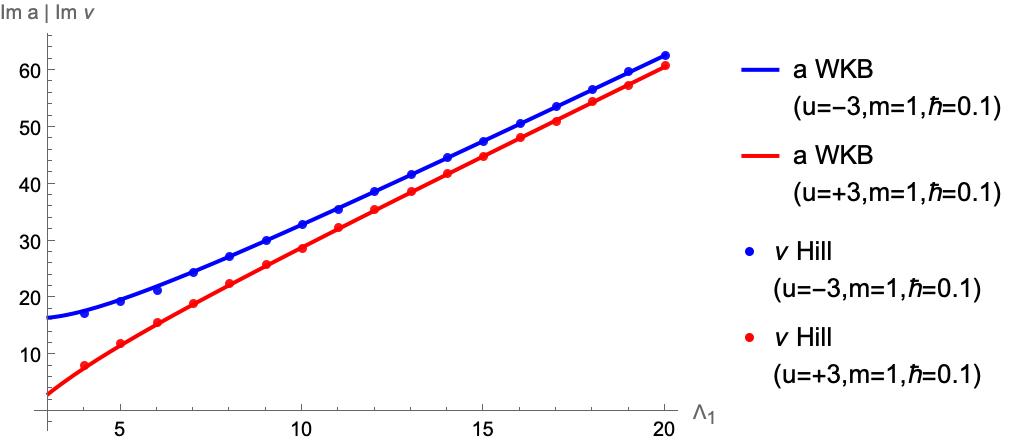}\includegraphics[width=0.51\textwidth]{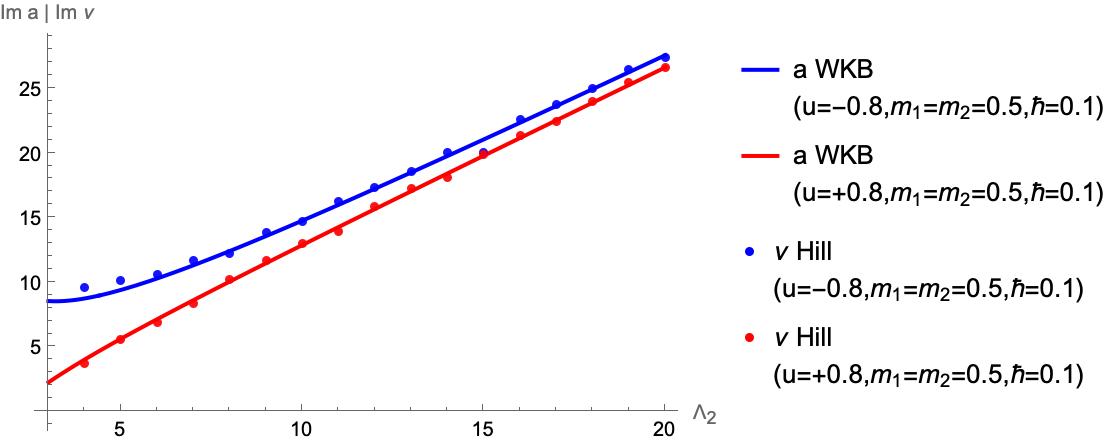}
\caption{A comparison between the Floquet exponent (computed through the Hill determinant, as explained in appendix \ref{appHill}) and the large $\Lambda_{N_f} \to + \infty$ WKB asymptotic \eqref{aLargeLNf1}, \eqref{aLargeLNf2} of the period $a/\hbar$ for the $N_f=1$ and $N_f=2$ gauge theories on the left and right respectively.}
\label{fig:anu2}
\end{figure}

We can check \eqref{anu} first in the $\theta \to + \infty$, that is $\Lambda_{N_f}/\hbar \to + \infty$ strong coupling regime, by using the WKB asymptotic expansion of the SW periods, given as elliptic integrals in \eqref{perEllipticNf1}, \eqref{perEllipticNf2}. They are the following, 
for $N_f=1$
\be
\ba \label{aLargeLNf1}
a(\hbar,u,m,\Lambda_1) \simeq \frac{a^{(0)}(u,m,\Lambda_1)}{\hbar}&\simeq i\frac{3 \sqrt{3} \sqrt{\pi } }{2 \Gamma \left(\frac{1}{6}\right) \Gamma \left(\frac{1}{3}\right)}\frac{\Lambda_1}{\hbar}+i\frac{-\frac{\left(\sqrt{3}+4 i\right) \sqrt{\pi }}{3 \Gamma \left(-\frac{1}{3}\right) \Gamma \left(\frac{5}{6}\right)}m^2+\frac{ \left(\sqrt{3}+2 i\right) \sqrt{\pi }}{\Gamma \left(-\frac{1}{3}\right) \Gamma \left(\frac{5}{6}\right)}u}{\hbar \Lambda_1}+O\left( \frac{1}{\hbar \Lambda_1^3}\right)\,,
\ea
\ee
and for $N_f=2$, setting $m_1=m_2=m$ for simplicity
\be
\ba
a(\hbar,u,m,m,\Lambda_2) \simeq \frac{a^{(0)}(u,m,m,\Lambda_2)}{\hbar}&\simeq i\frac{ \sqrt{\pi } }{\Gamma \left(\frac{1}{4}\right)^2}\frac{\Lambda _2}{\hbar}+i\frac{ \left(\frac{ \Gamma \left(\frac{1}{4}\right)^2}{2\pi ^{3/2}}+\frac{4\sqrt{\pi }}{\Gamma \left(\frac{1}{4}\right)^2}\right)m^2-\frac{  \Gamma \left(\frac{1}{4}\right)^2}{2\pi ^{3/2}}u}{\hbar \Lambda _2}+O\left( \frac{1}{\hbar \Lambda_2^3}\right)\,.
\ea\label{aLargeLNf2}
\ee
By comparing these expression for $a$ with the result of the Hill determinant for $\nu$, in plots \ref{fig:anu2} we verify \eqref{anu} in this regime.\footnote{The match has sometimes low accuracy, especially for $N_f=2$. This is partly due to the fact that we are truncating an asymptotic, rather than a convergent series. Moreover, there appears to be also some numerical instability (as evident random fluctuations around the fitting line) in the evaluation of the Hill determinant.} 

 \begin{table}
\centering
\begin{tabular}{c|c|c|c|c|c}
$\Lambda_1$ & $u$&$m$ &$\hbar$& $\nu$ \text{ Hill}& $a$ \text{ NS}\\
\hline
$0.04$&$1.1$&$0$&$1$ & $0.04880884830$ &$1+0.04880884830 $ \\
$0.08$&$1.1$&$0$&$1$ &$0.04880885678$&$1+0.04880885678$\\
$0.12$&$1.1$&$0$&$1$&$0.04880894630$&$1+0.04880894630$\\
$0.16$&$1.1$&$0$&$1$ & $0.04880939952 $&$1+0.04880939952$\\
\hline
$0.04$&$1.1$&$0.3$&$1$ & $0.04880750219$ &$1+0.04880750219 $ \\
$0.08$&$1.1$&$0.3$&$1$ &$0.04879808533$&$1+0.04879808533$\\
$0.12$&$1.1$&$0.3$&$1$&$0.04877256916$&$1+0.04877256912$\\
$0.16$&$1.1$&$0.3$&$1$ & $0.04872306409 $&$1+0.04872306354$
\end{tabular}
\caption{Comparison of $\nu$ as computed by the Hill determinant, as explained in appendix \ref{appHill} and $a$ for $N_f=1$ as computed from the instanton series \eqref{ainstex1}.}\label{tab:anu1}
\begin{tabular}{c|c|c|c|c|c|c}
$\Lambda_2$ & $u$&$m_1$&$m_2$&$\hbar$ & $\nu$ \text{ Hill}& $a$ \text{ NS}\\
\hline
$0.04$&$1.1$&$0$&$0$&$1$ & $0.04880882433$ &$1+0.04880882433 $ \\
$0.08$&$1.1$&$0$&$0$&$1$&$0.04880846678 $&$1+0.04880846679$\\
$0.12$&$1.1$&$0$&$0$&$1$&$0.04880691737 $&$1+0.04880691741$\\
$0.16$&$1.1$&$0$ &$0$&$1$& $0.04880274563$&$1+0.04880274601$\\
\hline
$0.04$&$1.1$&$0.2$&$0.2$&$1$ & $0.04880434367 $ &$1+0.04880434367 $ \\
$0.08$&$1.1$&$0.2$&$0.2$&$1$&$0.04879061871 $&$1+0.04879061867$\\
$0.12$&$1.1$&$0.2$&$0.2$&$1$&$0.04876703851 $&$1+0.04876703795$\\
$0.16$&$1.1$&$0.2$ &$0.2$&$1$& $0.04873254358$&$1+0.04873254036$
\end{tabular}
\caption{Comparison of $\nu$ as computed by the Hill determinant, as explained in appendix \ref{appHill}, and $a$ for $N_f=2$ as computed from the instanton series \eqref{ainstex2}.}\label{tab:anu2} 
\end{table} 
In the opposite weak coupling regime $\Lambda_{N_f}\to 0$, the gauge $a$ period can be computed through the Matone's relation
\be \label{matoneGen}
 u =  a^2 - \frac{2\Lambda_{N_f}}{4-N_f} \frac{\partial \mathcal{F}_{NS}^{inst}}{\partial \Lambda_{N_f}}  \,,
\ee 
where the NS prepotential $\mathcal{F}_{NS}^{inst}$ is given by the instanton series
\be
\mathcal{F}_{NS}^{inst} = \sum_{n=0}^\infty \Lambda_{N_f}^{(4- N_f)n} \mathcal{F}_{NS}^{(n)}\,.
\ee
The first two terms of these are, for $N_f=1$
\be 
\ba 
\mathcal{F}_{NS}^{(1)} &=- \frac{  m_1}{4(4 a^2- \hbar^2)} \\
\mathcal{F}_{NS}^{(2)} &=-\frac{4 m_1^2 \left(20 a^2+7 \hbar ^2\right)-3 \left(4 a^2-\hbar ^2\right)^2}{512 \left(a^2-\hbar ^2\right) \left(4 a^2-\hbar ^2\right)^3} \,,
\ea 
\ee
and for $N_f=2$
\small
\be 
\ba \label{FinstNf=2}
\mathcal{F}_{NS}^{(1)} &=- \frac{ m_1 m_2}{4(4 a^2- \hbar^2)}  \\
\mathcal{F}_{NS}^{(2)} &=-\frac{64 a^2(a^4+3 a^2(m_1^2+m_2^2)+5m_1^2 m_2^2)- \hbar^6+12\hbar^4(a^2+m_1^2+m_2^2)-16 \hbar^2[3a^4+6a^2(m_1^2+m_2^2)-7m_1^2m_2^2]}{2048 \left(a^2-\hbar ^2\right) \left(4 a^2-\hbar ^2\right)^3} \,.
\ea 
\ee
\normalsize
Then, inverting \eqref{matoneGen} with these expressions we obtain the following $\Lambda_{N_f}\to 0$ expansions for the $a$ period, for $N_f=1$
\be \label{ainstex1}
a(\hbar,u,m,\Lambda_1)=\sqrt{u}-\frac{ m}{\sqrt{u} \left(16 u-4 \hbar ^2\right)}\Lambda _1^3+\frac{ 3 u \left(\hbar ^2-4 u\right)^2-4 m^2 \left(60 u^2-35 u \hbar ^2+2 \hbar ^4\right) }{256 u^{3/2} \left(u-\hbar ^2\right) \left(4 u-\hbar ^2\right)^3}\Lambda _1^6\,,
\ee
and for $N_f=2$
\small
\be
\ba \label{ainstex2}
a(\hbar,u,m_1,m_2,\Lambda_2)&=\sqrt{u}-\frac{ m_1 m_2}{4 \sqrt{u} \left(4 u-\hbar ^2\right)}\Lambda _2^2 \\
&+\Biggl[\frac{192 m_1^2 u^3+192 m_2^2 u^3-96 m_1^2 u^2 \hbar ^2-96 m_2^2 u^2 \hbar ^2-960 m_1^2 m_2^2 u^2+12 m_1^2 u \hbar ^4}{1024 u^{3/2} \left(u-\hbar ^2\right) \left(4 u-\hbar ^2\right)^3}\\
&+\frac{12 m_2^2 u \hbar ^4+560 m_1^2 m_2^2 u \hbar ^2-32 m_1^2 m_2^2 \hbar ^4-64 u^4+48 u^3 \hbar ^2-12 u^2 \hbar ^4+u \hbar ^6}{1024 u^{3/2} \left(u-\hbar ^2\right) \left(4 u-\hbar ^2\right)^3} \Biggr]\Lambda_2^4\,.
\ea
\ee
\normalsize
Through these expressions, in tables \ref{tab:anu1} and \ref{tab:anu2} we verify the identification \eqref{anu} between the Floquet exponent and the gauge period. We notice the accuracy is remarkably high, thanks to the converging character of the $\Lambda_{N_f}\to 0$ instanton series. 

On the analytic side, we also notice that the first coefficients of the instanton series match the general mathematical analytical result (obtained through continued fractions tecnique) for the expansion of the eigenvalue $u$ of the Doubly Confluent Heun equation, given in~\cite{Ronveaux:1995} in terms of $\nu$ and which we report in~\eqref{ExpEigenDCHE}. This confirms relation \eqref{matoneGen}, with our gauge period-Floquet identification \eqref{anu}.

In conclusion, we presented strong numerical and analytical evidence for the $a$ period-Floquet identification~\eqref{anu}.  Then, by the Floquet-$T$ function identifications~\eqref{Ttnu1}-\eqref{TTnuNf=2}, new gauge-integrability basic connection formulas for the $T$ function and $a$ period ensue: for $N_f=1$
\be  \label{Ta1}
\ba
T_{+}(\th)&=2 \cos \frac{2 \pi  a}{\hbar}\\
\tilde{T}_{+}(\th )\tilde{T}_{-}(\th+i\frac{\pi}{3})&=2 \cos  \frac{2\pi  a}{\hbar} + 2 \cos \frac{2 \pi m}{\hbar} \,,
\ea
\ee 
and for $N_f=2$
\be \label{TTa2}
\ba
T_{+,+}(\th)T_{-,-}(\th)&=2 \cos  \frac{2 \pi  a}{\hbar} + 2 \cos \frac{2 \pi m_2}{\hbar} \\
 \tilde{T}_{+,+}(\th )\tilde{T}_{-,-}(\th)&=2 \cos  \frac{2\pi  a}{\hbar} + 2 \cos \frac{2 \pi m_1 }{\hbar}\,.
\ea
\ee 
In this way, the TBA - through the $Q$ and $T$ functions - allows the exact evaluation of also the gauge period $a$, at both strong and weak coupling, as well as in the intermediate coupling regime.

\section{Applications of gauge-integrability correspondence} \label{applications}

We now show some applications of the gauge-integrability correspondence, as new results on both sides. In particular, for gauge theory we find an interpretation of integrability's functional relations, as exact $R$-symmetry relations, never found before to our knowledge. For integrability instead we find new convenient formulae for the local integrals of motions in terms of the asymptotic gauge periods.

\subsection{Applications to gauge theory}

Let us consider first the $N_f=2$ gauge theory.
Now using the $TQ$ relation~\eqref{TQ2}, in the relation \eqref{TTa2} between $T$ and $a$, the LHS becomes
\be 
\ba\label{TTQQNf=2}
T_{++}(\theta)T_{--}(\theta)&=\frac{1}{Q_{++}(\theta)Q_{--}(\theta)}\Bigl[Q_{+-}(\theta+i\pi/2)Q_{-+}(\theta+i\pi/2)+Q_{+-}(\theta-i\pi/2)Q_{-+}(\theta-i\pi/2)\\&+e^{2i\pi q_{2}}Q_{+-}(\theta+i\pi/2)Q_{-+}(\theta-i\pi/2)+e^{-2i\pi q_{2}}Q_{+-}(\theta-i\pi/2)Q_{-+}(\theta+i\pi/2)\Bigr]\,.
\ea
\ee 
Instead the $T$ periodicity relation~\eqref{Tper2} 
inside~\eqref{TTa2} reads
\be  \label{TTTTNf2}
T_{++}(\th)T_{--}(\th) =T_{-+}(\th+i\pi/2)T_{+-}(\th+i\pi/2)\,.
\ee 
As a consequence of our identifications between $T$, $Y$ integrability functions and gauge periods $a$, $a_D$, relations \eqref{TTQQNf=2} and \eqref{TTTTNf2} become $\mathbb{Z}_2$ R-symmetry relation for the \emph{exact} gauge periods. Let us see how, by considering for simplicity the massless $SU(2)$ $N_f=2$ case, in which the periods are the same as for the $N_f=0$ theory~\cite{BilalFerrariQCD:1996}. If $u>0$ the leading asymptotic $\mathbb{Z}_2$ symmetry relations are
\be 
\ba \label{Z2SWNf=2massless}
a^{(0)}(-u,0,0)&= - ia^{(0)}(u,0,0)\\
a^{(0)}_D(-u,0,0)&= - i[a_D^{(0)}(u,0,0)-a^{(0)}(u,0,0)]\,.
\ea
\ee 
Now, we can see that expressing~\eqref{TTQQNf=2} and \eqref{TTTTNf2} in terms of gauge periods through~\eqref{TTa2} and~\eqref{epsaDNf=2} we get the same expressions~\eqref{Z2SWNf=2massless}. So, relations~\eqref{Z2SWNf=2massless} can be considered as \emph{derived} from the $TQ$ relation and the $T$ periodicity relations, coupled with our identifications with gauge periods. 

Similarly for $N_f=1$ case the $T$ periodicity is easily shown to be interpreted in gauge theory in the same way.
If $u>0$ and $m=0$ the other exact relation from the $T$ periodicity \eqref{Tper1} reduces to the $\mathbb{Z}_3$ symmetry in the asymptotic $\hbar \to 0$ (cf. \eqref{Z3SW})\footnote{We avoid here considering the $N_f=1$ $TQ $ relation since it requires some non-trivial analytic continuation of gauge-integrabiliy relations beyond the complex strip $\Im \th < \pi/3$ in which the TBA holds without analytic continuation.}
\be 
a^{(0)}(e^{-2\pi i/3}u,0) = - e^{2\pi i/3} a^{(0)}(u,0)
\ee
\be 
a^{(n)}(e^{-2\pi i/3}u,0) = - e^{2\pi i/3(1-n)} a^{(n)}(u,0)\,.
\ee

This simple derivations show that the new exact relations following from the integrability functional relations are actually a $\mathbb{Z}_3$, $\mathbb{Z}_2$ R-symmetry relations for $N_f=1,2$ respectively. To our knowledge, they were never found previously in the literature, except in their $\hbar \to 0$ asymptotic version in the massless case~\cite{BilalFerrariQCD:1996}.\footnote{A similar derivation for the simpler $N_f=0$ theory was given in our previous work~\cite{FioravantiGregori:2019}.}

\subsection{Applications to integrability}

Let us now show an inverse application of our two-sides correspondence from gauge to integrability.  
 Consider the large energy asymptotic expansion~\eqref{asyQLIM1} of $Q$ in terms of the LIMs. If we set $q=0$ so to recover the LIMs of Liouville $b = \sqrt{2}$. For this special case, the asymptotic expansion simplifies as
\be 
\ln Q (\th,p) \doteq- C_0 e^\th - \sum_{n=1}^\infty e^{\th(1-2n)}C_n \mathbb{I}_{2n-1}\,,\qquad \th \to +\infty \,,\qquad p \quad \text{finite}\,,
\ee 
the normalization constants being given by 
\be 
C_n=\frac{\Gamma \left(\frac{2 n}{3}-\frac{1}{3}\right)\Gamma \left(\frac{n}{3}-\frac{1}{6}\right) }{3 \sqrt{2 \pi } n!}\,.
\ee 
Crucially, we can also expand the LIMs $\mathbb{I}_{2n-1}$, as polynomials in $p^2$ with coefficients $\Upsilon_{n,k}$
\be 
\mathbb{I}_{2n-1} = \sum_{k=0}^n \Upsilon_{n,k} p^{2k }\,,
\ee 
where the leading and subleading coefficients are found to be~\cite{FioravantiGregori:2019}
\be \label{Upsnm}
\Upsilon_{n,n} = (-1)^n\,, \qquad \Upsilon_{n,n-1} =\frac{1}{24} (-1)^n n (2 n-1)\,.
\ee

Now, since in Seiberg-Witten theory $u$ is finite as $\theta \to +\infty$, to connect the IM $\th \to + \infty$ \textit{asymptotic} expansion, it is necessary to take the further limit 
\be 
p^2(\theta) =  4\frac{u}{\Lambda_1^2} e^{2 \theta} \to +\infty\,.\ee
In this double limit, an infinite number of LIMs $\mathbb{I}_{2n-1}(b=\sqrt{2})$, through their coefficients $\Upsilon_{n,k}$, are re-summed into a quantum gauge period's asymptotic mode. For instance, the leading order is obtained from the resummation of all $\Upsilon_{n,n}=(-1)^n$ terms as
\begin{align}
\ln Q^{(0)}(u,0,\Lambda_1)&=-  \sum _{n=0}^\infty \frac{\Gamma \left(\frac{2n }{3}-\frac{1}{3} \right) \Gamma \left(\frac{n}{3}-\frac{1}{6}\right)}{3 \sqrt{2\pi} n!} \left(-\frac{ 4u}{\Lambda_1^2}\right)^n \,,
\end{align}
and from it we can derive the higher orders as usual through differential operators \eqref{oper-higher}.
In particular, in the massless case the first simplify as
\be
\ba
\ln Q^{(1)}(u,0,\Lambda_1) &= \left(\frac{\Lambda_1}{2}\right)^2\left [ \frac{u}{6} \frac{\partial^2}{\partial u^2} + \frac{1}{12} \frac{\partial }{\partial u}\right]  \ln Q^{(0)}(u,0,\Lambda_1)\\
\ln Q^{(2)}(u,0,\Lambda_1) &= \left(\frac{\Lambda_1}{2}\right)^4\left [\frac{7}{360}u^2 \frac{\partial^4}{\partial u^4}+\frac{31}{360}u   \frac{\partial^3}{\partial u^3} + \frac{9}{160} \frac{\partial^2 }{\partial u^2}\right]  \ln Q^{(0)}(u,0,\Lambda_1)\\
\ln Q^{(3)}(u,0,\Lambda_1)&=  \left(\frac{\Lambda_1}{2}\right)^6 \left[\frac{31 u^3}{15120}\frac{\partial^6}{\partial u^6}+\frac{443 u^2}{18144}\frac{\partial^5}{\partial u^5} +\frac{43 u}{576}\frac{\partial^4}{\partial u^4} +\frac{557}{10368}\frac{\partial^3}{\partial u^3}\right]\ln Q^{(0)}(u,0,\Lambda_1)\,.
\ea
\ee
Indeed these expression match with the resummation of LIMs at higher orders \eqref{Upsnm}:
\begin{align}
\ln Q^{(1)}(u,0,\Lambda_1)&=  \left(\frac{\Lambda_1}{2}\right)^2 \sum _{n=0}^\infty \left [\frac{n}{12} + \frac{1}{24} \right] \frac{\Gamma \left(\frac{2n }{3}+\frac{1}{3} \right) \Gamma \left(\frac{n}{3}+\frac{1}{6}\right)}{3 \sqrt{2\pi} n!} \left(-\frac{4 u}{\Lambda_1^2}\right)^n\\
\ln Q^{(2)}(u,0,\Lambda_1)&= - \left(\frac{\Lambda_1}{2}\right)^4\sum _{n=0}^\infty \left [\frac{(14n+27)(2n+3)}{5760} \right] \frac{\Gamma \left(\frac{2n }{3}+1 \right) \Gamma \left(\frac{n}{3}+\frac{1}{2}\right)}{3 \sqrt{2\pi} n!} \left(-\frac{4 u}{\Lambda_1^2}\right)^n\\
\ln Q^{(3)}(u,0,\Lambda_1)&=  \left(\frac{\Lambda_1}{2}\right)^6\sum _{n=0}^\infty \left [\frac{1}{8}\frac{ [4 n (93 n+596)+3899](2 n+5)}{362880}\right] \frac{\Gamma \left(\frac{2n }{3}+\frac{5}{3} \right) \Gamma \left(\frac{n}{3}+\frac{5}{6}\right)}{3 \sqrt{2\pi} n!} \left(-\frac{4 u}{\Lambda_1^2}\right)^n\,.
\end{align}
So in general we find the relation
\be  \label{lnQUps}
\ln Q^{(k)}(u,0,\Lambda_1) = (-1)^{k+1} \left(\frac{\Lambda_1}{2}\right)^{2k} \sum_{n=0}^\infty \, \Upsilon_{n+k,n}\,\frac{\Gamma \left(\frac{k+n}{3}-\frac{1}{6}\right) \Gamma \left(\frac{2 (k+n)}{3}-\frac{1}{3}\right)}{3 \sqrt{2 \pi } (k+n)!}\left(\frac{4 u}{\Lambda_1^2}\right)^n\,.
\ee
The differential operators to obtain any order $\ln Q^{(n)}$ can be derived systematically as described in \cite{FioravantiGregori:2019}. Then we notice that this procedure can be a convenient way to compute the LIMs, through their $p^2$ coefficients $\Upsilon_{n+k,n}$.

Alternatively and equivalently, we can use the correspondence in the other direction, to compute the $k$-th mode of the
(alternative dual) quantum period in terms of the LIMs coefficients as follows
\begin{align} 
\frac{4 \pi}{\Lambda_1}a^{(k)}_1(u,0,\Lambda_1) &=- \sum_{k=0}^\infty   \Upsilon_{n+k,n}\,\frac{\Gamma \left(\frac{k+n}{3}-\frac{1}{6}\right) \Gamma \left(\frac{2 (k+n)}{3}-\frac{1}{3}\right)}{3 \sqrt{2 \pi } (k+n)!}   \,2 \sin \left(\frac{1}{3} \pi  (k+n+1)\right)\Bigl( \frac{4u }{\Lambda_1^2}\Bigr)^n \label{aDnLIM1}\,.
\end{align}  

\section{Gravitational correspondence and applications} \label{gravity}

In the previous sections we established a correspondence between gauge theory and integrable models. It has been possible ultimately because those theories were derived by certain ODEs which can map into each other. Now, it turns out the same ODEs appear also in black hole (BH) perturbation theory. In particular, the Doubly Confluent Heun equation (see appendix~\ref{appDCHE}) we have for the $SU(2)$ $N_f=1,2$ gauge theories and Generalized Perturbed Hairpin integrable model can be associated to extremal black holes. Thus, in this section we will see how our integrability-gauge correspondence can extend to include these gravitational systems too.

\subsection{Gravitational system for the $N_f=2$ theory}


In general, the $N_f=2$ gauge theory can be made to correspond to the gravitational background defined, in type IIB supergravity, as the intersection of four stacks of D3-branes. This geometry is characterised by four different charges $\mathcal{Q}_i$ which, if all equal, lead to an extremal Reissner-Nordstr\"om (RN) BH, that is maximally charged. For this reason we shall call this geometry Generalized Extremal Charged BHs. In details, using isotropic coordinates its line element writes as~\cite{Ikeda:2021,BianchiConsoliGrilloMorales:2021b}
\be  \label{lineintD3}
d s^2 = - f(r) d t^2 + f(r)^{-1}[d r^2 +r^2 (d \theta^2 + \sin^2 \theta d \phi^2)]\,,
\ee
where
\be
f(r) = \prod_{i=1}^4 \left(1 + \mathcal{Q}_i/r\right)^{-\frac{1}{2}}\,.
\ee
\normalsize
The ODE describing the scalar perturbation in this background is
\be  \label{ODEgrav2S}
\frac{d^2 \phi}{d r^2}+\left [ -\frac{(l+\frac{1}{2})^2-\frac{1}{4}}{r^2}+\omega^2\left(1+\sum_{k=1}^4 \frac{\Sigma_k}{r^k}\right)\right ]\phi = 0\,,
\ee 
\normalsize
\normalsize
with
\be
\Sigma_k =\sum_{i_1 <...<i_k}^4 \mathcal{Q}_{i_1}\cdots \mathcal{Q}_{i_k}\,.
\ee
We can map ODE \eqref{ODEgrav2S} into that for Generalized Perturbed Hairpin IM~\eqref{ODEint2}, by changing variables as follows
\be \label{DictIntD3branes}
\ba
r&=\sqrt[4]{\Sigma_4} e^{y} \qquad &\omega &= - \frac{i}{\sqrt[4]{\Sigma_4} } e^{\th}\,, \\ 
 p^2&=(l+\frac{1}{2})^2- \omega^2 \Sigma_2\,\qquad &q_j &= \frac{1}{2} \frac{\Sigma_{2j-1}}{\sqrt[4]{\Sigma_4}^{2j-1}} e^\th\,, \qquad j=1,2 \,.
\ea 
\ee 
\normalsize 

Through the map \eqref{DictIntD3branes}, the $Y$ system \eqref{Ysyst2} transforms into gravity variables as
\be  \label{Ysyst2grav}
\ba
&Y(\th+ \frac{i\pi}{2},-i\Sigma_1,-\Sigma_2,i\Sigma_3)Y(\th-\frac{i\pi}{2},-i \Sigma_1,-\Sigma_2,i\Sigma_3 )= [1 + Y(\th,\Sigma_1,\Sigma_2,\Sigma_3 )][1 + Y(\th,-\Sigma_1,\Sigma_2,-\Sigma_3 )]\,,
\ea
\ee 
where we omit $\Sigma_4$ since it remains fixed. Then, the $Y$ system~\eqref{Ysyst2grav} can be inverted into the following TBA:
\be \label{TBA2}
\ba 
\varepsilon_{\pm,\pm}(\th)&= \varepsilon_{\pm,\pm}^{(0)} e^{\th}- \varphi \ast (\bar{L}_{\pm \pm}+ \bar{L}_{\mp \mp})(\th) \\
\bar{\varepsilon}_{\pm,\pm}(\th)&=\bar{\varepsilon}_{\pm,\pm}^{(0)}e^{\th}- \varphi \ast (L_{\pm \pm}+ L_{\mp \mp})(\th)\,, \\
\ea 
\ee 
\normalsize
where we defined
 $\varepsilon_{\pm,\pm}(\th) =-\ln Y(\th,\pm\Sigma_1,\Sigma_2, \pm\Sigma_3,\Sigma_4)$, and $\bar{\varepsilon}_{\pm,\pm}(\th) = \varepsilon(\th,\pm i\Sigma_1,-\Sigma_2,\mp i \Sigma_3,\Sigma_4)$, while as usual $L = \ln [1+\exp \{ -\varepsilon \} ]$, $\varphi(\th) = ( \cosh(\th))^{-1}$. \normalsize 
Then, the forcing terms are computed as follows
\small
\be
\ba
\varepsilon^{(0)}_{\pm,\pm}&=\varepsilon^{(0)}(\pm\Sigma_1,\Sigma_2,\pm\Sigma_3,\Sigma_4) = -\ln Q^{(0)}(\Sigma_1,\Sigma_2,\Sigma_3,\Sigma_4) -\ln Q^{(0)}(-\Sigma_1,\Sigma_2,-\Sigma_3,\Sigma_4)\mp \frac{i \pi}{2} (\frac{\Sigma_1}{\Sigma_4^{1/4}}-\frac{\Sigma_3}{\Sigma_4^{3/4}})\\
\bar{\varepsilon}^{(0)}_{\pm,\pm}&=\varepsilon^{(0)}(\pm i \Sigma_1,\Sigma_2,\mp i\Sigma_3,\Sigma_4) = -\ln Q^{(0)}(i\Sigma_1,\Sigma_2,-i\Sigma_3,\Sigma_4) -\ln Q^{(0)}(-i\Sigma_1,\Sigma_2,i\Sigma_3,\Sigma_4)\pm \frac{ \pi}{2} (\frac{\Sigma_1}{\Sigma_4^{1/4}}+\frac{\Sigma_3}{\Sigma_4^{3/4}})\,,
\ea
\ee
\normalsize
with 
\small
\be \label{c0gravTBA2}
\ln Q^{(0)}(\mathbf{\Sigma})= \int_{-\infty}^\infty\left[\sqrt{2 \cosh(2 y) +\frac{\Sigma_1}{ \sqrt[4]{\Sigma_4}}e^y+ \frac{\Sigma_3}{\sqrt[4]{\Sigma_4^3}}e^{-y}+\frac{\Sigma_2}{\sqrt{\Sigma_4}}} -2\cosh y -\frac{1}{2} \frac{\Sigma_1}{\sqrt[4]{\Sigma_4}}\frac{1}{1+e^{-y/2}}-\frac{1}{2} \frac{\Sigma_3}{\sqrt[4]{\Sigma_4^3}}\frac{1}{1+e^{y/2}}\right]\, dy \,.
\ee
\normalsize
Similarly as with the integrability TBA in subsection \ref{subsecintTBA}, we should input the parameter $l$ through the boundary condition at  $\th \to  -\infty$: 
\be 
\varepsilon_{\pm,\pm}(\th) \simeq 4 (l+1/2) \th -2 D_2(l+1/2) \qquad \th \to -\infty\,.
\ee
The subleading constant $D_2$ is given by the $q_1 \to 0$, $q_2 \to 0$ limit of the integrability one \eqref{Cint2}
\be 
D_2(p) = C_2(p,0,0)=  \ln \left(\frac{2^{1-2 p} p \Gamma (2 p)^2}{\Gamma \left(p+\frac{1}{2}\right)^2}\right) \,,
\ee
as follows from the asymptotic of the ODE~\eqref{ODEint2} as explained in subsection \ref{subsecintTBA}. We emphasize one should pay special attention to the different change of variables from integrability to gauge theory \eqref{DictGau2} or to gravity \eqref{DictIntD3branes}: they imply different TBA equations, as first noted in~\cite{FioravantiGregori:2019}. In fact, we notice the gravity TBA \eqref{TBA2} mixes features of the gauge \eqref{ga-TBA2} and integrability TBAs \eqref{int-TBA2}: it has the same structure and $\theta \to + \infty$ forcing term of the former, but the $\theta \to - \infty$ boundary condition of the latter.

Now let us consider an important black hole observable: the quasinormal modes (QNMs) $\omega_n$. They can be expressed in terms of $ \theta_n$ according to the map \eqref{DictIntD3branes}. The remarkable finding of \cite{FioravantiGregori:2021} is that they are determined by the so-called \emph{Bethe root} condition on the integrability functions, that is
\be \label{quanteps2}
\bar{\varepsilon}_{+,+}(\th_{n} +i \pi/2) = - i \pi (2n+1) \,,\qquad Q_{+,+}(\th_n) = 0 \qquad n \in \mathbb{Z}\,.
\ee
Thanks to the gauge integrability correspondence developed in the previous sections, in particular the final identification between $Y$ and $a_D$~\eqref{gauint2}, it becomes clear that QNMs must also correspond to quantization conditions on the gauge periods:
\be  \label{quantPerNf2}
\frac{2 \pi i}{\hbar(\th_n)}a_D(\th_n,u,m_1,m_2,\Lambda_2)=- i \pi (2n+1)\,.
\ee
This observation constitutes an at least mathematical proof of the fundamental finding of~\cite{AminovGrassiHatsuda:2020} and the following literature (see the introduction).\footnote{A note of caution, though. Literature following~\cite{AminovGrassiHatsuda:2020} uses another definition of gauge period which we denote by $A_D$ which derives from the instanton expansion of the prepotential. As we explain in appendix~\ref{D3brane} the two definitions can be actually related by formulas like~\eqref{Q0GGM} for the $N_f=0$ theory.
Generalizations of formula~\eqref{Q0GGM}, already exist for the subcase of the $N_f=1$ gauge theory~\cite{GrassiHaoNeitzke:2021} (see next subsection) and so we expect them to exist also for the whole $N_f=2$ theory and even more generally. In this way we expect that in general the integrable Bethe roots condition, which we have shown to follow straightforwardly from BHs physics, in gauge theory indeed corresponds to the quantization of the gauge $A_D$ period as stated in~\cite{AminovGrassiHatsuda:2020}.}

\begin{figure}
\centering
\includegraphics[width=0.95\textwidth]{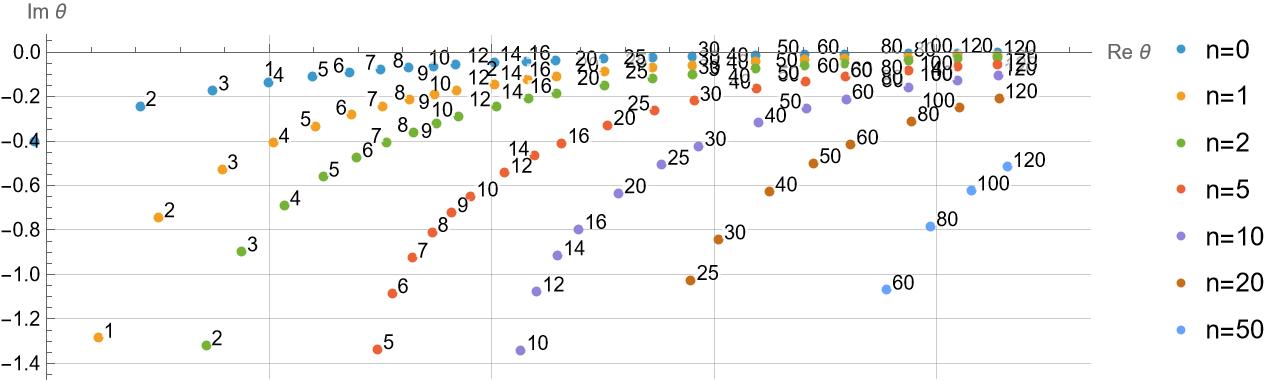}
\caption{Plot of the Bethe roots in the $\theta$ plane correspoding to quasinormal modes $\omega_n$, for several $n$ and $l$ of generalized RN BH ($N_f=2$ gauge theory) corresponding to $\Sigma_1=\Sigma_3=0.2$, $\Sigma_2 = 0.4$, $\Sigma_4=1$, as obtained from \eqref{quanteps2}.} \label{plotBetheRootsNf2}
\end{figure}

The numerical results in tables~\ref{tabQNMsNf=2-case1} and \ref{tabQNMsNf=2-case2} show the agreement between the QNMs computed from TBA and those computed through the standard continued fraction (Leaver) method and geodetic WKB approximation ($l \to \infty$)~\cite{Leaver:1985}. In figure \ref{plotBetheRootsNf2} we also show the Bethe roots $\theta_n$ - in the whole $- \pi/2 <\Im \theta<0$ strip - corresponding to the QNMs $\omega_n$, for various $l$ and $n$. We notice that in the case $\Sigma_1 \neq \Sigma_3$ and $\Sigma_4 \neq 1$ the Leaver method is not applicable, at least in its original version, since the recursion produced by the ODE involves more than $3$ terms (compare~\cite{BianchiConsoliGrilloMorales:2021,Leaver:1985}). Thus for this reason the TBA method may be regarded as convenient.\footnote{However, we point out that there exists a development of the Leaver method, the so-called matrix Leaver method which is still applicable~\cite{Leaver:1990,KumarBannonPribytokRodgers:2020}.}
\begin{table}[t]
\centering
\begin{tabular}{c|c|c|c|c|c|c|c|c}
$n$&$l $&$\Sigma_1$&$\Sigma_2$&$\Sigma_3$&$\Sigma_4$& TBA&Leaver& WKB  \\
\hline
$0$& $2$&$0.2$&$0.4$&$0.2$&$1$&$1.47799\, -0.36814 i$&$1.47789-0.36824 i$&$1.494 -0.3660 i$\\
$0$&$4$&$0.2$&$0.4$&$0.2$&$1$&$2.68035\, -0.36664 i$&$2.68035-0.36664 i$&$2.689-0.3660 i$\\
$0$&$5 $&$0.2$&$0.4$&$0.2$&$1$& $3.27959\, -0.36641 i$&$3.27959-0.36641 i$&$3.287\, -0.3660 i$\\
$0$&$6 $&$0.2$&$0.4$&$0.2$&$1$& $3.87833\, -0.36629 i$&$3.87833-0.36629 i$&$3.884\, -0.3660 i$\\
$0$&$8 $&$0.2$&$0.4$&$0.2$&$1$& $5.07501\, -0.36615 i$&$5.07501-0.36615 i$&$5.080\, -0.3660 i$\\
$1$& $2$&$0.2$&$0.4$&$0.2$&$1$&$1.21708\, -1.12174 i$&$1.22004-1.12585 i$&$ 1.494 \, - 1.0979 i$\\
$1$&$ 4$&$0.2$&$0.4$&$0.2$&$1$&$2.54243\, -1.10434 i$&$2.54246-1.10513 i$&$2.689\, - 1.0979 i$\\
$1$& $6$&$0.2$&$0.4$&$0.2$&$1$ &$3.78385\, -1.10090 i$&$3.78380-1.10122 i$&$ 3.885\, - 1.0979 i$\\
$1$& $8$&$0.2$&$0.4$&$0.2$&$1$ &$5.00305\, -1.09963 i$&$5.00300-1.09981 i$&$ 5.080\, - 1.0979 i$\\
\end{tabular}
\caption{Comparison of QNMs obtained from TBA~\eqref{TBA2}, through~\eqref{quanteps2}, Leaver method and WKB approximation.}\label{tabQNMsNf=2-case1}
\end{table}
\begin{table}[t]
\centering
$\begin{array}{c|c|c|c|c|c|c|c}
 n&l&\Sigma _1 & \Sigma _2 & \Sigma _3 & \Sigma _4 & \text{TBA} & \text{WKB} \\
 \hline
0& 2 &0.1 & 0.2 & 0.3 & 1 &  1.5308\, -0.39676 i & 1.55114\, -0.394579 i \\
0& 4 &  0.1 & 0.2 & 0.3 & 1 &2.78078\, -0.39525 i & 2.79206\, -0.394579 i \\
0&8 & 0.1 & 0.2 & 0.3 & 1 &  5.26765\, -0.394765 i & 5.27389\, -0.394579 i \\
 0&16 &0.1 & 0.2 & 0.3 & 1 &  10.2396\, -0.394738 i & 10.2376\, -0.394579 i \\
0&32 & 0.1 & 0.2 & 0.3 & 1 &  20.178\, -0.394749 i & 20.1649\, -0.394579 i \\
0&64 & 0.1 & 0.2 & 0.3 & 1 &  40.0473\, -0.394733 i & 40.0195\, -0.394579 i \\
0& 128 &0.1 & 0.2 & 0.3 & 1 &  79.7812\, -0.394726 i & 79.7288\, -0.394579 i \\
 0&256 &0.1 & 0.2 & 0.3 & 1 &  159.238\, -0.394721 i & 159.147\, -0.394579 i \\
\end{array}$
\caption{Comparison of QNMs obtained from TBA~\eqref{TBA2}, through~\eqref{quanteps2}, and WKB approximation.}\label{tabQNMsNf=2-case2}
\end{table}

This new characterization of QNMs and the corresponding numerical computation are a direct consequence of the identification we proved between the $\varepsilon = - \ln Y$ function and the dual gauge period $a_D$. However, we could also investigate implications of the identification between the $T$ function and gauge period $a$. First, by making considerations on the $TQ$ systems and the $QQ$ system~\eqref{QQ2}, we can derive a quantization condition on the $T$ function, in the case of equal masses $q_1=q_2\equiv q$ whereby:
\be \label{TTqNf=2}
T_{+,+}(\th_n)T_{-,-}(\th_n) = 4 \, \,.
\ee
Instead, we cannot conclude any similar quantization condition in the case of different masses. Relation \eqref{TTqNf=2} generalizes~\eqref{quantT0}, derived in~\cite{FioravantiGregori:2021} for the $N_f=0$ gauge theory, as reported in appendix~\ref{D3brane}

Let us now prove~\eqref{TTqNf=2}. From the $QQ$ system~\eqref{QQ2} we can write, for general $q_1,q_2$ and $\theta \simeq \theta_n$
\begin{align}
e^{i\pi q_1}Q_{-+}(\th-i \pi/2) &= c_0 \left[1 \pm i  e^{i \pi \frac{q_1-q_2}{2}}\sqrt{Q_{+,+}(\th)Q_{-,-}(\th)}\right] \label{Q2s1b}\\
e^{-i\pi q_2}Q_{+-}(\th+i \pi/2) &= \frac{1}{c_0} \left[1 \mp i  e^{i \pi \frac{q_1-q_2}{2}}\sqrt{Q_{+,+}(\th)Q_{-,-}(\th)}\right]\label{Q2s2b} \\
e^{i\pi q_2}Q_{+-}(\th-i \pi/2) &= \frac{1}{c_0'} \left[1 \pm i  e^{-i \pi \frac{q_1-q_2}{2}}\sqrt{Q_{+,+}(\th)Q_{-,-}(\th)}\right] \label{Q2s3b}\\
e^{-i\pi q_1}Q_{-+}(\th+i \pi/2) &= c_0' \left[1 \mp i  e^{-i \pi \frac{q_1-q_2}{2}}\sqrt{Q_{+,+}(\th)Q_{-,-}(\th)}\right] \,.\label{Q2s4b}
\end{align}
From the 2 $TQ$ system~\eqref{TQ2} at the Bethe roots we get the same relation
\be 
c_0(-q_1,q_2) = - c_0'(-q_1,q_2)\,.
\ee 
We can also exchange the masses in~\eqref{Q2s1b} and~\eqref{Q2s3b} to obtain the relation
\be 
c_0(-q_1,q_2) c_0(-q_2,q_1) = -1\,.
\ee 
In addition, considering real parameters, we have
\be 
c_0 = - c_0^*\,.
\ee 
However, we cannot fix $c_0$ completely in general, but only when $q_1 = q_2 = q$ we can say
\be 
c_0(q_1,q_2= q_1) = \pm i\,.
\ee 
We notice also that
\be 
Q_{+,-} = Q_{-,+} \qquad q_1  = q_2 = q\,.
\ee 
We can generalize the $N_f=0$ procedure by considering the $Y$ system instead of the $Q$ system.
\small
\be 
\ba \label{TTY}
T_{+,+}(\th) T_{-,-} (\th) Y_{+,+}(\th) &= [e^{i \pi q} Q_{+,-}(\th- i \pi/2) + e^{- i \pi q} Q_{+,-}(\th+ i \pi/2)][e^{i \pi q} Q_{-,+}(\th- i \pi/2) + e^{- i \pi q} Q_{-,+}(\th+ i \pi/2)]\\
&= Y_{+,-}(\th- i \pi/2) + Y_{-,+} (\th+ i \pi/2) + 2 + 2 Y_{+,+}(\th)\,.
\ea 
\ee 
\normalsize
Notice that we can write shifted $Y$ as
\be 
\ba
Y_{+,-}(\th- i \pi/2) &= e^{2 \pi i q} Q_{+,-}(\th-i\pi/2)Q_{-,+}(\th-i\pi/2)\\&=
-1 \mp 2 i  \sqrt{Q_{+,+}(\th)Q_{-,-}(\th)}+Q_{+,+}(\th)Q_{-,-}(\th)\\
&=-1 \mp 2 i  \sqrt{Y_{+,+}(\th)}+Y_{+,+}(\th)\,,
\ea
\ee
and
\be 
\ba
Y_{-,+}(\th+ i \pi/2) &= -1 \pm 2 i  \sqrt{Y_{+,+}(\th)}+Y_{+,+}(\th)\,.
\ea
\ee 
Inserting these shifted-$Y$ expressions in what we could call the $TY$ relation~\eqref{TTY} we find
\be 
\ba
T_{+,+}(\th) T_{-,-} (\th) Y_{+,+}(\th) 
&=   4 Y_{+,+}(\th)\,,
\ea 
\ee 
that is nothing but quantization relation on $T$~\eqref{TTqNf=2}.

Now, since we in previous sections we proved the $T$ function to be related to the $a$ gauge period, it is natural to ask whether the $T$ quantization we just proved implies a quantization on $a$ too. Let us observe that in the case of equal masses, since $T_{+,-}(\th+i \frac{\pi}{2})= T_{-,+}(\th+i \frac{\pi}{2})$, on plugging the $T$ periodicity relations~\eqref{Tper2} in the relation between $T$ and $a$ 
 \eqref{TTa2}, we get the simplification to only one $T$
\be \label{TnuNf=2simp}
\ba
\pm \sqrt{2 \cos 2\pi a + 2 \cos 2 \pi q_2} &= T_{+,+}(\th)\,,
\ea
\ee 
We notice also the symmetry property of the period $a(\theta,q_1,q_2) = a(\theta,-q_1,-q_2)$, 
for which the same relation holds for $T_{-,-}(\th)$. 
Then, from the $T$ quantization~\eqref{TTqNf=2} for $q_1=q_2=q$, which reads
\be 
T_{+,+}(\th_n) T_{-,-}(\th_n) =\pm2 \left[ \cos \left\{2\pi a (\th_n)\right\}+  \cos 2 \pi q \right] = 4\,,
\ee 
it follows a quantization condition on the \textit{combination} of $a$ and $q$:
\be  \label{nuqQuant}
 \cos \left\{2\pi a (\th_n)\right\}+ \cos  2 \pi q   =\pm  2 \, .
\ee
Therefore, from this derivation we do not expect that the alternative QNMs quantization condition on the gauge $a$ period \eqref{QuantA0} found in~\cite{BianchiConsoliGrilloMorales:2021} for $N_f=0$ generalizes to other gauge theories, both because the integrabilty $T$ function is not quantized generally (for different masses $q_1\neq q_2$) and because even when it is, it implies a quantization on only the combination of $a $ period and masses.

Finally, we can find also an integrability interpretation of the symmetry under Couch-Torrence transformation found in~\cite{BianchiDiRusso:2021} for this gravitational background.\footnote{The authors~\cite{BianchiDiRusso:2021} derived it as a consequence of identifications between certain scattering angles and the SW $a$ period.} This is the symmetry that exchanges infinity ($y \to + \infty$) and the (analogue) horizon ($y \to - \infty$), leaving the photon sphere ($y=0$) fixed. Then, in our ODE/IM approach, it corresponds to the following wave function property
\be  \label{invsym}
\psi_{+,0}(y) = \psi_{-,0}(-y)\,, \qquad (q_1=q_2)\,,
\ee 
which we notice is valid only for equal mass parameters. In this respect, under~\eqref{invsym} we have the $T$ and $\tilde{T}$ identity
\be 
\tilde{T}_{+,+}(\th) = T_{+,+}(\th)\, \qquad (q_1=q_2) \,,
\ee
as can be easily understood from their very definitions~\eqref{Tdef2}.

\subsection{Gravitational system for the $N_f=1$ theory}

Now, to get a gravitation counterpart of the $N_f=1$ gauge theory, we should take the limit from the $N_f=2$ theory, as explained in appendix~\ref{appLimit},which in gravity variables corresponds to 
\be 
\Sigma_4 \to 0\,.
\ee 
In terms of BH charges can be realized for instance with $\mathcal{Q}_4 \to 0$. The physical interpretation is that of a null entropy limit on the intersection of D3 branes~\cite{BianchiConsoliGrilloMorales:2021b}. Upon this limit, we get the following gravity-integrability parameters dictionary
\be \label{dictGravNf1}
 \omega \sqrt[3]{\Sigma_3} = - i e^{\th }  \qquad \frac{\Sigma_1}{\sqrt[3]{\Sigma_3}}=2 q e^{-\th }  \qquad p^2  = (l+\frac{1}{2})^2 - \omega^2 \Sigma_2\,,
\ee 
which transforms ODE \eqref{ODEint1} into
\be  
\frac{d^2 \phi}{d r^2}+\left [ -\frac{(l+\frac{1}{2})^2-\frac{1}{4}}{r^2}+\omega^2\left(1+\sum_{k=1}^3 \frac{\Sigma_k}{r^k}\right)\right ]\phi = 0\,.
\ee 
The $N_f=1$ $Y$ system in gravitational variables reads
\be
\ba 
&Y(\th+i \pi/2,- i \Sigma_1,-\Sigma_2)Y(\th-i\pi/2,-i \Sigma_1,-\Sigma_2) \\
&= [1 + Y(\th+i \pi/6,-i e^{-2 \pi i/3}\Sigma_1,-e^{2\pi i/3}\Sigma_2)][1 + Y(\th-i \pi/6,-i e^{2 \pi i/3}\Sigma_1,-e^{-2\pi i/3}\Sigma_2)]\,,
\ea
\ee 
which shows it is convenient to define the following $Y$ functions
\be 
\ba
Y_{0,\pm}(\th)&= Y(\th,\pm i \Sigma_1,-\Sigma_2) \,\\ Y_{1,\pm}(\th) &= Y(\th, \pm i e^{2\pi i/3}\Sigma_1,-e^{-2\pi i/3}\Sigma_2) \, \\ Y_{2,\pm}(\th)&= Y(\th,\pm i e^{-2\pi i/3}\Sigma_1,-e^{2\pi i/3}\Sigma_2) \,.
\ea
\ee 
\normalsize
The $Y$ system can be inverted in a TBA made of $3$ coupled equations as
\begin{align} \label{TBAG1}
\ve_{0,\pm}(\th) &= \ve_{0,\pm}^{(0)}e^\th- (\varphi_+ \ast L_{1,\pm})(\th)-(\varphi_- \ast L_{2,\pm})(\th) \\
\ve_{1,\pm}(\th) &= \ve_{1,\pm}^{(0)} e^\th- (\varphi_+ \ast L_{2,\pm})(\th)-(\varphi_- \ast L_{0,\pm})(\th) \\
\ve_{2,\pm}(\th) &= \ve_{2,\pm}^{(0)} e^\th- (\varphi_+ \ast L_{0,\pm})(\th)-(\varphi_- \ast L_{1,\pm})(\th)\,,
\end{align}
with the kernels \eqref{kern1} and as usual $L_{k,\pm} = \ln [1+ \exp \{-\varepsilon_{k,\pm} \}]$. 
Under change to gravity variables \eqref{dictGravNf1} $q(\th ) \propto e^{\th }$ and so the leading order is given by
\small
\be 
\ba
\varepsilon^{(0)}_{k,\pm} &= - e^{-i \pi/6} \ln Q^{(0)}(\mp i e^{\frac{2\pi i(1-k)}{3}} \Sigma_1,-e^{-\frac{2 \pi i(1+k)}{3}}\Sigma_2) - e^{i \pi/6} \ln Q^{(0)}(\mp i e^{\frac{2\pi i(-1-k)}{3}} \Sigma_1,-e^{-\frac{2 \pi i(-1+k)}{3}}\Sigma_2)\\
&\quad \mp \frac{2 \pi i}{3}\frac{ e^{-2 \pi i k/3}\Sigma_1}{\sqrt[3]{\Sigma_3}}\,,
\ea
\ee 
with
\normalsize
\be \label{c0TBAgrav1}
\ln Q^{(0)}( \Sigma_1, \Sigma_2,\Sigma_3) = \int_{-\infty}^\infty \left[\sqrt{e^{2y}+ e^{-y} + \frac{\Sigma_1}{\sqrt[3]{\Sigma_3}}e^y+\frac{\Sigma_2}{\sqrt[3]{\Sigma_3^2}}} -e^y -e^{-y/2} -\frac{1}{2} \frac{\Sigma_1}{\sqrt[3]{\Sigma_3}} \frac{1}{1+e^{-y/2}}\right]\, dy \,.
\ee 
\begin{table}[t]
\centering

$\begin{array}{c|c|c|c|c|c|c}
 n&l &\Sigma _1 & \Sigma _2 & \Sigma _3 &  \text{TBA} & \text{WKB} \\
 \hline
 0& 1&0.1 & 0.2 & 1  & 0.994542\, -0.319089 i & 1.018635-0.317055 i \\
 0& 2&0.1 & 0.2 & 1  & 1.68332\, -0.317781 i & 1.69772-0.31706 i \\
0& 4& 0.1 & 0.2 & 1  & 3.04791\, -0.317279 i & 3.05590-0.31706 i \\
 0& 8& 0.1 & 0.2 & 1  & 5.76803\, -0.317118 i & 5.77226-0.31706 i \\
 0& 16&0.1 & 0.2 & 1  & 11.2012\, -0.317032 i & 11.20498-0.31706 i \\
0& 32& 0.1 & 0.2 & 1  & 22.0861\, -0.31805 i & 22.0704-0.3171 i \\
0& 64 & 0.1 & 0.2 & 1 & 43.8454\, -0.31797 i & 43.8013-0.3171 i \\
0& 128& 0.1 & 0.2 & 1  & 87.3566\, -0.317921 i & 87.2631-0.3171 i \\
0& 256& 0.1 & 0.2 & 1  & 174.366\, -0.317942 i & 174.187-0.317 i \\
 \hline
0& 1& 0.2 & 0.4 & 1  & 0.941929\, -0.283424 i & 0.959220-0.281321 i \\
0& 2& 0.2 & 0.4 & 1  & 1.58836\, -0.282078 i & 1.59870-0.28132 i \\
0& 4& 0.2 & 0.4 & 1  & 2.87185\, -0.281554 i & 2.87766-0.28132 i \\
 0& 8&0.2 & 0.4 & 1  & 5.4339\, -0.281439 i & 5.43558-0.28132 i \\
0 & 16& 0.2 & 0.4 & 1 & 10.5673\, -0.283207 i & 10.55142-0.28132 i \\
0& 32& 0.2 & 0.4 & 1  & 20.8256\, -0.283026 i & 20.7831-0.2813 i \\
0& 64& 0.2 & 0.4 & 1  & 41.3399\, -0.28305 i & 41.2464-0.2813 i \\
0& 128& 0.2 & 0.4 & 1  & 82.3684\, -0.283005 i & 82.1732-0.2813 i \\
0& 256& 0.2 & 0.4 & 1  & 164.426\, -0.283012 i & 164.027-0.281 i \\
\end{array}$
\caption{Comparison of QNMs obtained from TBA \eqref{TBAG1}, through \eqref{quanteps1}, and WKB approximation.}
\end{table}\label{QNMsNf=1}
\begin{figure}[t]
\centering
\includegraphics[width=0.95\textwidth]{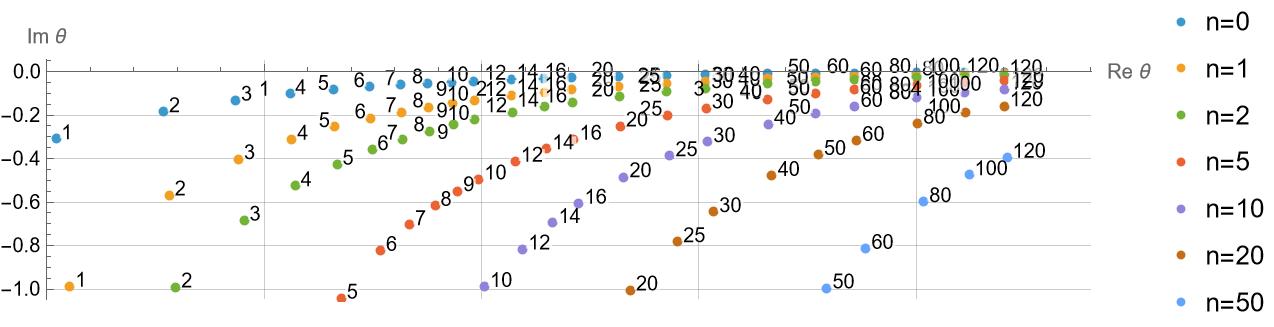}
\caption{Plot of the Bethe roots in the $\theta$ plane correspoding to quasinormal modes $\omega_n$, for several $n$ and $l$ of generalized RN BH ($N_f=1$ gauge theory) corresponding to $\Sigma_1=0.1$, $\Sigma_2 = 0.2$, $\Sigma_3=1$, as obtained from \eqref{quanteps1}} \label{plotBetheRootsNf1}
\end{figure}
As in the Perturbed Hairpin integrability TBA, also this gravity TBA does not contain explicitly $p\sim l$, so that is has to be solved through the following boundary condition 
\be 
\ve_{k,\pm}(\th )  \simeq 6(l+1/2) \th-2 D_1(l+1/2) \,,\quad \th \to -\infty\,,
\ee
with the subleading constant $D_1$ being the $q\to 0$ limit of the Hairpin one \eqref{Cint1}
\be 
D_1(p)=C_1(p,0)=\log \left(\frac{2^{-p-\frac{1}{2}} \Gamma (2 p) \Gamma (2 p+1)}{\sqrt{\pi } \sqrt{\Gamma \left(p+\frac{1}{2}\right)^2}}\right)\,.
\ee
From the general analysis of \cite{FioravantiGregori:2021} we can safely claim that the QNMs $\omega_n \propto e^{\theta_n}$ are given by zeros of $Q_{+}$, or the equivalent conditions on $\varepsilon$
\be  \label{quanteps1}
Q_{+}(\th_n) = 0\,,\qquad
\varepsilon_{0,+}(\th_n+ i \pi/2) = -i \pi (2n+1) \,,\qquad n \in \mathbb{Z}\,.
\ee 
Now from our gauge-integrability identification~\eqref{gaupseudocycles} we can prove a quantization on the gauge period $a_1$
\be \label{quantPerNf1}
 \frac{2\pi i}{\hbar(\theta_n)} a_1(\th_n,u,m) = -i \pi (2n+1) \qquad n \in \mathbb{Z}\,.
\ee 
We can now compare directly with the work~\cite{GrassiHaoNeitzke:2021} in which eq. 8.12 (in the first arXiv version) shows that zeros of $Q$ correspond to quantization conditions on the gauge periods, thus again recovering the characterization of QNMs of~\cite{AminovGrassiHatsuda:2020}.

Through the last relation we can actually compute the QNMs as we explained previously. 
We report their values we obtained in table \ref{QNMsNf=1}. Again, we find the Leaver method is not applicable to this case, at least in its original version~\cite{Leaver:1985}, so we compare only with the geodetic WKB approximation. In figure \ref{plotBetheRootsNf1} we also show the Bethe roots $\theta_n$ - in the whole $- \pi/3 <\Im \theta<0$ strip - corresponding to various $l$ and $n$. 

Applying the $N_f=1$ $TQ$ system~\eqref{TQ1} to also this background, we find the same limitations as for $N_f=2$ in finding quantization conditions for $T$ and $a$ as explained below~\eqref{TTqNf=2} and~\eqref{nuqQuant}.

\section{Conclusions and perspectives} \label{conclusions}

In summary, from the properties of the ODEs \eqref{ODEgau1}-\eqref{ODEgau2} governing $N_f=1,2$ $SU(2)$ NS gauge theories and the corresponding gravitational systems, 
we have derived functional and integral equations for their central and lateral connection coefficients, that is the $Q$, $T$ and $Y$ functions: these form a consistent IM. In particular we have derived formulae \eqref{gauint1}-\eqref{gauint2} which connect the integrability $Y$ function to the gauge periods $a,a_D$. This has been proven first in the $\theta \to + \infty$ and $\theta \to - \infty$ regime and then exactly by verifying the equivalence of integrability \eqref{int-TBA}-\eqref{int-TBA2} and gauge TBAs \eqref{ga-TBA}-\eqref{ga-TBA2}. Remarkably, similar identifications \eqref{Ta1}-\eqref{TTa2} hold for (quadratic combinations of) the $T$ functions. To prove them, we have extended the ODE/IM correspondence by connecting its basis to the Floquet basis, which provides relations between the $T$ functions and the Floquet exponent. Then we have checked the equality of this with the $a$ period \eqref{anu}, in both $\theta \to + \infty$ and $\theta \to - \infty$ regimes. We have also found some applications of our integrability-gauge correspondence: an interpretation of the integrability functional relations as exact $R$-symmetry relations for the gauge periods \eqref{TTQQNf=2}, \eqref{TTTTNf2}, as well as a new method of computation of the LIMs \eqref{lnQUps} or gauge periods \eqref{aDnLIM1}. Finally we have applied this to black hole perturbation theory: we have verified the QNMs of these correspond to quantization conditions on $Q$ and $Y$ functions \eqref{quanteps1}-\eqref{quanteps2} and so can be computed through the TBAs \eqref{TBAG1}-\eqref{TBA2}. Then, we have explored quantization conditions on the $T$ functions and found they also correspond to QNMs \eqref{TTqNf=2}, in the case of equal mass parameters $q_1=q_2$. By virtue of our previous identifications between integrability functions and gauge theory periods these facts have allowed us to prove QNMs correspond also to quantization conditions on $a_D$ \eqref{quantPerNf1}-\eqref{quantPerNf2}, but not on $a$ in general. All these considerations  show how integrability structures give valuable insights into several aspects of the new gauge-gravity correspondence \cite{AminovGrassiHatsuda:2020,BianchiConsoliGrilloMorales:2021,BianchiConsoliGrilloMorales:2021b,BonelliIossaLichtigTanzini:2021}.

Our main computational tool has been the TBA, of which we have shown several instances in integrability \eqref{int-TBA}-\eqref{int-TBA2}, gauge theory \eqref{ga-TBA}-\eqref{ga-TBA2} and gravity  \eqref{TBAG1}-\eqref{TBA2}. For each $N_f=1,2$ theory these are all equivalent to each other, under appropriate change of variables from integrability to gauge \eqref{DictGau1}-\eqref{DictGau2} or to gravity \eqref{dictGravNf1}-\eqref{DictIntD3branes}, as explained in subsection \ref{subsecYex}. In general the TBA has the advantage of delivering exact numeric computations in all regimes, overcoming the limitations of the instanton and WKB asymptotic expansions. We have derived the TBA equations directly from the ODE (which in its turn stems from the 4-dimensional theory), through the extension of the methods of the ODE/IM correspondence. Moreover, we have independently derived the SW spectrum \eqref{ZNf1}. Viceversa, supposing the spectrum, one may think it is possible to derive 
the same TBA equations  formally by the prescription of Gaiotto, Moore and Neitzke (in a ``conformal" set-up) as in \cite{GaiottoMooreNeitzke:2008,GaiottoMooreNeitzke:2009,Chen:2010pk,GaiottoOpers,GrassiGuMarino}. Yet we remark that although their prescription can be used very generally, it is for this very reason arguably more conjectural than our ODE/IM approach. 
At present, we also have some technical limitations, which should be overcome in the future. Namely, analytic continuations of the TBAs in the moduli seem necessary to obtain some special gravity parameters (like for extremal Kerr black holes). 

The present work vastly extends and completes our previous \cite{FioravantiGregori:2019,FioravantiPoghossian:2019,FioravantiGregori:2021}, by showing how in general 2D integrable models find a natural connection to (NS-deformed) $\mathcal{N}=2$ 4D supersymmetric gauge theories and to black hole perturbation theory, shedding light also on the relation between these two. This new triple correspondence, or \textit{triality}, besides being interesting in itself, allows also to
derive new results on all three sides and at the exact or non-perturbative level. 

On these new directions, much extension work in either breadth and depth could be done. Further developments of the gauge-integrability correspondence could be explored, especially the extension of the basic identifications upon wall crossing in gauge theory. For gravity, we have shown that the $SU(2)$ $N_f=0,1$ and $N_f=2\equiv(1,1)$ (symmetrical) gauge theories correspond to BH perturbation theory in a gravitational background given by generalised extremal charged BHs. However, other $SU(2)$ $N_f=(2,0)$ (asymmetrical), $N_f=3,4$ and quiver gauge theories correspond to many other gravitational backgrounds, such as Schwarzschild, Kerr, AdS BHs, in remarkable generality~\cite{AminovGrassiHatsuda:2020,BianchiConsoliGrilloMorales:2021b}. We have not yet related them to IMs, but from the generality of the ODE/IM correspondence construction it is manifest that our method should apply to them too. 
We notice that much of the BH theory seems to go in parallel to the ODE/IM construction and its 2D integrable field theory interpretation, beyond the determination of QNMs. So it is very intriguing to investigate also other applications of integrability to BHs. For example the greybody factor seems to be ratio of $Q$s\footnote{This can be understood by considering its absorption coefficient role in 1D quantum mechanics. We hope to write more details on this in the future.} \cite{BonelliIossaLichtigTanzini:2021}. Rotating BHs such as Kerr have their perturbations governed by a separable PDE (similarly for Kerr-Newman BH with scalar perturbation), so that the problem can be still tackled through the ODE/IM correspondence. Moreover, when the PDE is not separable (such as in the case of Kerr-Newman BH with non scalar perturbation), a possible extension of the (connection coefficients and monodromy) theory from ODEs to PDEs could be pursued, which is very interesting in itself and for applications. Importantly, these new applications of $\mathcal{N}=2$ supersymmetry gauge theory and quantum integrability to black holes physics allow new non-perturbative characterizations and computations of QNMs  and other BH observables\footnote{Nonetheless at present we always remain within black hole \textit{perturbation} theory. What we mean is that within that theory we can give some new exact characterisations which through standard methods could have been just perturbative, (say ``perturbative at second order").}. This constitutes a remarkable transfer of methods for exact solution to a new very physical research field and has the potential to illuminate aspects of classical and quantum gravitational theories that could be difficult to access through traditional methods, giving a deeper mathematical grasp to gravitational waves observations \cite{LIGOScientific:2016,CardosoPani:2017,BianchiConsoliGrilloMorales:2021}.

\vspace{10mm}
{\bf Acknowledgements} We thank M. Bianchi, D. Consoli, A. Grassi, A. Grillo, F. Morales, H. Poghossian, K. Zarembo for discussions and suggestions. This work has been partially supported by the grants: GAST (INFN), the MPNS-COST Action MP1210, the EC Network Gatis and the MIUR-PRIN contract 2017CC72MK\textunderscore 003. DG and HS thank Nordita for warm hospitality.
\normalsize

\newpage

\appendix 

\section{Quantum Seiberg-Witten theory with fundamental matter} \label{sec:qSWNf}

\subsection{Quantum SW curves}
\label{sec:qSWNf1}

The Seiberg-Witten (SW) curve for $\mathcal{N}=2$ $SU(2)$ with $N_f$ fundamental matter flavour hypermultiplets is given by
\be 
K(p) - \frac{\bar{\Lambda}}{2} (K_+(p) e^{i x} + K_-(p) e^{- i x} ) = 0
\ee
where
\be 
\bar{\Lambda} = \begin{cases}
\Lambda_0^2   \quad &N_f = 0\\
\Lambda_1^{3/2}  \quad &N_f = 1\\
\Lambda_2^1 \quad &N_f = 2
\end{cases}
\ee 
\be 
K(p) = \begin{cases}
p^2-u   \quad &N_f = 0\\
p^2-u  \quad &N_f = 1\\
p^2-u + \frac{\Lambda_2^2}{8}\quad &N_f = 2 
\end{cases}
\ee 
\be 
K_+(p) = \prod_{j=1}^{N_+} (p + m_j) \,, \quad K_-(p) = \prod_{j=N_++1}^{N_f} (p + m_j) \,.
\ee 
$u$ is the Coulomb moduli parameter and $m_i$ are the masses $1\leq N_+\leq N_f$. By introducing $y_{SW}= \bar{\Lambda} K_+ (p) e^{i x} - K(p)$ we get the SW curve in standard form 
\be \label{SWcurve4}
y^2_{SW} = K(p)^2 -\bar{\Lambda}^2 K_+(p) K_-(p)\,.
\ee
The SW differential is then 
\be 
\lambda = p d \ln \frac{K_-}{K_+} - 2 \pi i p \, d x\,,
\ee 
and defines a symplectic form $d \lambda = d p \wedge d x$, which doubly integrated gives the SW periods~\cite{ItoKannoOkubo:2017}
\be 
a = \oint_A p(x) \, dx \qquad a_D = \oint_B p(x) dx\,.
\ee 
The quantum SW curve is obtained by letting $p$ become the differential operator $- i \hbar \frac{d}{dx}$\cite{ItoKannoOkubo:2017}:
\be  \label{genqswode}
\left( K(- i \hbar \partial_x)) -\frac{\bar{\Lambda}}{2} ( e^{i x/2} K_+ (- i \hbar \partial_x ) e^{i x/2} + e^{-i x/2} K_- (- i \hbar \partial_x ) e^{-i x/2} \right) \psi(x) = 0\,.
\ee 

Let us now write formula~\eqref{genqswode} explicitly for the $N_f=0,1,2$ cases of relevance for this paper. Let $N_f= 0$ and $x = -i y $. We get
\be 
-\hbar^2  \frac{d^2}{d y^2} \psi + ( \Lambda_0^2 \cosh y + u) \psi = 0\,.
\ee 
Let $N_f= 1$ and $x = -i y $. We get
\be 
-\hbar^2  \frac{d^2}{d y^2} \psi + \left[\frac{1}{16} \Lambda_1^3 e^{2y} +\frac{1}{2}\Lambda_1^{3/2}e^{-y}  +\frac{1}{2} \Lambda_1^{3/2} m_1 e^y +u\right] \psi = 0\,.
\ee Let $N_f= 1$ and $x = -i y $, $y \to y- \frac{1}{2} \ln \Lambda_1 +\ln 2$. We get
\be 
-\hbar^2  \frac{d^2}{d y^2} \psi + \left[\frac{1}{4} \Lambda_1^2 (e^{2y} +e^{-y} ) + \Lambda_1 m_1 e^y +u\right] \psi = 0\,.
\ee 
Let $N_f=2$, $N_+=1$ and $x = - i y$. We get
\be 
-\hbar^2  \frac{d^2}{d y^2} \psi + \left[\frac{1}{16} \Lambda_2^2 (e^{2y} +e^{-2y} ) + \frac{1}{2}\Lambda_2 m_1 e^y + \frac{1}{2}\Lambda_2 m_2 e^{-y} +u\right] \psi = 0\,.
\ee 
Let $N_f=2$, $N_+=2$ and $x = - i y$. We get
\be \label{qsw20noshift}
-\hbar^2 \frac{d^2}{dy^2} \psi + \frac{e^{2 y} \Lambda_2^2\left( m_1- m_2\right)^2+e^y \left(\Lambda_2^3-2 \Lambda_2 \hbar^2+8 \Lambda_2 m_1 m_2-8 \Lambda_2 u\right)+16 u-6 \Lambda_2^2+8 \Lambda_2 e^{-y}}{4  \left(\Lambda_2 e^y-2\right)^2} \psi=0\,.
\ee 

In this paper to relate to BHs and IMs we need to consider only the first realization $N_+=1$ (symmetric) for $N_f=2$. We notice also that the second realization $N_+=2$ (asymmetric) has a rather different singular structure: two regular and one irregular singularities instead of two irregular singularities. Therefore it is a Confluent Heun equation rather than a Doubly Confluent Heun equation as all the others considered in this paper (see appendix~\ref{appDCHE}). We refer though to~\cite{BianchiConsoliGrilloMorales:2021b} for a dictionary with BHs also for this second realization.

\subsection{$N_f=1,2$ Seiberg-Witten periods} \label{app:Nf1SWper} \label{appSWper}

In this subsection we define and give some relations for the Seiberg-Witten periods for the $SU(2)$ $N_f=1,2$ theories, that is, the leading $\hbar \to 0$ of the quantum (or deformed) exact periods, which in sections \ref{secY} and \ref{secT} we prove to be connected to integrability exact $Y$ and $T$ functions.

\subsubsection{Massless $N_f=1$ SW periods}

Let us consider first the massless cases. Since the massless $N_f=2$ gauge periods are just the $N_f=0$ gauge periods already dealt with in~\cite{FioravantiGregori:2019}, we consider here only the (much more complex) $N_f=1$ massless $m=0$ case, following and extending~\cite{BilalFerrariQCD:1996}. In that case the low energy effective action has three finite $\mathbb{Z}_3$ symmetric singularities, corresponding to dyon BPS particles becoming massless. If we set $\Lambda_1 = \Lambda_1^*$ with
\be 
\Lambda_1^* = \sqrt[6]{\frac{256}{27}}\,,
\ee 
those singularities are situated at 
\be 
u_0 = - 1 \qquad u_1 = - e^{2\pi i/3} \qquad u_2 = - e^{-2\pi i/3}\,.
\ee 
The massless $m=0$ $N_f=1$ SW curve is
\be 
y_{SW}^2(u,\Lambda_1)= x^3-u x^2 - \frac{\Lambda_1^6}{64}\,,
\ee 
and it gives the SW periods through the integrals
\be 
\begin{pmatrix}a^{(0)}(u,\Lambda_1) \\a^{(0)}_D(u,\Lambda_1) \end{pmatrix}
=\frac{1}{4\pi} \oint_{A,B} dx\, \frac{2u - 3 x}{\sqrt{x^3-ux^2- \frac{\Lambda_1^6}{64}}} \,.
\ee 
It can be shown then that $\Pi^{(0)}=a^{(0)},a^{(0)}_D$ satisfy the SW Picard-Fuchs equation 
\be
\left(\frac{27 \Lambda_1^6}{256}+u^3 \right) \frac{\partial^2 \Pi^{(0)}(u)}{\partial u^2}+\frac{u}{4}\Pi^{(0)}(u)=0\,,
\ee
with boundary condition as $u/\Lambda_1^2 \to \infty$ as
\be
\ba
a^{(0)}(u, \Lambda_1) &\simeq   \sqrt{ u } \qquad u/\Lambda_1^2  \to \infty \\
a_D^{(0)}(u, \Lambda_1)&\simeq -i\left[\frac{1}{2\pi}a^{(0)}(u,0,\Lambda_1) \left(-i\pi - 3 \ln \frac{16 u}{\Lambda_1^2}\right)+\frac{3}{\pi}\sqrt{ u }\right]\qquad u/\Lambda_1^2  \to \infty\,.
\ea
\ee
The massless SW Picard-Fuchs equation can be mapped into an hypergeometric equation and then explicit formulas for $a^{(0)}, a^{(0)}_D$ follow:
\be 
\ba
a^{(0)}(u,\Lambda_1) &=  \sqrt{ u } \, _2F_1\left(-\frac{1}{6},\frac{1}{6};1;-\frac{27 \Lambda_1^6}{256 u^3}\right) \\
a_D^{(0)}(u,\Lambda_1) &= \begin{cases}
-a^{(0)}(u,\Lambda_1)+e^{-i \pi/3}f_D(u,\Lambda_1)\qquad0< \arg (u)\leq\frac{2 \pi }{3}\\
 f_D(u,\Lambda_1)-2 a^{(0)}(u,\Lambda_1)\qquad\frac{2 \pi }{3}< \arg (u)\leq\pi \\
 a^{(0)}(u,\Lambda_1)-f_D(u,\Lambda_1)\qquad-\pi < \arg (u)<-\frac{2 \pi }{3}\\ 
 \exp \left(-\frac{2 \pi  i}{3}\right) f_D(u,\Lambda_1)\qquad-\frac{2 \pi }{3}\leq \arg (u)\leq0
\end{cases}\,,
\ea
\ee
(the sectors are given assuming $\Lambda_1 >0$), where
\be
f_D(u,\Lambda_1)= \frac{   \Lambda_1 \left(\frac{256 u^3}{27 \Lambda_1^6}+1\right) \, _2F_1\left(\frac{5}{6},\frac{5}{6};2;\frac{256 u^3}{27 \Lambda_1^6}+1\right)}{4  \sqrt[3]{2}  \sqrt{3}}\,.
\ee
Under these definitions, $a^{(0)} $ has a branch cut for $u<0$ (due to the square root and three other cuts from the origin $u=0$ to $u_0$, $u_1$ and $u_2$ (due to the hypergeometric function). Similarly, $a^{(0)}_D$ has a branch cut for $u<0$ and from $u=0$ to $u_2$.

\subsubsection{$\mathbb{Z}_3 $ R-symmetry}

By direct use of the explicit formulae above, we find the following $\mathbb{Z}_3 $ R-symmetry relations
\begin{equation}
\begin{aligned}  \label{Z3SW}
a^{(0)}(e^{2 \pi i/3} u ) &= - e^{- 2 \pi i/3} a^{(0)}(u) \,\, &-  \pi < \arg u \leq  \pi/3 \\
a^{(0)}(e^{2 \pi i/3} u ) &=  e^{- 2 \pi i/3} a^{(0)}(u) \,\, & \pi/3 < \arg u \leq \pi \\
a^{(0)}(e^{-2 \pi i/3} u )& = - e^{2 \pi i/3} a^{(0)}(u) \,\, &-  \pi /3 < \arg u \leq  \pi  \\
a^{(0)}(e^{-2 \pi i/3} u )& = e^{2 \pi i/3} a^{(0)}(u) \,\, &-  \pi < \arg u \leq - \pi /3\\
a_D^{(0)}(e^{2 \pi i/3} u ) &= - e^{- 2 \pi i/3} \left [ a_D^{(0)}(u) -a^{(0)}(u) \right ]  \qquad &- \pi  < \arg u \leq \pi /3\\
a_D^{(0)}(e^{-2 \pi i/3} u ) &= - e^{ 2 \pi i/3} \left [ a_D^{(0)}(u) + a^{(0)}(u) \right ]  \qquad&-  \pi /3 < \arg u \leq \pi\,.
\end{aligned}
\end{equation}

\subsubsection{Massive $N_f=1,2$ SW periods}

The massive $N_f=1$ SW curve is~\cite{BilalFerrari-massive:1997}
\be
y_{SW}^2=x^3-ux^2+\frac{\Lambda_1^3}{4} m_1 x - \frac{\Lambda_1^6}{64}\,,
\ee
while the SW differential is
\be 
\lambda =  \frac{1}{2\pi} \left[-\left(3x-2u + \frac{\Lambda_1^3}{4} \frac{m}{x} \right) \frac{d x}{2 y_{SW}} \right]\,.
\ee 
The SW periods $a^{(0)}_1$, $a^{(0)}_2$ are given by the integrals
\be  \label{perEllipticNf1}
a_i^{(0)}(u,m,\Lambda_1)=\int_{\gamma_i} \lambda = \frac{1}{2 \pi} \left[ u I_1^{(i)} - 3I_2^{(i)} - \frac{\Lambda_1^3}{4} m I_3^{(i)} \left(-\frac{u}{3} \right) \right]\,.
\ee 
Now let us define $e_k$ as the roots of the Seiberg-Witten curve in canonical form
\be 
\ba
y_{SW}^2(x=\xi + \frac{u}{3})&=(\xi - e_1) (\xi - e_2) (\xi - e_3)\\
&=-\frac{\Lambda_1 ^6}{64}+\xi  \left(\frac{\Lambda_1 ^3 m}{4}-\frac{u^2}{3}\right)+\frac{1}{12} \Lambda_1 ^3 m u+\xi ^3-\frac{2 u^3}{27}\,,
\ea 
\ee
Then it can be proven that basic integrals over the cycle $\gamma_1$ are given by
\be 
\ba 
I_1^{(1)} &= 2 \int_{e_3}^{e_2} \frac{d \xi}{\eta} = \frac{2}{(e_1-e_3)^{1/2}}K(k) \\
I_2^{(1)} &= 2 \int_{e_3}^{e_2} \frac{\xi d \xi}{\eta} = \frac{2}{(e_1-e_3)^{1/2}}[e_1 K(k)+(e_3-e_1) E(k)]\\
I_3^{(1)} &= 2 \int_{e_3}^{e_2} \frac{d \xi}{\eta(\xi-c)} = \frac{2}{(e_1-e_3)^{3/2}}\left[\frac{1}{1-\tilde{c}+k'}K(k) +\frac{4 k'}{1+k'} \frac{1}{(1-\tilde{c})^2k^{'2}} \Pi_1\left(\nu(c), \frac{1-k'}{1+k'}\right) \right]\,,
\ea 
\ee
with
\be 
\ba
k^2&= \frac{e_2-e_3}{e_1-e_3} \quad k^{'2} = 1- k^2 \\
\tilde{c} &= \frac{c-e_3}{e_1-e_3} \quad \nu(c) = - \left( \frac{1-\tilde{c}+k'}{1-\tilde{c}-k'} \right)^2 \left( \frac{1-k'}{1+k'}\right)^2\,,
\ea
\ee 
that is, in terms of elliptic integrals of the first, second and third kind:
\be 
\ba 
K(k) &= \int_0^1 \frac{dx}{[(1-x^2)(1-k^2 x^2)]^{1/2}} \\
E(k) &= \int_0^1 dx\left(\frac{1-k^2 x^2}{1-x^2}\right)^{1/2} \\
\Pi_1(\nu,k) &= \int_0^1 \frac{dx}{[(1-x^2)(1-k^2 x^2)]^{1/2}(1+\nu x^2)}\,.
\ea 
\ee 
The corresponding integrals $I_i^{(2)}$ over the cycle $\gamma_2$ are obtained by exchaning in $I_i^{(1)}$ $e_1$ and $e_3$~\cite{BilalFerrari-massive:1997}.

For $N_f=2$ we have similarly (in the cubic SW curve conventions~\cite{BilalFerrari-massive:1997})
\be 
y_{SW}^2= x^3- u x^2  - \frac{\L_2^4}{64}(x-u) + \frac{\L_2^2}{4} m_1 m_2 x - \frac{\L_2^4}{64} (m_1^2+m_2^2)\,,
\ee 
\be 
\ba
\lambda &= - \frac{1}{2 \pi} \frac{d x}{y_{SW}} \left[ x-u - \frac{\L_2^2}{16}\frac{(m_1-m_2)^2}{x-\frac{\L_2^2}{8}}+\frac{\L_2^2}{16}\frac{(m_1+m_2)^2}{x+\frac{\L_2^2}{8}} \right]\,\\
&= - \frac{1}{2\pi} \frac{y_{SW}\, d x}{x^2- \frac{\Lambda_2^4}{64}}\,,
\ea
\ee 
\be \label{perEllipticNf2}
\ba
a_i^{(0)}&(u,m_1,m_2,\Lambda_2)=\int_{\gamma_i} \lambda \\&= \frac{1}{2 \pi} \left[ \frac{4}{3} u I_1^{(i)} - 2 I_2^{(i)} + \frac{\L_2^2}{8} (m_1-m_2)^2 I_3^{(i)}\left(\frac{\L_2^2}{8} -\frac{u}{3} \right) -\frac{\L_2^2}{8} (m_1+m_2)^2 I_3^{(i)}\left(-\frac{\L_2^2}{8} -\frac{u}{3} \right) \right] \,.
\ea
\ee 

\subsubsection{Relations between alternatively defined periods}

Importantly, we notice that the periods $a^{(0)}$ and $a^{(0)}_D$ so defined are in principle different from the periods $a_1^{(0)}$ and $a_2^{(0)}$ defined as integrals. They are in fact linear combinations of each other, which also possible separate mass term contribution.

Let us now show the relation between $a^{(0)}, a^{(0)}_D$ and $a_1^{(0)},a_2^{(0)}$ in the massless $N_f=1$ case. Assuming $u > 0$ and with small $|u|$ we have
\be 
\ba 
a^{(0)}(u) &= a^{(0)}_1(u)\qquad \Re{a^{(0)}(u)}>0 \\
a^{(0)}_D(u) &=-a^{(0)}_2(u))\qquad \Re{a^{(0)}_D(u)}<0  \\
a^{(0)}(e^{2\pi i/3}u)&= a^{(0)}_1(e^{2\pi i/3}u)-a^{(0)}_2(e^{2\pi i/3}u) \\
a^{(0)}_D(e^{2\pi i/3}u) &= -a^{(0)}_1(e^{2\pi i/3}u)+2a^{(0)}_2(e^{2\pi i/3}u) \\
a^{(0)}(e^{-2\pi i/3}u)&=a^{(0)}_1(e^{-2\pi i/3}u)-a^{(0)}_2(e^{-2\pi i/3}u) \\
a^{(0)}_D(e^{-2\pi i/3}u)&=-a^{(0)}_2(e^{-2\pi i/3}u)\,,
\ea 
\ee
with their inverses
\be 
\ba  \label{a12aaD}
a^{(0)}_1(u) &= a^{(0)}(u)\qquad &\Re{a^{(0)}_1(u)}>0 \\
a^{(0)}_2(u) &= - a^{(0)}_D(u) \qquad &\Re{a^{(0)}_2(u)}>0\\
a^{(0)}_1(e^{2\pi i/3}u) &=  a^{(0)}_D(e^{2\pi i/3} u)+2 a^{(0)} (e^{2\pi i/3}u)\qquad &\Re{e^{2\pi i/3}a^{(0)}_1(e^{2\pi i/3}u)}<0  \\
a^{(0)}_2(e^{2\pi i/3}u) &=  a^{(0)}_D(e^{2\pi i/3} u)- a^{(0)} (e^{2\pi i/3}u)\qquad &\Re{e^{2\pi i/3}a^{(0)}_2(e^{2\pi i/3}u)}>0  \\
a^{(0)}_1(e^{-2\pi i/3}u) &= a^{(0)}(e^{-2\pi i/3}u) - a^{(0)}_D(e^{-2\pi i/3}u)\qquad &\Re{e^{-2\pi i/3}a^{(0)}_1(e^{-2\pi i/3}u)}<0 \\
a^{(0)}_2(e^{-2\pi i/3}u)&=-a^{(0)}_D(e^{-2\pi i/3}u)\qquad &\Re{e^{-2\pi i/3}a^{(0)}_2(e^{-2\pi i/3}u)}>0\,.
\ea
\ee 
Also
\be 
\ba 
a^{(0)}(-u) &= -a^{(0)}_1(-u)+a^{(0)}_2(-u) \\
a^{(0)}_D(-u) &=3a^{(0)}_1(-u)-2a^{(0)}_2(-u) \\
a^{(0)}(-e^{2\pi i/3}u)&= a^{(0)}_2(-e^{2\pi i/3}u)  \\
a^{(0)}_D(-e^{2\pi i/3}u) &=-a^{(0)}_1(-e^{2\pi i/3}u)+a^{(0)}_2(-e^{2\pi i/3}u)   \\
a^{(0)}(-e^{-2\pi i/3}u)&= a^{(0)}_2(-e^{-2\pi i/3}u)  \\
a^{(0)}_D(-e^{-2\pi i/3}u)&= a^{(0)}_1(-e^{-2\pi i/3}u)-2a^{(0)}_2(-e^{-2\pi i/3}u) \,.
\ea 
\ee

In the massive case, similar relations can be found by looking at the large $u$ asymptotics and, if the small $u$ region is of interest, also to the continuous behaviour of the functions involved.

\section{Derivation of TBA forcing terms} \label{appForcing}

In this appendix, we explicitly compute the forcing terms and boundary conditions for the TBAs.

We can simply compute concretely the integrals $\ln Q^{(0)}$ \eqref{lnQ0gauint} and \eqref{lnQ0gauint2} by expanding the square root integrand in multiple binomial series for small parameters, producing simple Beta function integrals. In particular, for $N_f=1$ we get
\be \label{lnQ0gau}
\ln Q^{(0)}(u,m,\Lambda_1)=\sum_{n=0}^{\infty}\sum_{l=0}^{\infty}\binom{1/2}{n}\binom{1/2-n}{l}B_1(n,l)\left(\frac{4m}{\Lambda_1}\right)^{n}\left(\frac{4 u}{\Lambda_1^2}\right)^{l}\,,
\ee
with
\be
\ba
B_1(n,l)&=\frac{1}{3}B \left(\frac{1}{6} (2 l+4 n-1),\frac{1}{3} (2 l+n-1)\right) \qquad (n,l)\neq(1,0) \\
B_1(1,0)&=\frac{2 \ln (2)}{3}\,,
\ea
\ee
and for $N_f=2$ we obtain
\be
\ba 
\ln Q^{(0)}(u,m_1,m_2,\Lambda_2)&= \sum_{l,m,n=0}^\infty \binom{\frac{1}{2}}{l} \binom{\frac{1}{2}-l}{m} \binom{-l-m+\frac{1}{2}}{n} B_2(l,m,n) \left(\frac{8 m_1}{\Lambda_2}\right)^n \left(\frac{16 u}{\Lambda_2^2}\right)^m \left(\frac{8 m_2}{\Lambda_2}\right)^l \,,
\ea
\ee
with
\be 
\ba
B_2(l,m,n)&=\frac{\Gamma \left(\frac{1}{4} (3 l+2 m+n-1)\right) \Gamma \left(\frac{1}{4} (l+2 m+3 n-1)\right)}{4 \Gamma \left(l+m+n-\frac{1}{2}\right)} \\
B_2(1,0,0)&=\frac{1}{2} (\ln 2-1)  \quad  B_2(0,0,1)= \frac{1}{2}\ln 2\,.
\ea 
\ee
Of course, when $u,m,\Lambda_1$ ($u,m_1,m_2,\Lambda_2$) are such that the leading order~\eqref{eps0gau} computed through~\eqref{lnQ0gauint} has a negative real part, the TBA~\eqref{ga-TBA} no longer converges. In general, we find the convergence region to correspond to $u,m$ ($u,m_1,m_2$) finite but small with respect to $\Lambda_1$ ($\Lambda_2$). For instance in the $N_f=1$ massless case, this region corresponds on the real axis of $u$ precisely to the strong coupling region $-3\Lambda_1^2/2^{8/3}<u<3\Lambda_1^2/2^{8/3}$. For $N_f=2$ massless instead it corresponds to the region $-\Lambda_2^2/8<u<\Lambda_2^2/8$~\cite{BilalFerrariQCD:1996}.

Similarly for the $N_f=1$ gravitational TBAs, we can compute the integral \eqref{c0TBAgrav1} analytically as a double binomial series for small $\Sigma_1, \Sigma_2$
\be
\ln Q^{(0)}( \Sigma_1, \Sigma_2,\Sigma_3)
=\sum_{n=0}^{\infty}\sum_{l=0}^{\infty}\binom{1/2}{l}\binom{1/2-l}{n}B_1(n,l)\left(\frac{\Sigma_1}{\sqrt[3]{\Sigma_3}} \right)^{n}\left(\frac{\Sigma_2}{\sqrt[3]{\Sigma_3^2}}\right)^{l}\,.
\ee
Similarly \eqref{c0gravTBA2} for the gravitational TBA for $N_f=2$ can be expressed either through a triple power series for small parameters as 
\be
\ba 
\ln Q^{(0)}(\Sigma_1,\Sigma_2,\Sigma_3,\Sigma_4)
&= \sum_{l,m,n=0}^\infty \binom{\frac{1}{2}}{l} \binom{\frac{1}{2}-l}{m} \binom{-l-m+\frac{1}{2}}{n} B_2(l,m,n) \left(\frac{\Sigma_1}{\sqrt[4]{\Sigma_4}}\right)^n \left(\frac{\Sigma_2}{\sqrt{\Sigma_4}}\right)^m \left(\frac{\Sigma_3}{\sqrt[4]{\Sigma_4^3}}\right)^l \,.
\ea 
\ee

Following \cite{ItoMarinoShu:2018,GrassiGuMarino}, we can also prove the expression for the $\theta \to - \infty$ boundary conditions. In fact, under boundary condition \eqref{ga-bdy-cond}, the  dilogarithm trick leads to the ``effective central charge'' associated with the TBA equations (\ref{ga-TBA}) for $N_f=1$
\be
\label{ceff}
c_{\rm eff}=\frac{6}{\pi^2}\int d\theta e^\theta \sum_{j=0}^2\ve^{(0)}_{\pm,j} L_{\pm,j}(\theta)=3\,,
\ee
which coincides with the numeric test and thus tests the validity of our boundary condition. Similarly for $N_f=2$ the boundary condition \eqref{ga-bdy-cond-2} follows consistently by the fact that the effective central charge is 
\be
\label{ceff2}
c_{\rm eff}=\frac{6}{\pi^2}\int d\theta e^\theta \sum_{\pm } \big(\varepsilon^{(0)}_{\pm,\pm} L_{\pm,\pm}(\theta)+ \bar{\varepsilon}_{\pm,\pm}^{(0)} \bar{L}_{\pm,\pm}(\theta)\big)=4\,.
\ee

Let us show also how to compute the $\theta \to - \infty$ boundary conditions \eqref{bdyint1}, \eqref{bdyint2} for the  integrability TBAs. We can change variable in the ODE \eqref{ODEint1} in two different ways. First let $y_+ = y + \theta$, to obtain
\be
- \frac{d^2}{dy_+^2} \psi(y_+) + (e^{2 y_+} + q e^{y_+}+ e^{3 \theta -y_+} +p^2) \psi(y_+)=0\,,
\ee 
whose $\theta \to - \infty$ asymptotic solution is given in terms of confluent hypergeometric function as
\be
\psi_{+,0}(y_+)\simeq 2^p e^{p y_+ -e^{y_+}} U\left(q+p+\frac{1}{2},2 p+1,2 e^{y_+}\right)\,,\qquad \theta \to - \infty\,.
\ee
Then let $y_- = y -2\theta$, to obtain
\be
- \frac{d^2}{dy_-^2} \psi(y_-) + (e^{6 \theta +2 y_-}+ e^{3 \theta + y_-}+e^{- y_-} +p^2) \psi(y_-)=0\,,
\ee
whose asymptotic solution for $\theta \to - \infty$ is given in terms of modified Bessel function as
\be
\psi_{-,0}(y_-)\simeq \sqrt{\frac{2}{\pi }} K_{2 p}\left(2 e^{-\frac{y_-}{2}}\right)\,,\qquad \theta \to - \infty.
\ee
Changing back to $y$, we can verify these solutions have the correct asymptotic behaviours \eqref{asyreg1} for $y \to \pm \infty$. Then we can compute their Wronskian, to obtain $Q$ as
\be
Q(\theta)= W[\psi_{+,0},\psi_{-,0}]\simeq \frac{2^{\frac{1}{2}-p} p e^{-3 \theta  p} \Gamma (2 p)^2}{\sqrt{\pi } \Gamma \left(p+q+\frac{1}{2}\right)} \,, \qquad\theta \to - \infty\,,
\ee
and finally $Y$, by \eqref{intYdef1}, as
\be
Y(\theta) \simeq e^{6 p \theta} e^{i \pi q} \left(\frac{2^{\frac{1}{2} (1-2 p)} p \Gamma (2 p)^2}{\sqrt{\pi } \sqrt{\Gamma \left(p-q+\frac{1}{2}\right)} \sqrt{\Gamma \left(p+q+\frac{1}{2}\right)}}\right)^2 =e^{6 p \theta} e^{i \pi q} e^{2 C_1(p,q)}\,,
\ee
which reproduces exactly boundary condition \eqref{bdyint1} and constant $C_1$ \eqref{Cint1}. A similar derivation for $N_f=2$ leads to the boundary condition \eqref{bdyint2} and constant $C_2$ \eqref{Cint2}.

\section{R\'esum\'e of $N_f=0$ theory}

\subsection{Gauge-integrability identification for $Q$} \label{proofNf=0}\label{appProof}

We report here the proof of the gauge integrability equivalence between $Q = \sqrt{Y}$ and $a_D$ in the $SU(2)$ $N_f=0$ gauge theory case. This hopefully can illuminate the proofs in the main text for the more complex $N_f=1,2$ theories. The starting point is the $N_f=0$ $SU(2)$ quantum SW curve, that is the following modified Mathieu equation
\begin{align} \label{mMathieuSU2}
- \hbar^2   \frac{d^2}{dy^2} \psi(y) +[ \Lambda_0^2 \cosh{y} + u] \psi(y)&=0 \,\, .
\end{align}
We consider in first subsection the leading $\hbar \to 0$ asymptotic SW order, while in the second subsection the exact generalization. 

\subsubsection{Asymptotic proof}

The leading order of the quantum momentum for \eqref{mMathieuSU2} is
\be \label{SWlqm}
\CP_{-1}=-i \Lambda_0   \sqrt{ \cosh{y'}+\frac{u}{\Lambda_0^2}} \, .
\ee
The SW gauge periods are then explicitly
\begin{align}
 a^{(0)}(u, \Lambda_0)
&=\frac{1}{2 \pi} \int_{-\pi}^{\pi} \sqrt{u-\Lambda_0^2 \cos{z} } \, dz =\Lambda_0\sqrt{ u/\Lambda_0^2+1}\,\, {}_2F_1(-\frac{1}{2},\frac{1}{2},1;\frac{2}{1+u/\Lambda_0^2}) \,\, , \label{a0}  \\
 a^{(0)}_D(u, \Lambda_0)
&= \frac{1}{2\pi} \int_{- \arccos(u/\Lambda_0^2)-i0}^{ \arccos(u/\Lambda_0^2)-i0} \sqrt{u-\Lambda_0^2 \cos{z} } \, dz    =  - i \Lambda_0 \frac{(u/\Lambda_0^2-1)}{2\sqrt{2}} \,\, {}_2F_1(\frac{1}{2},\frac{1}{2},2;\frac{1-u/\Lambda_0^2}{2})  \, . \label{a0D} 
\end{align}
We need to compare this with the leading $\hbar \to 0$ order for $\ln Q$. To compute it as an integral on the real $y$ line, we need to regularize the leading quantum momentum~\eqref{SWlqm}. Since, in the limits $y \to \pm \infty$, we have $\CP_{-1} = - i \frac{\Lambda_0}{\sqrt{2}\h} e^{\pm y/2}+ O(e^{\mp y/2})$, it follows that the Seiberg-Witten regularized momentum is
\be
\CP_{reg,-1}(y)= \CP_{-1}(y) +  i \sqrt{2}\Lambda_0 \cosh \frac{y}{2}   =-i \Lambda_0 \Bigl [  \sqrt{  \cosh{y'}+ \frac{u}{\Lambda_0^2}} - \sqrt{2} \cosh \frac{y'}{2} \Bigr] \,.
\label{Preg-1}
\ee 
The leading order of $\ln Q$ is then~\cite{FioravantiGregori:2019}
\begin{align}
\ln{Q}^{(0)}( u,\Lambda_0) &=\int_{-\infty}^\infty i \CP_{reg,-1}(y) \,dy=  \Lambda_0 \int_{-\infty}^\infty  \Bigl [  \sqrt{  \cosh{y}+ \frac{u}{\Lambda_0^2}} - \sqrt{2} \cosh \frac{y}{2} \Bigr] dy \,.    \label{QSWreg} 
\end{align}
\begin{figure}[t]
\centering
\includegraphics[width=0.85\textwidth]{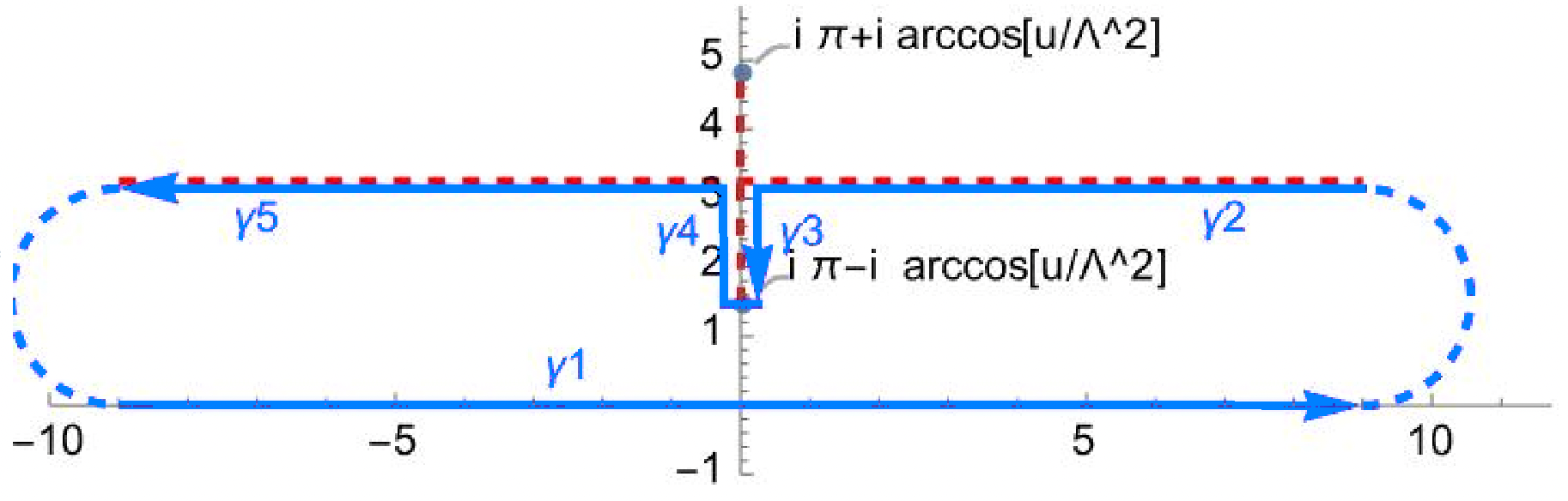}
\caption{A region of the $y$ complex plane, where in yellow we show the contour of integration of SW differential for the $SU(2)$ $N_f=0$ theory we use for the proof equality of the dual SW period $a_D^{(0)}$ and the leading $\hbar \to 0 $ order of the logarithm of the Baxter's $Q$ function $\ln Q^{(0)}$. In red are shown the branch cuts of the SW differential.}\label{figcontourNf=0}
\end{figure}
We assume $u < \Lambda_0^2$. Let us consider the integral of $i \CP_{reg,-1}(y)$ on the (oriented) closed curve which runs along the real axis, slightly below the cut and closes laterally. We can decompose it as $\gamma = \gamma_1 \cup \gamma_{lat,R} \cup \gamma_2 \cup  \gamma_3 \cup \gamma_4 \cup \gamma_5 \cup \gamma_{lat,L}$, with $\gamma_1 =(-\infty , +\infty)$, $\gamma_2 = (+\infty + i \pi - i 0  , 0^+ + i \pi - i 0)$ , $\gamma_3 = (0^+ +  i \pi  - i 0, 0^+ +i \pi - i\arccos (u/\Lambda_0^2)$, $\gamma_4 = (0^- + i \pi - i \arccos (u/\Lambda_0^2), 0^- + i \pi - i 0 )$, $\gamma_5 = (0^- + i \pi - i 0  , -\infty + i \pi -i 0 )$, and $\gamma_{lat,L}$ $\gamma_{lat,R}$ are the lateral contours which close the curve (see figure \ref{figcontourNf=0}). 
We expect the integral of $ \mathcal{P}_{reg,-1}(y)$ on $\gamma$ to be zero, since the branch cuts are avoided and no singularities are inside the curve.
By expanding the square root for $\Re{y} \to \pm \infty$, $| \Im y | < \pi$, we get the asymptotic behaviour:
\begin{align}
  \frac{\hbar }{\Lambda_0}i\mathcal{P}_{reg,-1}(y) &= - \frac{u/\Lambda_0^2+1}{\sqrt{2}}e^{-y/2} +o(e^{-y/2} ) \qquad &\Re{y} &\to  + \infty \label{exp+} \\
   \frac{\hbar}{\Lambda_0}i \mathcal{P}_{reg,-1}(y) &= -\frac{u/\Lambda_0^2+1}{\sqrt{2}}e^{y/2} +o(e^{y/2} ) \qquad &\Re{y} &\to  -\infty \label{exp-}\,,
\end{align} 
from which, we deduce that the integrals on the lateral contours $\gamma_{lat,L/R}$ are exponentially suppressed.
For $\g_2$ and $\g_5$, we consider $\CP_{reg,-1}(t + i \pi - i 0)$ for $t \in \mathbb{R}$: 
\be
  \frac{\hbar }{\Lambda_0}i \CP_{reg,-1}(t + i \pi - i 0) =\sqrt{- \cosh{t}+\frac{u}{\Lambda_0^2}}-\sqrt{2} i \sinh \frac{t}{2}\,.
\ee
 Since for $t=0$ it is necessary to cross a cut, we find the oddness property $\CP_{-1}(t + i \pi - i 0) = -\CP_{-1}(-t + i \pi - i 0)  $. Besides also the regularizing part is odd and therefore, for $t \in \mathbb{R}$ we have
\be
\CP_{reg,-1}(t + i \pi - i 0) = -\CP_{reg,-1}(-t + i \pi - i 0) \,.
\ee 
As a consequence, the integrals on $\g_2$ and $\g_5$ cancel each other.
The integrals on $\g_3$ and $\g_4$, around the cut, can be better taken into account in the variable $z=-i y - \pi$. There is no contribution from the regularizing part, which has no cut. Instead $\CP_{-1}$, which is
\be
\CP_{-1}(z - i 0)=\Lambda_0\sqrt{- \cos{(z-i0)}+\frac{u}{\Lambda_0^2}}\,,
\ee
has the oddness property 
\be
\CP_{-1}(-z + i 0)= - \CP_{-1}(z - i 0) \qquad z \in \mathbb{R}\,.
\ee 
It follows that the integrals on $\g_3$ and $\g_4$ add to each other
\be
\int_{-\arccos (u /\Lambda_0^2)} ^{0}  \CP_{-1}(z-i0) \, dz +  \int_{0}^{-\arccos (u /\Lambda_0^2)}  \CP_{-1}(z+i0) \, dz  =\int_{-\arccos (u /\Lambda_0^2) -i0}^{ +\arccos (u /\Lambda_0^2) -i0}  \CP_{-1}(z ) \, dz \,.  
\ee
In conclusion, we find a relation between the integrals on $\g_1 $ and on $\g_3$ and $\g_4$:
\be
 \int_{-\infty }^{+\infty }i \mathcal{P}_{reg,-1}(y) \, dy =\int_{-\arccos (u /\Lambda_0^2) -i0}^{ +\arccos (u /\Lambda_0^2) -i0} i \CP_{-1}(z ) \, dz \,,
\ee
which in terms of physical quantities is
\be
\ln Q^{(0)}(u,\Lambda_0) = 2\pi i a_D^{(0)}(u,\Lambda_0)  \, .
\ee

\subsubsection{Exact analytic proof}

\begin{figure}[t]
 \centering
 \includegraphics[width=0.7\textwidth]{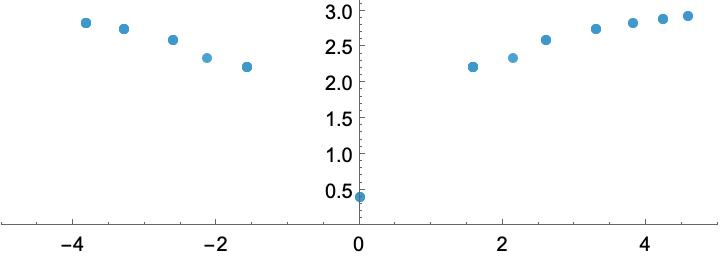}
 \caption{Poles for the exact quantum SW differental $\mathcal{P}(y)$ (here with integrability parameters  $\theta = 0$, $p=0.1$) for the $SU(2)$ $N_f=0$ theory, whose set we denote by $B$.}\label{fig:polesNf=0}
 \end{figure}

We can also construct an $\hbar$-exact analytic proof of the relation between the Baxter's $Q$ function and $a_D$ period, which then reads
\be
\label{idQaD}
Q(\th,P) = \exp  \frac{2 \pi i \,a_D(\hbar ,u,\Lambda_0)}{\hbar}\,.
\ee 
The exact proof follows along the lines of the $\hbar \to 0$ asymptotic proof, by using Cauchy theorem to relate the exact integral for the Baxter's $Q$ function and $a_D$ period. In particular, $\ln Q$ is defined as the $y$ integral over $(-\infty,+\infty)$ of ($i$ times) the regularised NS momentum (as in~\eqref{lnQPi}-\eqref{Q2++}, but see also~\cite{FioravantiGregori:2019})
\be
\CP_{reg}(y)= \CP(y) +\sqrt{2} i e^{\theta}\cosh \frac{y}{2} - \frac{i}{4} \tanh y \,\, .
\label{Preg'}
\ee 
Let us consider then the integral of $i \CP_{reg}(y)$ on the (oriented) closed curve with the poles we computed numerically as in figure~\ref{fig:polesNf=0}.
We can define the exact dual periods as the exact integrals of $\mathcal{P}(y) = - i \frac{d}{dy} \ln \psi(y) $ as in~\eqref{qperint}. However, we can now also write them as the sum over residues at the poles which as $\hbar\to 0$ reduce to the classical cycles (branch cuts), as shown in figure~\ref{fig:polesNf=0} \footnote{It may appear at first that the choice of poles for the two cycles is not well defined. However, on one hand we numerically find that the period $a$ as given precisely as the integral from $-i \pi$ to $i\pi$ as required by the equality $a= \nu$. On the other hand, the choice of poles for the period $a_D$ is unambiguous because it includes all of them.}
\begin{equation}
 \frac{1}{\hbar}a_D(\hbar,u,\Lambda_0) \doteq \oint_{B} \mathcal{P}( y,\hbar,u,\Lambda_0)\, d y =2\pi i\sum_{n}\text{Res}\mathcal{P}(y)\biggr |_{y_n^{B}} \,.
 \end{equation}
In this way we prove precisely the exact relation~\eqref{idQaD}. A further numerical proof can be given by showing the equivalence of the gauge and integrability TBAs, similarly to~\eqref{numTBAeq}.

\subsubsection{Relation with other gauge period}
 
A parallel finding was the following relation between the $Q$ function and
the gauge periods $A_D,a$~\cite{GrassiGuMarino} 
\be  \label{Q0GGM}
Q(\hbar,a, \Lambda_0) =i \frac{\sin \frac{1}{\hbar}A_D(\hbar,a,\Lambda_0)}{\sin  \frac{2\pi }{\hbar} a}\,. 
\ee  
We can easily check numerically this relation, as shown in table \ref{tabQAD0}, by computing the l.h.s. by the Liouville TBA for $b=1$~\eqref{TBA0} and the r.h.s. relies on the expansion of the prepotential $\mathcal{F}$ in $\Lambda_0$ \cite{Nekrasov:2002,NekrasovOkounkov:2003}: the period $a$ is related to the moduli parameter $u$ (or $P$) through the Matone's relation~\cite{Matone:1995,FlumeFucitoMoralesPoghossian:2004} and the dual one is given by $A_D=\partial{\mathcal{F}}/{\partial a} $, namely  
\be  \label{MatoneInst0}
 u =  a^2 - \frac{\Lambda_0}{2} \frac{\partial \mathcal{F}_{NS}^{\mathrm{inst}}}{\partial \Lambda_0} = a^2+\frac{\Lambda_0^4}{ 4 a^2-\hbar^2 }+O(\Lambda_0^8)\,,
\ee 
\be 
\ba \label{ADinst0}
\frac{A_D}{\hbar} =   \frac{\partial \mathcal{F}_{NS}}{ \partial a}=\frac{4 a}{\hbar} \ln \frac{\sqrt{2} \hbar}{\Lambda_0}  + \ln \frac{\Gamma(1 + \frac{2 a}{\hbar})}{\Gamma(1 - \frac{2 a}{\hbar})} + \frac{1}{\hbar}\frac{ 8 a}{(4 a^2 - \hbar^2)^2} \Lambda_0^4 + O(\Lambda_0^8)\,,
\ea 
\ee 
where the instanton prepotential is given by the infinite perturbative series
\be
\ba 
\mathcal{F}_{NS}^{\mathrm{inst}} = \sum_{n=0}^\infty \Lambda_0^{4n} \mathcal{F}_{NS}^{(n)}\,,
\ea 
\ee
with first terms 
\be 
\ba 
\mathcal{F}_{NS}^{(1)} &=- \frac{1}{4(4 a^2- \hbar^2)} \\
\mathcal{F}_{NS}^{(2)} &=- \frac{20 a^2+7 \hbar^2}{128(a^2-\hbar^2)(4a^2-\hbar^2)^3} \\
\mathcal{F}_{NS}^{(3)} &=- \frac{144 a^4+232 a^2\hbar^2+29 \hbar^4}{768(4a^2-\hbar^2)^5(4a^4-13 a^2\hbar^2+9\hbar^4)}\,.
\ea 
\ee
In this respect we noticed that only the first instanton contributions are easily accessible and summing them up (naively) is accurate as long as $|\Lambda_0|/\hbar \ll 1$. Moreover, $A_D(\hbar,u) $ so defined is different from our dual cycle $a_D(\hbar,u) $: it is not a cycle integral but is defined as the derivative of the prepotential (logarithm of the partition function) coming from instanton counting \eqref{ADinst0}.
Thus, thanks to \eqref{idQaD}, relation \eqref{Q0GGM} becomes a relation between the two definition of dual cycles
\be
 i \frac{\sin \frac{1}{\hbar}A_D(\hbar,a,\Lambda_0)}{\sin  \frac{2\pi }{\hbar} a}  = \exp \frac{2 \pi i \,a_D(\hbar ,u)}{\hbar} \label{ADaD} \,.
\ee
This relation means that the two cycles $a_D$ and $A_D$ differ by non-perturbative terms in $\h$. From the gauge theory point of view, they are precisely respectively the dyon and monopole period in the strong coupling region.
\begin{table}
\centering
\begin{tabular}{c|c|c|c|c}
$\Lambda_0$&$ p$&$\hbar $ &$-\frac{1}{2} \varepsilon(\th,p)$& $\ln i \sin A_D/\sin (2 \pi a/\hbar)$ \\
\hline 
$\frac{\Gamma^2(\frac{1}{4})}{16 \sqrt{\pi}}$&$2$&$-i$ &$9.27325$ & $9.273204$ \\
$ \frac{\Gamma^2(\frac{1}{4})}{16 \sqrt{\pi}}$&$3$&$-i $ &$18.7522$ & $18.752173$ \\
$ e^{-1}\frac{\Gamma^2(\frac{1}{4})}{16 \sqrt{\pi}}$&$2$&$-i$ &$17.2829$ & $17.282910$ \\
$ e^{1}\frac{\Gamma^2(\frac{1}{4})}{16 \sqrt{\pi}}$&$2$&$-i$ &$1.04849$ & $1.04235$ \\
\end{tabular}
\caption{Numerical check of formula~\eqref{Q0GGM}, through TBA \eqref{TBA0} and instanton formul{\ae} \eqref{ADinst0} and \eqref{MatoneInst0}, which we truncated at second order $O(\Lambda_0^8)$.}
\label{tabQAD0}
\end{table}

\subsection{D3 brane quantization relations} \label{D3brane}

As shown in~\cite{FioravantiGregori:2021}, the physical QNMs condition translates into
\be  \label{QNMBethe}
Q(\th_n) =0 \,.
\ee 
In other words, the QNMs are precisely the zeros of the Baxter's $Q$ function, which are the Bethe roots~\cite{BazhanovLukyanovZamolodchikov:1997}. Now, from \eqref{Q0GGM} we directly prove that condition~\eqref{QNMBethe} is equivalent to the quantization condition of the dual gauge period
\be  \label{QuantAD0}
\frac{1}{\hbar}A_D(a,\Lambda_{0,n},\hbar) = i\pi  n \,,\qquad n \in \mathbb{Z}\,.
\ee 
as conjectured heuristically in~\cite{AminovGrassiHatsuda:2020}. 

Eventually, the $QQ$ system~\eqref{QQ0}, with $Y=Q^2$, characterizes the QNMs as $Y(\th_n+i\pi/2)=-1$, {\it i.e.} the {\it TBA quantization condition} 
\be \label{quanteps0}
\varepsilon(\th_{n} +i \pi/2) = - i \pi(2n+1) \,, \qquad n \in \mathbb{Z}\,,
\ee 
which can be easily implemented by using the TBA
\begin{align} 
\varepsilon(\theta)   &=  \frac{16\sqrt{\pi^3}}{\Gamma(\frac{1}{4})^2} e^{\theta} - 2 \int_{-\infty}^\infty\frac{ \ln \left [ 1 +  \exp \{ - \varepsilon(\theta') \} \right ] }{\cosh (\theta-\theta' ) } \frac{d \theta'}{ 2 \pi} \, .
\label{TBA0} 
\end{align} 

The $TQ$ system
\begin{equation}
T(\theta)Q(\theta) = Q(\theta - i \pi /2) +Q(\theta + i \pi /2) \,, \label{TQ0}
\end{equation} 
and the $QQ$ relation 
\begin{equation} 
 Q(\theta + i \pi/2) Q(\theta -i \pi/2)= 1+ Q(\theta)^2  \, , \label{QQ0}
\end{equation}
impose
\be \label{quantQ0}
Q(\th_n \pm i \pi/2) = \pm i \,.
\ee
Again (\ref{QQ0}) around $\th_n$ forces $Q(\th + i \pi/2) =  i \pm   Q(\th)+\dots$ and $Q(\th - i \pi/2) =   - i \pm Q(\th)+\dots$ up to smaller corrections (dots). Therefore, the $TQ$ system imposes another quantization condition on the $T$ function 
\be \label{quantT0}
T(\th_n) =\pm 2\,.
\ee 
Similarly as in section \ref{secT}, we can derive also the relation
\be  \label{Tnua}
T(\th) = 2 \cos \left\{\frac{2 \pi }{\hbar}a \right\} \, .
\ee 
from which it follows that the period $a$ is also quantised
\be  \label{QuantA0}
\frac{1}{\hbar(\theta_n)}a(\th_n) = \frac{n}{2} \,, \qquad n \in \mathbb{Z}\,\,.
\ee 
This is exactly the condition used by \cite{BianchiConsoliGrilloMorales:2021}. Yet, here we have fixed the general limits of its validity as relying on specific forms of the $TQ$ and $QQ$ systems \eqref{TQ0} and (\ref{QQ0}) for the $N_f=0$ theory. It does not work in general, but we show in section \ref{gravity} the specific conditions for its validity for the $N_f=2$ theory.

\section{Floquet exponent through Hill determinant} \label{appHill}

Let us explain how to compute the Floquet exponent $\nu$ through the Hill determinant method. It suffices to consider the more general $N_f=2$ equation and change variable as $z = i y$, to get the general form
\be 
\frac{d^2}{dz^2} \psi + [\th_0 + \th_2  e^{2 iz}+\th_{-2} e^{-2 iz} + \th_1 e^{iz}+ \th_{-1} e^{-i z}]\psi = 0\,,
\ee 
with coefficients
\be 
\th_0 = p^2 \qquad \th_{\pm 2} = e^{2\th}\qquad \th_{\pm 1} = 2e^\th q_{1,2}\,.
\ee 
We search for Floquet solutions, with the property that
\be 
\psi_+(z+ 2\pi ) = e^{2 \pi \nu} \psi_+(z) \qquad \psi_-(z+2\pi ) = e^{-2 \pi \nu} \psi_-(z)\,,
\ee 
which implies they can be expanded in Fourier series as
\be 
\psi(z) = e^{\nu z} \sum_{n=-\infty}^\infty b_n e^{n i z}\,.
\ee 
From the equation we get the recursion
\be 
(\nu + i n)^2 b_n + \sum_{m=-2}^2 \th_m b_{n-m}=0\,.
\ee 
Dividing by $\th_0-n^2$ we get the following matrix with convergent determinant
\be
\left(\begin{array}{cccccccc}
\vdots &  &  &   &  & & \vdots\\
\cdots &\xi_{n,n-1} & 1 & \xi_{n,n+1} &\xi_{n,n+2} &0 & \cdots\\
\cdots  & \xi_{n+1,n-1} & \xi_{n+1,n} & 1 & \xi_{n+1,n+2} &\xi_{n+1,n+3}& \cdots\\
\cdots & 0 & \xi_{n+2,n}& \xi_{n+2,n+1}   & 1 &\xi_{n+2,n+3}& \cdots\\
\cdots & 0 & 0 & \xi_{n+3,n+1}  & \xi_{n+3,n+2} & 1&\cdots\\
\vdots &  &  &   &  & &\vdots\\
\\
\end{array}\right)\left(\begin{array}{c}
\vdots\\
b_{n-1}\\
b_{n}\\
b_{n+1}\\
b_{n+2}\\
\vdots\\
\\
\end{array}\right)=0\,,
\ee
with matrix elements
\be 
\xi_{m n} = \frac{- \th_{m-n}}{(m- i \nu)^2 - \th_0} \qquad \xi_{m,m} = 1\,.
\ee 
Defining $\mathcal{A}_n$ as the finite $2n+1 \times 2n +1 $ submatrix
\be
{\cal A}_{n}={\tiny{\left(\begin{array}{ccccccccccccccc}
1 & \xi_{-n,-n+1} & \xi_{-n,-n+2}  \\
\xi_{-n+1,-n}& 1 & \xi_{-n+1,-n+2}  \\
 \xi_{-n+2,-n} &\xi_{-n+2,-n+1} & 1  \\
\\
 &  &  & \vdots & \cdots&\\
\\
 &  &      & &  \chi_{-1} & 1 & \xi_{-1} & 0 & 0&\\
 &  &     & & \xi_{0,-2} & \xi_{0,-1} & 1 & \xi_{0,1} & \xi_{0,2}&\\
 &  &      && 0 & \xi_{1,-1}  &\xi_{1,0} & 1 & \xi_{1,2}&\xi_{1,3}\\
 &  &      &  &  &  &  &  &  & \cdots & \vdots&\\
\\
 &  &      &  &  &  &  &  &  &  & &   &  1 & \xi_{n-2,n-1} & \xi_{n-2,n}\\
 &  &    &  &  &  &  &  &  &  & &  &\xi_{n-1,n-2}  & 1 & \xi_{n-1,n} \\
 &  &      &  &  &  &  &  &  &  &  &  &  \xi_{n,n-2} &\xi_{n,n-1}  & 1
\end{array}\right)}}
\ee
\normalsize
and 
\be
\Delta(i \nu) = \lim_{n\to \infty} \mathrm{det}\, \mathcal{A}_n\,,
\ee
then by ordinary methods~\cite{whittaker_watson_1996} we arrive at the following relation
\be 
\Delta(i \nu) = \Delta(0) - \frac{\sin^2 (\pi i \nu)}{\sin^2 \pi \sqrt{\th_0}}\,.
\ee
The Floquet exponent is finally given as the root of the equation
\be 
\sin^2 ( \pi i \nu) = \Delta(0) \sin^2 \pi \sqrt{\th_0}\,,
\ee
or equivalently
\be 
\cosh (2 \pi \nu) = 1- 2 \Delta(0) \sin^2 \pi p\,.
\ee 
In particular, for $N_f=2$ the matrix elements of $\mathcal A_n$ are given by
\be 
\xi_{m,m\mp 2}^{(2)} = - \frac{e^{2\th}}{(m-i \nu)^2- p^2} \qquad \xi_{m,m\mp 1}^{(2)} = - \frac{2e^\th q_{1,2}}{(m-i \nu)^2- p^2}\,,
\ee 
while for $N_f=1$
\be 
\xi_{m,m-2}^{(1)} = - \frac{e^{2\th}}{(m-i \nu)^2- p^2} \qquad \xi_{m,m+1}^{(1)}  = - \frac{e^{2\th}}{(m-i \nu)^2- p^2} \qquad \xi_{m,m- 1}^{(1)}  = - \frac{2e^\th q_{1 }}{(m-i \nu)^2- p^2}\,.
\ee  

\section{Doubly Confluent Heun equation} \label{appDCHE}

\subsection{Equation maps}

Let us now show how the quantum SW ODEs \eqref{ODE0}, \eqref{ODEgau1}, \eqref{ODEgau2}, for $N_f=0,1,2$, can be reduced to particular cases of the \emph{doubly confluent Heun} equation:
\be 
\frac{d^2 w}{d z^2}+\left(\frac{\gamma}{z^2}+ \frac{\delta}{z}+ \epsilon \right)\frac{d w}{d z}+\frac{\alpha z - \bar{q}}{z^2}w = 0\,,
\ee 
whose general solution is given by Mathematica as\footnote{In Mathematica's notation, we let $\delta \leftrightarrow \gamma$ and set $\epsilon=1$.}
\be 
w = c_1 \text{HeunD}[\bar{q},\alpha ,\gamma ,\delta ,\epsilon ,z]+c_2 z^{2-\delta } e^{\frac{\gamma }{z}-z \epsilon } \text{HeunD}[\delta +\bar{q}-2,\alpha -2 \epsilon ,-\gamma ,4-\delta ,-\epsilon ,z]\,.
\ee 
By changing the independent variable as $z = e^y$
\be 
\frac{d^2 w}{dy ^2}+(\delta +\gamma  e^{-y}+e^y \epsilon -1)\frac{d w}{dy}+(\alpha  e^y-\bar{q}) w = 0\,,
\ee
and the dependent variable as
\be 
\psi(y) = \exp \left\{ \frac{1}{2} \left(\gamma  e^{-y}+(1-\delta ) y-\epsilon e^y  \right) \right\} w(y)\,,
\ee 
we obtain the following form
\be  \label{dchexp}
\frac{d^2 \psi}{d y^2}-\frac{1}{4}  \left[\gamma ^2e^{-2 y}+2 \gamma  (\delta -2) e^{-y}+ \left(2 \gamma  \epsilon +(\delta -1)^2+4 \bar{q}\right)+e^{ y} (2 \delta  \epsilon -4 \alpha )+\epsilon ^2 e^{2 y} \right] \psi(y) = 0\,.
\ee 

Now by comparing \eqref{dchexp} with the quantum SW curve for $N_f=2$ \eqref{ODEgau2}, we get an identification under the parameter dictionary
\be 
\ba 
\gamma &= \pm \frac{\L_2}{2\hbar} \qquad \epsilon=  \frac{\L_2}{2\hbar} \\
\delta &= \frac{2(1\pm m_2)}{\hbar} \\
\alpha &=\frac{1}{2\hbar^2}(\L_2\hbar-m_1 \L_2\pm m_2 \L_2) \\
\bar{q}&= \frac{1}{8\hbar^2}[-2 \hbar^2+ 8 u -8 m_2^2 \mp  8 m_2 \hbar \mp \L_2^2]\,,
\ea 
\ee 
or
\be 
\ba 
\gamma &= \pm \frac{\L_2}{2\hbar} \qquad \epsilon=  -\frac{\L_2}{2\hbar} \\
\delta &= \frac{2(1\pm m_2)}{\hbar} \\
\alpha &=\frac{1}{2\hbar^2}(-\L_2\hbar-m_1 \L_2\mp m_2 \L_2) \\
\bar{q}&= \frac{1}{8\hbar^2}[-2 \hbar^2+ 8 u -8 m_2^2 \mp  8 m_2 \hbar \pm \L_2^2]\,.
\ea 
\ee 
Also by comparing \eqref{dchexp} with the quantum SW curve for $N_f=1$ \eqref{ODEgau1}, with $y \to - y$, we get the parameter dictionary
\be 
\ba 
\gamma &=\pm \frac{\L_1}{\hbar}\\
\epsilon &= 0\\
\delta &=\frac{2 (\hbar\pm m_1)}{\hbar} \\
\alpha &= - \frac{\L_1^2}{4}\\
\bar{q} &= \frac{1}{4\hbar^2}[-\hbar^2+4 u -4 m_1^2\mp 4 m_1 \hbar]\,.
\ea 
\ee 
Finally by comparing \eqref{dchexp} with the quantum SW curve for $N_f=0$ \eqref{ODE0}, after also change of variable $y \to y/2$\footnote{Notice though that as for the $N_f=1,2$ theories in this paper, with respect to $N_f=0$ in~\cite{FioravantiGregori:2019} we use make the rescaling $\hbar \to \sqrt{2}\hbar$.},
we get the parameter dictionary
\be 
\ba 
\gamma &= \pm\frac{2\sqrt{2}\L_0}{\hbar}\qquad \epsilon=\frac{2\sqrt{2}\L_0}{\hbar}\qquad \alpha= \frac{2\sqrt{2}\L_0}{\hbar}\\q &= \frac{1}{4\hbar^2} [-\hbar^2\mp16\L_0^2 + 16u]\qquad \delta =2\,,
\ea 
\ee 
or
\be 
\ba 
\gamma &= \pm\frac{2\sqrt{2}\L_0}{\hbar}\qquad \epsilon=-\frac{2\sqrt{2}\L_0}{\hbar}\qquad
\alpha= -\frac{2\sqrt{2}\L_0}{\hbar} \\
\bar{q} &= \frac{1}{4\hbar^2} [-\hbar^2\pm16\L_0^2 + 16u]\qquad \delta = 2\,.
\ea 
\ee

\subsection{Eigenvalue expansion}

Another form for the doubly confluent Heun equation is possible, namely~\cite{Ronveaux:1995}
\be  \label{dchealt}
z \frac{d}{dz} z \frac{d}{dz} w+ \alpha \left(z+\frac{1}{z} \right) z \frac{d}{dz} w +\left[(\beta_1 + \frac{1}{2} ) \alpha z + \left( \frac{\alpha^2}{2} -\gamma\right)+(\beta_{-1}-\frac{1}{2}) \frac{\alpha}{z} \right] w = 0\,.
\ee
Transforming it in normal form, then changing independent variable as $z = e^y$ and transforming again into normal form, we get
\be 
-\frac{d^2}{dy^2}\psi + \left (\gamma +\frac{1}{4} \alpha ^2 e^{-2 y}+\frac{1}{4} \alpha ^2 e^{2 y}-\alpha  \beta_{-1} e^{-y}-\alpha  \beta_1 e^y \right)\psi = 0\,,
\ee
with the relation between the depend variable being
\be 
w(z) = e^{-\frac{\alpha}{2} \left(z-\frac{1}{z} \right) } \psi(y)\,.
\ee 
Then we have the following parameter map to \eqref{ODEgau2} for $N_f=2$
\be  \label{dchealtparmapNf2}
\alpha = \pm\frac{\L_2}{2 \hbar} = \pm 2 e^\th\qquad \beta_1 = \mp \frac{m_1}{\hbar}=\mp q_1 \qquad \beta_{-1} = \mp\frac{m_2}{\hbar}=\mp q_1\qquad \gamma = \frac{u}{\hbar^2} = p^2\,.
\ee 
To connect precisely to the authors~\cite{Ronveaux:1995}, we choose the lower sign convention and then get also the following relation between dependent variables
\begin{align} 
w_{\infty,1}(y) &\simeq (-2 e^{\th+y})^{-(\frac{1}{2}+q_1) } \simeq e^{e^{\th+y}}e^{-i \pi(\frac{1}{2}+q_1) }\psi_{+,0}(y) \qquad &y &\to +\infty \\
w_{\infty,2}(y)&\simeq e^{2 e^{\th+y}}(-2 e^{\th+y})^{q_1-\frac{1}{2}}\simeq e^{e^{\th+y}} \psi_{+,1}  \qquad &y &\to + \infty\,,
\end{align}
with
\be 
W[w_{\infty,2},w_{\infty,1}]=1\,.
\ee

For the study of eigenvalues of \eqref{dchealt} we can define the following parameter 
\be 
\lambda = \gamma - \alpha^2/2 \,.
\ee 
The DCHE has a countable number of eigenvalues, denoted as $\lambda_\mu(\alpha,\beta)$ with $
\mu \in \nu + \mathbb{Z}$, where $\nu$ is the Floquet characteristic exponent. Then, the eigenvalues have the expansion
\be 
\lambda_\mu(\alpha,\beta) = \mu^2 + \sum_{m=1}^\infty \lambda_{\mu,m}(\beta) \alpha^{2 m}	\,.
\ee
The first coefficient is explicitly~\cite{Ronveaux:1995}
\be \label{ExpEigenDCHE}
\lambda_{\mu ,1}(\beta) = - \frac{1}{2} + \frac{2 \beta_{-1}\beta_1}{4\mu^2-1}\,,
\ee
and, as explained in subsection~\ref{subsec:Ta}, it turns out to have the precise same expression as the leading instanton term in the gauge theory Matone relation \eqref{matoneGen}, under the parameter map~\eqref{dchealtparmapNf2}.

\section{Limit to lower flavours gauge theories} \label{appLimit}

Starting from some higher flavour $SU(2)$ gauge theory, it is possible to obtain a lower flavour gauge theories, under a suitable limit. In this appendix we show how to do it, for the $N_f=0,1,2$ theories considered in this paper. Moreover, we make some considerations also on the limits of the gauge theory periods.

\subsection{Limit from $N_f=1$ to $N_f=0$}

We notice that the Seiberg-Witten curve for $N_f=1$
\be 
y^2_{SW,1} =x^2(x-u) + \frac{\Lambda_1}{4}m_1 x -\frac{\Lambda_1^6}{64}\,,
\ee 
in the limit
\be 
\Lambda_1 \to 0 \, \qquad m_1 \to \infty \, \qquad \text{with}\quad  \Lambda_1^3 m_1 = \Lambda_0^4\,,
\ee 
flows to the Seiberg-Witten curve for $N_f=0$
\be 
y^2_{SW,0}= x^2(x-u) + \frac{\Lambda_0^4}{4}x\,.
\ee 
Similarly the $N_f=1$ quantum Seiberg-Witten curve:
\be 
-\hbar^2  \frac{d^2}{d y_1^2} \psi + \left[\frac{1}{16} \Lambda_1^3 e^{2y_1} +\frac{1}{2}\Lambda_1^{3/2}e^{-y_1}  +\frac{1}{2} \Lambda_1^{3/2} m_1 e^{y_1} +u\right] \psi = 0 \,,
\ee
if we let 
\be 
y_1 = y_0-\frac{1}{2}\ln m_1 \to - \infty\,,
\ee
becomes
\be 
-\hbar^2  \frac{d^2}{d y_0^{2}} \psi + \left[\frac{1}{16} \frac{\Lambda_1^3}{m_1} e^{2y_0} +\frac{1}{2}\Lambda_1^{3/2}m_1^{1/2}e^{-y_0}  +\frac{1}{2} \Lambda_1^{3/2} m_1^{1/2} e^{y_0} +u\right] \psi = 0\,,
\ee
that is, it precisely reduces to the $N_f=0$ equation:
\be 
-\hbar^2  \frac{d^2}{d y_0^2} \psi + ( \Lambda_0^2 \cosh y_0 + u) \psi = 0 \,.
\ee

We can also consider the limit on the integrability equation as follows. The Perturbed Hairpin IM ODE/IM equation is 
\be  \label{inteqNf=1}
-\frac{d^2}{d{y_1}^2} \psi(y_1) +[ e^{2\theta_1} (e^{2y_1}+e^{-y_1}) +2q e^{\theta_1}e^{y_1} + p^2_1]\psi(y_1) = 0 \,
\ee 
and it must reduce to the ODE/IM equation for the Liouville model studied in~\cite{FioravantiGregori:2019}
\be \label{inteqNf=0}
-\frac{d^2}{d{y_0}^2} \psi(y_0) +\{ e^{2\theta_0} [e^{y_0}+e^{-y_0}] + p^2_0\}\psi(y_0) = 0\,.
\ee 
In order for~\eqref{inteqNf=1} to go into~\eqref{inteqNf=0} we need to impose
\be 
\ba 
 e^{2\th_1+2 y_1} &\to 0
\\
 e^{2\th_1-y_1} = e^{2\th_0-y_0} \qquad
2 q e^{\th_1+y_1}&= e^{2\th_0+y_0} \qquad
p_1 = p_0\,,
\ea 
\ee 
or
\be
\ba 
q &= \frac{1}{2}\frac{e^{4\th_0}}{e^{3\th_1}}\\
y_1 &= y_0 -2 \th_0+2 \th_1 \,.
\ea
\ee
Now the limit requires $\th_1+y_1 \to -\infty$, that is 
\be 
 \qquad \th_1 \to - \infty\,,
\ee
and as a consequence 
\be 
q \sim e^{-3 \th_1} \to \infty \qquad \th_1 \to - \infty\,.
\ee 
We now consider also the limit on gauge periods. 
Through numerical experiments, we find the following relations, for $u,m_1,\Lambda_1 >0$, $\Lambda_1 \to 0$, $m_1 \to \infty$, $\Lambda_1^3 m_1 = \Lambda_0^4$
\begin{align}
    a_{1,1}^{(0)}(u,m_1,\Lambda_1) &\to -a_{0,D}^{(0)}(u,\Lambda_0) \\
    a_{1,1}^{(0)}(-u,m_1,\Lambda_1) &\to -a_{0,D}^{(0)}(-u,\Lambda_0)+a_{0}^{(0)}(-u+i 0,\Lambda_0)\\
    &=-i a_{0,D}^{(0)}(u,\Lambda_0) \\
     a_{1,2}^{(0)}(\pm u,m_1,\Lambda_1)+\frac{m_1}{\sqrt{2}} &\to \frac{1}{2}a_{0}^{(0)}(\pm u,\Lambda_0) \\
      a_{1,1}^{(0)}( e^{\pm 2\pi i/3}u,e^{\mp2\pi i/3}m_1,\Lambda_1)-\frac{e^{\mp2\pi i/3}m_1}{\sqrt{2}} &\to \frac{1}{2}a_{0}^{(0)}( u,e^{\mp i \pi/6}\Lambda_0) \\
        a_{1,1}^{(0)}( -e^{+ 2\pi i/3}u,e^{-2\pi i/3}m_1,\Lambda_1)-\frac{e^{-2\pi i/3}m_1}{\sqrt{2}} &\to e^{-2\pi i/3}[a_{0,D}^{(0)}( -u,\Lambda_0)- \frac{1}{2}a_{0}^{(0)}( -u+ i 0,\Lambda_0) ]\\
        a_{1,1}^{(0)}( -e^{- 2\pi i/3}u,e^{2\pi i/3}m_1,\Lambda_1)-\frac{e^{2\pi i/3}m_1}{\sqrt{2}} &\to e^{2\pi i/3}[-\frac{1}{2}a_{0}^{(0)}(- u+i 0,\Lambda_0) ]\,.
\end{align}

\subsection{Limit from $N_f=2$ to $N_f=1$}

Staring from the $N_f=2$ quantum Seiberg Witten curve
\be 
-\hbar^2  \frac{d^2}{d y_2^2} \psi + \left[\frac{1}{16} \Lambda_2^2 (e^{2y_2} +e^{-2y_2} ) + \frac{1}{2}\Lambda_2 m_1 e^{y_2} + \frac{1}{2}\Lambda_2 m_2 e^{-y_2} +u\right] \psi = 0\,,
\ee
under the limit
\be  \label{Nf2Nf0lim}
m_2 \to \infty \qquad \Lambda_2 \to 0 \qquad\Lambda_2^2 m_2 = \Lambda_1^3  \,,
\ee
we can set
\be 
y_2 = y_1+\frac{1}{2} \ln m_2 \to + \infty\,,
\ee
so that the equation becomes
 \be 
-\hbar^2  \frac{d^2}{d y_1^2} \psi + \left[\frac{1}{16} \Lambda_2^2\left ( m_2 e^{2y_1} +\frac{1}{m_2}e^{-2y_1} \right) + \frac{1}{2}\Lambda_2 \sqrt{m_2}m_1  e^{y_2} + \frac{1}{2}\Lambda_2 \sqrt{m_2} e^{-y_2} +u\right] \psi = 0\,,
\ee
and in the limit \eqref{Nf2Nf0lim} it reduces precisely to the $N_f=1$ quantum Seiberg-Witten curve equation:
\be 
-\hbar^2  \frac{d^2}{d y_1^2} \psi + \left[\frac{1}{16} \Lambda_1^3 e^{2y_1} +\frac{1}{2}\Lambda_1^{3/2}e^{-y_1}  +\frac{1}{2} \Lambda_1^{3/2} m_1 e^{y_1} +u\right] \psi = 0 \,.
\ee
In integrability variables, we impose the conditions that allow the limit of the differential equations
\be 
e^{2 \th+2y_2} = e^{2 \th_1+2 y_1} \,, \quad e^{\th_2+y_2} q_1= e^{\th_1+ y_1} q_1\,,\quad
2e^{\th_2-y_2} q_2 = e^{2 \th_1-y_1} \,,\quad e^{2 \th_2 - 2 y_2} \to 0\,,\quad p_2^2 = p_1^2\,.
\ee 
from which we deduce that we have to take the following limit
\be 
y_2 = - \th_2 + \th_1 + y_1 \qquad \th_2 \to - \infty \qquad M_2 = \frac{1}{2} e^{3 \th_1-2 \th_2} \to \infty\,.
\ee

\bibliographystyle{elsarticle-num}


\end{document}